\documentclass[12pt,twoside]{report}
\usepackage{epsfig,latexsym}
\begin{document}
\begin{titlepage}
\title{\bf Non-Equilibrium Field Theory Dynamics in Inflationary
Cosmology} 
\author{by \\ \\ Daniel Cormier}
\date{\vskip 2in Submitted in partial fulfillment of the degree \\ 
of Doctor of Philosophy at the \\ Department of
Physics, Carnegie Mellon University \\ April, 1998}
\maketitle
\end{titlepage}
\begin{abstract}
For simple inflationary models, we provide a consistent and complete 
scheme by which the macro-physical 
details of early universe inflation may be determined explicitly from the
underlying micro-physical theory.
We examine inflationary dynamics within the context of a fully 
renormalized, non-perturbative, and non-equilibrium
quantum field theory using the closed time path formalism.  
We study the non-perturbative dynamics using the self-consistent 
Hartree approximation and the large $N$ limit. 
In addition, we include gravitation dynamically by means
of the semi-classical approximation to Einstein gravity,
allowing the cosmological geometry to be determined self-consistently
by the evolution of the particle fields.
We develop a simple and numerically implementable regularization
and renormalization scheme for which the renormalization counterterms
contain no explicit dependence on the initial state.  

We concern ourselves first with the dynamics of new inflation models,
showing how the dynamics become quantum fluctuation driven.  We then describe
how these quantum fluctuations may be reassembled into an effective field
which behaves classically and acts as the source for the gravitational
background.  This allows us to determine the spectrum of primordial
density perturbations in a consistent manner.
We also examine explosive particle production after inflation through
the processes of spinodal decomposition and parametric amplification,
showing that the non-perturbative backreaction on the inflaton field
prevents catastrophic particle production.  We find a strong dependence
of particle production on the expansion rate and on the symmetry 
properties of the inflaton.

\end{abstract}
\subsection*{Preface}

Cosmology, the study of the beginnings of the Universe, has been
a subject of study and speculation for thousands of years.  
We now live in a remarkable era in which 
cosmology has become an analytical subject of research to be studied,
observed, and understood by science.  

Observations of the universe today have led to a description
of the universe back to times of less than a tiny fraction 
of a second after the beginning of time.  The most important
advance in the study of our universe has been the construction 
of the Standard Cosmological Model.  In analogy to the 
Standard Model of Particle Physics, this Model provides
the framework around which all known aspects of Cosmology
fit.  

The Standard Cosmology states that we live in an expanding
universe in which, on the largest of scales, each point is
like any other.  While the details of the cosmology, such 
as the exact size, shape, and total content of matter,
are yet to be determined, the uniformity of our universe 
has been verified to remarkable accuracy by relatively
recent observations.  

The remarkable success of this model leads one to ask a
very simple question: Why is the universe so uniform?
It is an incredible fact that science has arrived at an
answer on the basis of what is known about high energy
physics and gravitation.  The answer: Inflation.

The solution to such a profound question deserves scrutiny
and attention.  The purpose of this dissertation is
to examine what is known about inflation, to question
its assumptions, and to carefully analyze its consequences
and predictions.  

\vskip .5in

{\bf Acknowledgments}

This work was made possible by the hard work, dedication, 
and support of a large network of individuals.  At the 
focal point of this network are a number of people
I would like to acknowledge for their extraordinary efforts
on my behalf.  First and foremost I thank my advisor, Rich Holman, 
not only for leading me on the road to becoming a scientist,
but also for pointing out all the overgrown paths
and encouraging their exploration, even when it may
not be the safest route.  Rich has a true love of 
knowledge, which becomes contagious when combined with his
outstanding ability to communicate ideas at all levels.

I have been very lucky to have not just one advisor, but two,
as Dan Boyanovsky has often taken on that role.  Dan is full
of knowledge, but what has most impressed me is his 
eagerness to share his ideas and to listen to
the ideas of others.  He has uncommon patience, and the
ability to transform a vague concept into a detailed project.

I have also benefited a great deal from collaboration with Hector
de Vega.  My extensive email communications with Hector led to 
a much greater understanding of many of the details in this thesis.
I have enjoyed the opportunities I have had to work with him, and
I hope to continue our collaborations into the future.

I would also like to thank Ling-Fong Li and Martin Savage for the
important roles they played in my education.  I thank Bob Griffiths
for his curiosity and for making me explain even those aspects of
cosmology of which I am unsure, and Bob Nichol for providing the
perspective that only an observational cosmologist can provide.
I would be amiss if I did not thank the great staff of the physics
department, in particular Nancy Watson, Mary Shope, Lori Marcus,
Chuck Gitzen, Cheryl Wehrer, Pricilla Jarrett, Terri Jones, and Mary Placeway.  
A special bit of thanks goes to Joe Rudman, who was invaluable in
helping me recover my thesis after I deleted it from the hard drive.
For their discussions, I thank Prem Kumar and Anupam Singh, as well
as most of the other graduate students in the department who have 
provided a stimulating atmosphere in which to work.

Particular thanks go to good friends, who have provided so much 
support -- Mark Mattson, Phil Koran, Eldon Decker, and especially
James Walden and Elaine Kirkpatrick.  My greatest thanks go to my
wife Johanna who makes each and every day the best day of my life.

\noindent{\tiny G'dalm K'ding}

\baselineskip=25pt
\tableofcontents
\chapter{Introduction}

The ultimate goal of studies of the early universe is the ability
to accurately predict the macro-physical properties of the universe
from a micro-physical description.  This is a daunting task which
requires the relation of the full dynamics of the underlying quantum 
field theory to
the geometrical dynamics of gravity.  This evolution must then be
followed over an extraordinary dynamic range before the results may be
interpreted to yield concrete predictions for observable properties of
today's universe.

Such a program requires a number of features:
\begin{itemize}
\item The ability to evolve quantum fields dynamically without any
assumptions of an equilibrium state.
\item The ability to treat states of extremely high (non-perturbative)
energy densities.
\item The ability to evolve gravity in a manner self-consistent with
the evolution of the quantum fields.
\item The ability to interpret the results of the quantum field theory
dynamics.
\end{itemize}
In addition to these fundamental features, one must be concerned
with the renormalization aspects of the field theory and also have
means to explicitly evaluate the dynamics through analytical and
numerical means.  

For a class of simple models, we have developed such a program.  
In the following pages, we apply
this program to inflationary cosmology. 

\section{The Need for a Non-equilibrium Description}

Non-equilibrium field theory\cite{neq}, which describes the dynamical
evolution of quantum fields, has become an important tool in particle
physics and cosmology.  The ability to compute observables in highly
dynamical situations in which the results of ordinary equilibrium
field theory fail, is necessary in order to gain
a better understanding of a wide number of
phenomena, including phase transitions in condensed matter systems and
at particle accelerators\cite{neqnocosmo}.  
Perhaps the most important applications of non-equilibrium field
theory, however, are to phenomena of the early 
universe\cite{neqcosmo,paz,ringwald,leutwyler,weiss}.  

The reason
it is necessary to go beyond traditional equilibrium techniques in
describing cosmology is the extreme environment of our universe at
early times\cite{rocky,peebles}.  It is now well accepted that early in its history, our
universe was very hot with energy densities approaching the Planck
scale, while the universe was rapidly cooling due to
expansion.  As a result of the rapid cooling, it is expected that
matter goes through a number of phase transitions before we reach the
present state of our universe\cite{phasetrans,vilenkin1,mazenko,guthpi}.  

Unless the matter fields involved in these phase transitions interact
at a rate much greater than the cooling rate of the universe, the
evolution proceeds out of equilibrium.  This non-equilibrium
evolution may in turn have a number of important physical
consequences.  One example is baryogenesis\cite{baryogen}.  If one assumes that the
universe began in a state with equal numbers of baryons and
anti-baryons, then it is necessary to have some process in the history
of our universe which results in a biased production of baryons over
anti-baryons.  Such a process, which may have occurred during the
electroweak phase transition or at an earlier time, requires a
departure from equilibrium since baryons and anti-baryons are produced
in equal numbers in any equilibrium process.

Another example of a non-equilibrium process in the early universe is
the production of topological defects during a phase transition\cite{topo}.  Just
as the domain structure in a ferromagnet is very different when the
sample is cooled adiabatically through the Curie temperature compared
to a sample which is rapidly cooled, the production of topological
defects depends strongly on the cooling rate of the universe\cite{defform}.  Again,
equilibrium analyses will not provide an accurate picture of defect
formation.  

The phenomena to be studied here,
inflation\cite{guth,rev1,rev2,kolbook,lindebook} 
and the subsequent phase
of parametric amplification\cite{landau,branden,kls}, are likewise phenomena of a
non-equilibrium nature\cite{us1,frw1,desitter,boylee,frwbig,grav}.  
Inflation, and the near exponential expansion
which characterizes it, can only occur if the state of the inflating
universe differs from its equilibrium state.  Furthermore, it is the
dynamical evolution which allows inflation to end.  Parametric
amplification similarly relies on the rapid evolution of a quantum
field to produce a parametric resonance of field quanta.  A field in
equilibrium shows no such behavior.

Despite the non-equilibrium nature of these phenomena, the bulk of the
work treating the processes has utilized equilibrium techniques, most
notably the effective potential\cite{fieldtheory}.  The effective potential is the name
given to the equilibrium free energy of a system and while it is a very
useful quantity in determining a number of important properties, such
as the equilibrium ground state of the system, it is {\it
ineffective} in describing the dynamics of the
system\cite{ineffective}.  
It therefore
becomes necessary to move beyond equilibrium techniques and to analyze
the dynamics of these phenomena using non-equilibrium field theory.


\section{Previous Research in Non-equilibrium Quantum Field Theory}

With the recognition of the importance of non-equilibrium techniques,
a significant literature has developed which deals with techniques,
approximations, and applications of non-equilibrium field theory as
applied to the early universe.

The work of Hu and Calzetta and their collaborators\cite{neqcosmo} has focused
primarily on the formal issues of non-equilibrium field theory, with
an emphasis on statistical field theories.  The Langevin approach
derived from the influence functional has been shown to be a useful
tool in studying the early universe\cite{neqcosmo}.  It should be noted, however,
that statistical field theory remains an approximation to the full
quantum field theory.  Other important contributions of this research
group include the study of the non-perturbative $1/N$ expansion for
models with overall $O(N)$ symmetry.  In the context of the closed
time path treatment of full quantum field theory, this work has
included an analysis which places the large $N$ and other
approximations within the context of the computation of the full
two-particle irreducible (2PI) Green's functions\cite{huN,ramseyhu1,ramseyhu2}.  Much of their work
has been completed in expanding spacetimes and has included the issues
of renormalization as well as important applications.  Of particular
interest here is the work of Ramsey and Hu which treats the $O(N)$
theory in Friedmann-Robertson-Walker (FRW) spacetime including
renormalization\cite{ramseyhu1}, and applies the formalism to post-inflationary
parametric resonance\cite{ramseyhu2}, reproducing and verifying many of the results of
Refs.~\cite{frwprl,frwbig}, which are included here in Ch.~5.

Another research group with important contributions to the
subject is the Field Theory group at the Los Alamos National
Laboratory\cite{motola,largen,lasym,cooper}.  Their contributions include work on the $1/N$ expansion,
including a determination of the next-to-leading order contributions
in Minkowski spacetime as well as work on the leading order
approximation in FRW spacetime\cite{largen}.  Other important contributions are
provided by their non-equilibrium study of symmetry breaking\cite{lasym}.  Also of
importance to the present work is their analysis of the
non-perturbative Hartree and large $N$ approximations as compared to
exact results in a toy quantum mechanics problem, which provides some
indication as to the reliability of the approximations we use in our
analysis\cite{cooper}. 

The research which forms the foundation of the present work, however
is that of Boyanovsky, de Vega, and Holman and their collaborators\cite{us1,frw1,boylee,rolling}.
This important work includes the non-equilibrium study of symmetry
breaking\cite{rolling} and domain formation in scalar field theories\cite{boylee} and also
includes important contributions to the formalism of non-equilibrium
quantum field theory and its renormalization in FRW spacetime\cite{frw1}.  In
addition, the ground breaking work in collaboration with Lee and
Singh on reheating\cite{reheat} is an important piece of the puzzle, which along
with the present work, paints a complete picture of inflation from its
earliest stages to its end, and on to the subsequent phases of the
universe.

Of course, all of the recent work in this area is grounded in the
early pioneering work on non-equilibrium field theory\cite{neq} and the study of
quantum fields in curved spacetime\cite{birrell} which will be reviewed in further
detail in Ch.~2.


\section{Content of the Thesis}
In this work we develop the techniques of non-equilibrium quantum field
theory in conformally flat FRW spacetime, including non-perturbative
approximation techniques and the introduction of a novel
renormalization scheme which provides a simple interpretation of the
renormalizations and allows for numerical analysis of the dynamics of
quantum fields in both fixed and dynamically evolving FRW 
spacetimes\cite{frw1,frwprl,frwbig,desitter,grav}.
We then provide two new applications of these techniques to important
phenomena of the early universe:  new inflation\cite{desitter,grav} and post-inflationary
parametric amplification\cite{frwprl,frwbig}.

We begin our study in Ch.~2 with a review of FRW 
spacetime\cite{rocky,peebles,wald} and the
basics of field theory in an expanding universe\cite{birrell}.  We describe the
observational support for an FRW description including evidence for
the important properties of homogeneity and isotropy and also provide
the basics of the geometry of such a universe as
described by general relativity.  Next, we examine potential shortcomings of
this standard picture, in the horizon, flatness, and
relic problems, before moving on to the concept of inflation.  Here,
we describe how inflation provides solutions to the problems of FRW
cosmology and discuss the elements necessary to a successful
implementation of inflation.  As we shall see, this will lead to the
introduction of quantum field theory which in turn provides the
particularly interesting prospect that the same elements yielding
inflation may also result in the very metric perturbations from which
the large scale structure of our universe formed.  We then introduce
the basic concepts of field theory in FRW spacetime, including the
introduction of conformal coordinates and a short discussion of
renormalization in FRW spacetime.  Finally, we complete the chapter
with a short discussion of the formalism and physics of metric
perturbations\cite{mukhanov,bardeen}.  

We follow our discussion of field theory in FRW spacetime with an in
depth study of the techniques of non-equilibrium field theory in
Ch.~3.  First, we introduce the basic formalism before detailing the
models and approximations we will find useful in
our studies of the early universe.  We pay particular attention to the
non-perturbative Hartree\cite{hartree,frw1} and large 
$N$\cite{largeN,avan,mazzitelli,largen,desitter,ramseyhu1} 
approximations, deriving the
appropriate equations of motion and describing in detail the important
properties and limitations of these tools.  Next, we examine the
initial conditions which will be used to begin the evolution of the
dynamics when we apply our formalism to practical problems.  This oft
neglected detail will be seen to be of particular significance to both
formal and practical issues of non-equilibrium field theory in FRW
spacetime\cite{initcond}.  The renormalization of the field theory is another issue
of importance, especially when it will become necessary to numerically
solve the equations of motion in a particular model.  We show in
detail how we arrive at our renormalization scheme and discuss the
usefulness of our approach, while pointing out the important
connection between the initial conditions for the field theory and its
renormalization.  Finally, we provide a description of the
semi-classical approximation to gravity\cite{birrell}, which allows one to determine
the dynamics of the scale factor describing the classical expansion of
the universe self-consistently in terms of the evolution of the
quantum fields of that universe.  Such a description will also require
us to make a number of additional renormalizations.

Having dispensed with the details of the formalism of non-equilibrium
field theory in FRW spacetime, Chs.~4 and 5 deal with applications of
the techniques developed in Chs.~2 and 3.  First, we examine the
process of new inflation\cite{newinfl,newinfl2,desitter,grav}, beginning with a discussion of what the
basic scenario is, why it was introduced, and what the problems with
the scenario are.  This leads into the introduction of the particular
model of new inflation we study as well as a discussion of initial
conditions.  Next, we provide the analytical results of the dynamics of
new inflation, followed by the details and results of our numerical
study\cite{desitter,grav}.  The chapter is completed by a discussion of the dynamical
assembly of a classical object which may be interpreted as driving the
inflationary phase, followed by the details of the formation of
metric perturbations and the de-coherence properties of the quantum
field theory\cite{grav}.

Our second application, discussed in Ch.~5, is the analysis of the
phase of explosive particle production which may have occurred after a phase
of inflation
\cite{branden,kls,us1,mink,erice,par,frwprl,frwbig,tkachev,son,symrest,kaiser,yoshimura}.
We begin with a description of the basic scenario of
chaotic inflation.  We then analyze within our models the dynamics of phases
of particle production due to the phenomena of spinodal decomposition and 
parametric amplification, 
including both analytical and numerical results\cite{frwprl,frwbig}.  
We conclude with a discussion of the significance of our results in the
context of reheating.

Finally, we present our primary conclusions in Ch.~6, discussing the
importance of this work to our understanding of cosmology and its
interactions with particle physics.  We also discuss a number of other
applications such as baryogenesis and defect formation which may
benefit from a fully non-equilibrium analysis.

\chapter{Field Theory in FRW Spacetime}

\section{Introduction to FRW Spacetime}

The so-called standard cosmology presents a remarkably successful
picture of our universe\cite{peebles,weinberg}.  It is based upon 
two primary assumptions:
\begin{enumerate}
\item The universe is homogeneous and isotropic on large scales.
\item The geometry of the universe is determined by general
relativity.
\end{enumerate}
Taken together, these properties yield a Friedmann-Robertson-Walker
(FRW) cosmology \cite{friedmann,robertson,walker}\footnote{This definition 
of an FRW cosmology is somewhat
more general than that originally described by 
Friedmann\cite{friedmann}, for
which he made the additional assumption that the matter content of the
universe is a pressureless, ideal fluid.  While the properties of
homogeneity and isotropy require the matter to act as an ideal fluid,
we will allow for a more general equation of state.}.

The conditions of homogeneity and isotropy are very restrictive.  If
we define $h_{ij}$  to be the metric on the three-dimensional surface
defined by the spatial sections on the full four-dimensional
spacetime, such that the differential spatial distances are given by
\begin{equation}
d{\vec{l}}^2 = h_{ij}dx^i dx^j \; ,
\label{spatialmetric}
\end{equation}
then we can define the spatial Riemann curvature tensor 
$R^{(3) \; l}_{ijk}$.
Given the connection corresponding to the metric $h_{ij}$
$$
\Gamma^{k}_{ij} \equiv \frac12 h^{kl}\left[\partial_i h_{jl} +
\partial_j h_{il} - \partial_l h_{ij} \right] \; ,
$$
the Riemann tensor is
\begin{equation}
R^{(3) \; l}_{ijk} = \partial_j \Gamma^l_{ik} - \partial_i \Gamma^l_{jk}
+ \Gamma^m_{ik} \Gamma^l_{mj} - \Gamma^m_{jk} \Gamma^l_{mi} \; .
\label{spatialcurvature}
\end{equation}
The property of isotropy, which in fact, includes the property of
homogeneity if required at every spatial point, requires that the
spatial curvature defined in this way is constant.  That is,
\begin{equation}
R^{(3) \, kl}_{ij} = {\cal K} \left[\delta^k_i \delta^l_j - \delta^k_j
\delta^l_i \right] \; ,
\end{equation}
where the inverse metric $h^{ij}$ has been used to raise indices.

The constant ${\cal K}$ may take on any value, but there are just three
different cases corresponding to different topologies for the
universe.  If ${\cal K}$ is negative, then the spatial sections take on a
hyperbolic form.  These may be defined as the three-dimensional
surfaces embedded in a four dimensional Minkowski spacetime given by
the relation
$$
t^2 - x^2 - y^2 - z^2 = R^2 \; ,
$$
where $R$ is an arbitrary constant.  The second case is ${\cal K}=0$, which
means that at every point in the space, the curvature is zero.  The
spatial sections are flat and are therefore given by the three
dimensional Euclidean metric.  Finally, ${\cal K}$ may be positive,
corresponding to spatial sections of spherical form.  We may define
these by embedding them in a four dimensional Euclidean space.  They
are given by the relation
$$
w^2 + x^2 + y^2 + z^2 = R^2 \; .
$$
Again, $R$ is an arbitrary constant.

These three cases may be conveniently combined in the full four
dimensional spacetime metric
\begin{equation}
ds^2 = g_{\mu\nu}dx^\mu dx^\nu = dt^2 - a^2(t) d\vec{l}^2 \; ,
\label{metric}
\end{equation}
where $d\vec{l}$ is the differential element giving the spatial
sections and the scale factor $a(t)$ is, for now, an arbitrary
function of time.  The spatial metric may be written in spherical
coordinates as
\begin{equation}
d\vec{l}^2 = \frac{dr^2}{1-Kr^2} + r^2 d\theta^2 + r^2 \sin^2(\theta)
d\phi^2 \; .
\label{space}
\end{equation}
Here, $K\equiv {\cal K}/|{\cal K}|$ takes on the discrete values 
$K=-1,0,+1$ corresponding to
the cases of hyperbolic, flat, and spherical spatial sections
respectively.  We see that in the flat case of $K=0$, the spatial
metric may be written in Cartesian coordinates as
$$
d\vec{l}^2_{(K=0)} = dx^2 + dy^2 + dz^2 \; .
$$

The metric of the form given by (\ref{metric}) and (\ref{space})
describing a homogeneous and isotropic cosmology is referred to as a
Robertson-Walker metric\cite{robertson,walker}.  For a formal 
derivation of this important
piece of the standard model, we refer the reader to the book by
Wald\cite{wald}.  Further discussions are given in the books by
Weinberg\cite{weinberg} and Peebles\cite{peebles}.  The latter
reference also provides a nice discussion of the many equivalent
coordinatizations of the Robertson-Walker metric including conformal
coordinates, which are of particular importance and will be discussed
in Sec.~2.4.

The second of the primary assumptions of the standard cosmology
determines the dynamics of the scale factor $a(t)$ in terms of the
matter content of the universe.  General relativity\cite{weinberg,wald} 
is based upon
Einstein's equation relating the spacetime curvature to the
energy-momentum $T_{\mu\nu}$.  The full Riemann curvature tensor,
$R_{\alpha\beta\gamma}^{\delta}$, for the four dimensional spacetime
is defined in terms of the spacetime metric $g_{\mu\nu}$ in complete
analogy to the spatial curvature (\ref{spatialcurvature}).  
The connection on the spacetime
manifold is defined as
$$
\Gamma^{\rho}_{\mu \nu} \equiv \frac12 g^{\rho \sigma}\left[
\partial_\mu g_{\nu \sigma} + \partial_\nu g_{\mu \sigma} - 
\partial_\sigma g_{\mu \nu} \right] \; ,
$$
and the curvature tensor is
$$
R_{\mu \nu \rho}^{\sigma} = \partial_\nu \Gamma^\sigma_{\mu \rho} -
\partial_\mu \Gamma^\sigma_{\nu \rho}
+ \Gamma^\alpha_{\mu \rho} \Gamma^\sigma_{\alpha\nu} -
\Gamma^\alpha_{\nu \rho} \Gamma^\sigma_{\alpha \mu} \; .
$$
Appearing in Einstein's equation is the Einstein curvature tensor  
$G_{\mu\nu}$, constructed from the Ricci tensor 
$$
R_{\alpha\beta} \equiv R_{\alpha\mu\beta}^{\mu}
$$
and the Ricci scalar
$$
{\cal R} \equiv R^{\alpha}_{\alpha} \; .
$$
With these definitions, Einstein's equation reads
\begin{equation}
G_{\mu\nu} \equiv R_{\mu\nu} - \frac12 g_{\mu\nu} {\cal R} = -8\pi G_N T_{\mu\nu}
\; ,
\label{einstein}
\end{equation}
where the proportionality constant $G_N$ is Newton's gravitational
constant, which in natural units is related to the Planck mass by the
equation $G_N = 1/m_{Pl}^2$.

Given the metric (\ref{metric}) with (\ref{space}), Einstein's
equation reduces to just two independent equations of motion for the
scale factor:
\begin{eqnarray}
G^0_0 &=& -3\frac{\dot{a}^2}{a^2} - 3\frac{K}{a^2} = -8\pi G_N T^0_0
\; , \label{energyeqn} \\
G^1_1 &=& -2\frac{\ddot{a}}{a} - \frac{\dot{a}^2}{a^2} - \frac{K}{a^2}
= -8\pi G_N T^1_1 \; .
\label{pressureeqn}
\end{eqnarray}
Note that the non-diagonal elements of both $G_{\mu\nu}$ and
$T_{\mu\nu}$ are identically zero while the remaining diagonal
elements simply reproduce Eq.~(\ref{pressureeqn}).  We therefore see
that $T_{\mu\nu}$ must have the form of an ideal fluid with energy
density $\rho \equiv T^0_0$ and pressure $p \equiv - T^1_1$ in order
to be consistent with the homogeneity and isotropy conditions.

We can make a number of conclusions from the equations of motion
(\ref{energyeqn}) and (\ref{pressureeqn}).  First, if we assume that
$\rho > 0$, we see that $\dot{a}^2/a^2 > 0$ and the universe is
necessarily expanding (or contracting)\footnote{If $K=+1$, however, 
there is the
possibility of a single moment in time satisfying $8\pi G_N \rho/3 =
k/a^2$ for which $\dot{a}/a = 0$.}.  Second, if $p \geq 0$ then
$\ddot{a}/a < 0$ and the universe decelerates with time\footnote{In
fact, the universe decelerates for any equation of state satisfying $p
> -\rho/3$.}

We should note that for ordinary matter and radiation, the energy and
pressure densities, which satisfy the respective equations of state
$p=0$ and $p=\rho/3$, are functions of the scale factor and therefore
also of time.  In particular, ordinary matter has an energy density
inversely proportional to the physical volume and therefore changes
with the scale factor as
$$
\rho_m(t) \propto \frac{1}{a^3(t)} \; .
$$
Similarly the energy density of radiation changes as
$$
\rho_r(t) \propto \frac{1}{a^4(t)} \; ,
$$
which may be interpreted as a change in photon (or other massless
particle) density proportional to $1/a^3(t)$ with an additional
factor of $1/a(t)$ due to the redshifting of the photon frequency.

Given the properties of a dynamic, decelerating universe, if the
universe is presently expanding, as we will provide support
for below, then physical length scales were much smaller at earlier
times, with the expansion rate increasing as one goes back.  If one
extrapolates such a scenario of expansion far enough back, then one
arises at the conclusion that there is an instant of time at which all
physical length scales are reduced to zero.  This is the source of the
concept of the {\it big bang}, which should not be interpreted as an
explosion of matter into space as its name implies, rather it
represents the rapid expansion of space itself from an initial state
approaching zero size and infinite energy density.

Another property is that if the geometry of the universe is flat or hyperbolic with
$K=0$ or $K=-1$ respectively, then an expanding universe expands forever.  
However, in the spherical case with $K=+1$, then at some instant in
time $8\pi G_N \rho/3 = K/a^2$ and the expansion halts after which the
universe begins to contract.  In this case, the universe eventually
will return to a state approaching zero size and infinite energy density called
the {\it big crunch}.

A final property of the expanding universe described by the standard
cosmology that will be of particular interest is the existence of
particle horizons.  These exist because particles can travel at most
at the speed of light.  In particular, the furthest physical distance a light
signal may have traveled at any time is given by 
\begin{equation}
d_H(t) = a(t) \int^t \frac{dt^{'}}{a(t^{'})} \; .
\label{horizon}
\end{equation}
This quantity is called the horizon distance.  As we shall see, the
important point is that if the scale factor grows no faster than
linearly in $t$ as is the case in matter or radiation dominated 
cosmologies, then the horizon distance
decreases faster than the scale factor as one goes to earlier times.
This means that points which are in causal contact today may have been
causally disconnected at earlier times.  We will return to this
important point in the next section.

\subsection*{Observational Support for the Standard Cosmology}
The picture presented here of the standard cosmology has received
substantial observational support, which has elevated the simple model
to the level of a standard model\footnote{The terminology of the
standard model of cosmology was introduced by S.~Weinberg\cite{weinberg} 
in analogy
to the incredibly successful standard model of particle physics.  A
standard model has come to mean a simple, yet complete, theoretical
model which has held up under significant experimental or
observational testing.}.  The support comes in three different areas:
\begin{enumerate}
\item Support for the theory of general relativity, especially on
large scales.
\item Support for the expansion of the universe.
\item Support for the large scale homogeneity and isotropy of the
universe.  
\end{enumerate}
We now review the most significant evidence supporting a standard
cosmology description of our universe.  Complete discussions 
may be found in Refs.~\cite{weinberg,peebles,wald}.

Somewhat surprisingly, the element of support for the standard model
which is perhaps the least certain is the theory of general
relativity, largely because of the difficulties of verifying the
theory on large scales.  Of course, the success of general relativity
on small scales is well known, beginning with the prediction and
verification of the precession of the planet Mercury\cite{einstein,newcomb}.
To some extent, however, we must simply assume that general relativity
is the correct theory of gravity on all cosmological scales.

Observational evidence for the expansion of the universe, on the other
hand, is very extensive.  The expansion was first observed by 
Hubble\cite{hubble},
who found that light emanating from distant cosmological objects was
Doppler shifted toward the red part of the visible spectrum.  Such
redshifting indicates movement of objects away from our galaxy.
Furthermore, Hubble observed that the amount of redshifting increases
with the distance of the object.  In particular, Hubble determined
that there was a near linear relation between an objects redshift $1+z$,
defined as the ratio of the observed wavelength to the expected
wavelength for that source, and the distance $d$.  This relationship 
is expressed as Hubble's Law:
$$
z \simeq H_0 d \; ,
$$
where $H_0$ is Hubble's constant.  

While deficiencies in the ability of Hubble to accurately measure the
distances of cosmological objects led to a value of $H_0$ an order of
magnitude larger than recent determinations, Hubble's Law itself is
now well verified\footnote{This relation is in fact just the first term
in a more complete expansion of the form $z=H_0d+(q_0-1)(H_0d)^2/2+\cdots$.
Hubble's Law is accurate for distances $d < 1/H_0$.}.  We therefore 
know quite definitively that the
universe is undergoing expansion.

The final piece of evidence in support of the standard cosmological
model is direct observation of the homogeneity and isotropy of the
large scale universe.  The primary and most convincing observation in
this regard is provided by the Cosmic Background Radiation (CBR) which
was found to have an almost perfect blackbody spectrum at a
temperature $T=2.73K$ with little variation in this temperature with
the direction of the observation.  In fact, if one subtracts the
dipole component of the temperature variation, which may be explained
by the movement of our reference frame with respect to the rest frame
of the CBR, the temperature variation over the sky is of order $\delta
T/T \sim 10^{-4}$.  This means that at the time the photons making up
the CBR last interacted with ordinary matter, the universe was
homogeneous and isotropic to one part in 10,000.  To better understand
what this means, it helps to understand the source of the CBR.

Our standard picture of the early universe indicates that its
temperature was once many orders of magnitude hotter
than it is today.  At high enough temperatures, we know that electrons
dissociate themselves from protons.  At such high temperatures, the
radiation in the universe continually interacts with the charged
electrons and protons, and photons are in a state of thermal
equilibrium.  However, as the universe cools, the electrons combine
with the protons during a time called recombination (despite the fact
that these particles were never combined before this time).  Without
the charged particles on which to scatter, the photons become
effectively non-interacting.  As a result, they simply free stream
until they reach observers today.

In the meantime, the photons have redshifted with the expansion of the
universe, and since a redshifted thermal (blackbody) spectrum remains
thermal, we observe this cosmic background radiation as a near perfect
blackbody spectrum at a temperature $T_\gamma$ given in terms of the photon
temperature at recombination $T_{RC}$ by the relation
$$
T_\gamma = \frac{T_{RC}}{(1+z)} \; .
$$
The redshift at recombination is approximately $z \sim 1200$.

Since the ratio $\delta T/T$ is a constant for the cosmic radiation
background during its evolution from recombination to today, it is
clear that the recent observations of an isotropic and homogeneous CBR
indicate that the universe was very isotropic and homogeneous at the
time of recombination.

There are a number of other observations lending support to the model
of an isotropic and homogeneous universe.  The X-ray background, due
to unresolved high redshift sources has been determined to be
isotropic to about 5\%.  Similarly, the distribution of faint radio 
sources shows no preferential direction.  Furthermore, the peculiar
velocity field of galaxies on scales 50--100$Mpc$, which gives a
direct measure of density variation through the movement of galaxies
relative to the Hubble flow, provides a density contrast 
$\delta \rho/\rho \sim 0.1$.

Finally, galaxy distribution itself is found to be generally
isotropic, although it is somewhat disconcerting that we have observed
structure at even the largest observable scales.  If the universe is
truly isotropic on the largest scales then at some point we should
reach a limit beyond which there are no new, larger structures.

\section{Shortcomings of the Standard Cosmology}

As we have seen, the standard cosmology is very well founded.
However, it is not without problems.  Obviously, the universe is not
perfectly homogeneous since there are a number of individual 
structures ranging from planets to superclusters.  These presumably
form over time through gravitational instability, acting on the
initially small density variations we observe in the CBR.  However,
the standard cosmology does not account for these density
fluctuations, and it is fair to ask about their origin.

One of the most interesting challenges to the standard cosmology, on
the other hand, is the question of how the universe could be so
homogeneous in the first place.  This would seem like an empty
question given our assumptions of homogeneity were it not for the
concept of particle horizons.  As we saw in the preceding section,
the horizon size in a matter or radiation dominated cosmology decreases
relative to physical length scales as one goes to earlier times.  This
means that the horizon size at the time of recombination was much
smaller than it is today, and there would have been no way in which
correlations in the CBR in different directions could have formed.  

In fact, each causal region at the time of recombination subtends less 
than a single degree of the presently observable universe and one 
might expect each of these causal regions to have evolved somewhat 
independently, resulting in a microwave background temperature
spectrum which varies significantly from one region to the 
next.  The actually observed homogeneity may, of course, be
explained by simply assuming that the universe began in an
almost exactly homogeneous state independent of particle 
interactions.  However, we will see that there is a much more eloquent 
solution to the horizon problem based on a marriage of 
particle physics and cosmology.

A third problem of the standard cosmology is the relative flatness of
our current universe.  From the Friedmann equation (\ref{energyeqn}),
we see that there are two terms contributing to the expansion rate of
the universe.  The first is due to the energy density of matter and
fields and the second is due to the curvature of the universe.  It is
convenient to introduce the critical density:
$$
\rho_c(t) \equiv \frac{3}{8\pi G_N} \frac{\dot{a}^2(t)}{a^2(t)} \; .
$$ 
If $\rho=\rho_c$, then we necessarily have a critical, spatially flat
geometry while $\rho < \rho_c$ and $\rho > \rho_c$ correspond
respectively to the open, hyperbolic geometry ($K=-1$) and the
closed, spherical geometry ($K=+1$).  This behavior is often
summarized by the introduction of the single parameter $\Omega(t)
\equiv \rho(t)/\rho_c(t)$, with $\Omega = 1$ corresponding to the
critical universe.

While the present value $\Omega_0$ is not well determined, through
indirect measurement as a result of the successes of Big Bang
Nucleosynthesis combined with a number of more direct measurement
methods, it is restricted to the modest range $0.01 \leq \Omega_0 \leq
2.0$, while most cosmologists would agree that $\Omega_0$ is very likely
to lie in the range $0.2 \leq \Omega_0 \leq 1$\footnote{In fact the 
determination of $\Omega_0$ is an exciting and fast moving area of
observational research.  The current status of this research is 
perhaps best illustrated by two recent determinations.  The first
measures the acceleration of distant Type Ia supernovae, which are
assumed to be standard candles, and favors values of $\Omega_0 < 0.5$ while 
effectively ruling out the possibility of $\Omega_0=1$ unless a cosmological 
constant is added\cite{lowOmega}.  However, another observation which uses
number counts of distant x-ray clusters strongly favors 
$\Omega_0 > 0.5$\cite{highOmega}.  While progress is being made, there 
clearly is yet to be a consesus on $\Omega_0$.}.  

If, in addition to $\Omega$, we define the quantity $\Omega_R(t)
\equiv - K/\dot{a}^2(t)$, Friedmann's equation
(\ref{energyeqn}) becomes
\begin{equation}
\Omega(t) + \Omega_R(t) = 1 \; .
\label{omegaeqn}
\end{equation}
This equation displays a remarkable coincidence:  given the accepted
range of $\Omega$ and assuming $k \neq 0$ (that is, $\Omega \neq 1$), 
we see that the
contributions of $\Omega$ and $\Omega_R$ to unity in
Eq.~(\ref{omegaeqn}) are within an order of magnitude or so of each
other.  What makes this so remarkable is that these two terms have a
very different time dependence since $\Omega_R \propto 1/\dot{a}^2$
while $\Omega \propto 1/[\dot{a}^2(t)a(t)]$ for a matter dominated
cosmology and $\Omega \propto 1/[\dot{a}^2(t)a^2(t)]$ for a radiation
dominated cosmology.  For these two quantities, which have drastically
different time dependence, to be of the same
order at this instant of time despite the huge variation of $a(t)$ of
many, many orders of magnitude during the history of our universe is
extraordinary.  This coincidence is referred to as the flatness problem
since it does not exist if the geometry of the universe is flat since 
that would mean $\Omega_R=0$.  

Another way of posing this problem is to note that since $\Omega(t)$
decreases faster with expansion of the universe than $\Omega_R(t)$,
the condition that these quantities now be within a couple of orders
of magnitude of unity extrapolate back to the condition moments after
the Big Bang of $\Omega(t_{Pl}\simeq 10^{-43}s) \sim 1 \pm 10^{-60}$.  This indicates
an extreme fine tuning of the initial state of the universe.

The final problem of the standard cosmology arises as a result of the
Grand Unified Theories (GUT's) of particle physics.  These theories, which
unify all the known, non-gravitational forces of the universe,
typically predict the production of topological defects.  In
particular, all known GUT's predict the production of large numbers of
monopoles which would quickly come to dominate the energy density of the
universe.  Since these objects are topologically stable, they would
continue to dominate for all time, and the cosmology we observe today
could not exist.  A somewhat similar problem arises with the possible
production of domain walls in early universe phase transitions, which
would be expected to disrupt the isotropy of the universe.  Together,
such potential problems constitute the relic problem.

\section{Introduction to Inflation}

It is a remarkable fact that through the marriage of particle physics
and cosmology, a single mechanism has been discovered which addresses
all of the problems discussed in the preceding section:  Inflation.

Inflation\cite{guth} is a rapid expansion of the universe, characterized by
acceleration of the scale factor:
\begin{equation}
\frac{\ddot{a}(t)}{a(t)} > 0 \; .
\label{infcond}
\end{equation}
The proto-typical inflationary universe expands at an exponential
rate with $\dot{a}/a = H$ with $H$ constant. This is the so called de Sitter
universe\cite{wdesitter}.  

We now describe how such a universe solves the horizon, flatness, and
relic problems; we will reserve the fluctuation problem for later.  We
will use the exponentially expanding de Sitter universe with $a(t) =
a_0 \exp(Ht)$ as an example, although the arguments carry over, albeit
to a lesser effectiveness, for any case satisfying (\ref{infcond}).

For the de Sitter universe, the horizon size (\ref{horizon}) is a constant
$$
d_H^{(dS)} = a_0 e^{Ht} \int^t \frac{dt^{'}}{e^{Ht}} = \frac{1}{H} 
\; .
$$
Clearly, is such a situation, the physical length scales decrease
faster than the horizon distance as one goes back in time.  For a
concrete example, let us assume that the de Sitter phase lasts a time
$\tau$.  An initial region of radius $r$ is therefore expanded to size
$r \exp(Ht)$.  If we take this region to be homogeneous and causally
connected initially, then the expanded region will also be
homogeneous.  If we make $\tau$ large enough, we can therefore inflate
an initially small region of the universe creating a homogeneous
region which is large enough to fully contain our present universe.

We therefore see how inflation solves the horizon problem.  The
process ensures that distant regions of our observable universe which
could not have possibly interacted during the matter and radiation
dominated phases of expansion had interacted at a much earlier time
before inflation took place.  In this sense, inflation acts as a
mechanism to set the initial conditions for the standard cosmology
such that the present universe is homogeneous on large scales.  Note
that the horizon problem requires of the order of 60 e-folds of
inflation to take place in order to produce a homogeneous universe of
the present size.  In the exponential case, we therefore require that
$H\tau > 60$.

The simplest solution to the flatness problem is to remove the
contribution of $\Omega_R$ entirely from Eq.~(\ref{omegaeqn}), such
that $\Omega = 1$.  Such a contribution may be removed by reducing the
relative contribution of the curvature term in Friedmann's equation
(\ref{energyeqn}) with respect to the energy density.  The requirement
to do so is to have a component of the energy density which decreases
more slowly than $1/a^2(t)$.  If this requirement is satisfied then
sufficient expansion of the universe will reduce the contribution of
$\Omega_R$ relative to $\Omega$.  Having the universe dominated by
energy density satisfying this requirement is equivalent to the
condition $\ddot{a}/a > 0$, precisely the condition for inflation.

In simple terms, inflation solves the flatness problem by taking an
initial region which is small enough to have negligible curvature and
expanding it to the size of our present universe.  We also see that in
solving the flatness problem, inflation provides a powerful
prediction: $\Omega = 1$\footnote{There are a number of open inflation 
models which allow $\Omega < 1$ through a mechanism of false vacuum
tunneling.  However, these models are somewhat awkward in their 
construction and implementation.  See \cite{open1,open2}.}.

The relic problem is solved in a similar way to the flatness problem.
As monopoles act as ordinary matter, their energy density decays with
expansion at least as fast as $1/a^3(t)$.  It is therefore clear that
the same conditions required to solve the flatness problem will
necessarily dilute the number of monopoles to a negligible level.  

If it is also necessary to dilute the energy density in domain walls,
then this is also possible, although it may require a dominant
contribution to the energy density which falls more slowly then the
$1/a(t)$ dependence of the energy density of infinite domain walls.

Now that it is clear how inflation solves the horizon, flatness, and
relic problems of the standard cosmology, it is important to determine
under what conditions inflation occurs.  From Einstein's equations
(\ref{energyeqn}) and (\ref{pressureeqn}), we see that the requirement
$\ddot{a}/a > 0$ translates to the condition on the equation of state
of the universe of 
\begin{equation}
\rho + 3p < 0 \; .
\end{equation}
This requires that the pressure density of the universe is negative
with $p < -\rho/3$.

At first, such an equation of state (which breaks the strong energy
condition\cite{wald}) seems odd.  However, it is relatively easy to produce such
a situation in field theory.  
As an example, let there be a scalar field in Minkowski space with
Lagrangian 
$$
L = \partial_\mu \phi \partial^\mu \phi - V(\phi) \; .
$$
The energy density due to $\phi$ is
$$
\rho_\phi = \frac12 \dot{\phi}^2 + \frac12 (\vec{\nabla}\phi)^2 
+ V(\phi) \; ,
$$
while the pressure density is
$$ 
p_\phi = \frac12 \dot{\phi}^2 - \frac16 (\vec{\nabla}\phi)^2 
- V(\phi) \; .
$$
We see that if a situation arises in which the contributions of the
potential $V(\phi)$ dominate over the kinetic ($\dot{\phi}^2$) and
gradient ($\vec{\nabla}\phi^2$) terms in these quantities, we arrive at
an equation of state $p=-\rho$.  Such is the case when $\phi$ is
constant and different from its vacuum value.

This example also provides us with an example of the importance of
quantum field theory in producing inflation, since a
hydrodynamic description of the universe containing {\it classical} matter
does not allow an equation of state $p < 0$.  

It should be noted that the effect of a potential energy contribution
to the energy density is similar to that of a cosmological constant
$\Lambda$ in Einstein's equation:
$$
G_{\mu\nu} - \Lambda g_{\mu\nu} = -8\pi G_N T_{\mu\nu} \; .
$$
A cosmological constant dominating over the energy density of matter
will therefore produce inflation.  However, even in this case, quantum
field theory will be necessary to produce an end to inflation.  This
will be important in our study of new inflation in Ch.~4.

Finally, we return to the one problem of the standard cosmology we
have yet to address: the existence of small primordial density
fluctuations.  While inflation has the effect of homogenizing the
universe, it surprisingly also provides density fluctuations on
cosmological scales.  

The basic mechanism, which we will describe in more detail in
Sec.~2.5 after introducing the basics of field theory, is that
inflation takes quantum fluctuations on microscopic scales and
inflates them to cosmic scales.  It is these quantum fluctuations
which constitute the density fluctuations from which the structure of
our universe formed.

\section{Field Theory in FRW Spacetime}

Quantum field theory in curved spacetime is a rich and interesting
topic.  Here we restrict the discussion to the areas of primary
interest to the present study.  We present the basic elements of
scalar field quantization in FRW spacetime, providing the field
equations and describing the mode decompositions of the field.  We 
pay particular attention to the concept of conformal coordinates as
this will be of importance in constructing our renormalization scheme.
We complete this section with a short review of renormalization in
curved spacetime.  Extensions to fields of higher spin and to more
general spacetimes are provided in the literature\cite{birrell}.

The Lagrangian for a general scalar field in curved spacetime is 
\begin{equation}
L = \frac12 \nabla^\mu \Phi(x) \nabla_\mu \Phi(x) -
V\left[\Phi(x);x\right] \; ,
\label{genlagrange}
\end{equation}
where $\nabla_\mu$ is the usual covariant derivative and we allow for
a potential $V(\Phi;x)$ which may depend explicitly on the spacetime
coordinates due to the existence of curvature scalars such as the Ricci
scalar.  In FRW cosmologies, this explicit dependence is restricted to 
the time variable.  We will be primarily interested in the $\lambda
\Phi^4$ theory, which represents the simplest, non-trivial, interacting
scalar field theory.  In this case, the potential has the form
\begin{equation}
V\left[\Phi(x);t\right] = \frac12 \left[m^2 + \xi {\cal R}(t)\right]
\Phi^2(x) + \frac{\lambda}{4!}\Phi^4(x) \; .
\label{lphi4pot}
\end{equation}
Here, we have included a term which couples the scalar field directly
to the scalar curvature ${\cal R}(t)$ through the coupling constant $\xi$.
There are two values of $\xi$ which receive particular attention:
\begin{itemize}
\item minimal coupling with $\xi = 0$, and
\item conformal coupling with $\xi = 1/6$.
\end{itemize}
The importance of the former is obvious from its simplicity, while the
importance of the latter will be discussed when we introduce conformal
coordinates.  Of course, like any other coupling constants, without
any strong physical principle to tell us the value of $\xi$, we must
assume that this constant may take on any value and allow it to be
determined by measurement.  

The action for a scalar field in curved spacetime is simply
\begin{equation}
S = \int d^4x \sqrt{-g} L \; ,
\label{genaction}
\end{equation}
where $g \equiv \det g_{\mu\nu}$.  Variation of this action with
respect to the field $\Phi$ provides the field equation
\begin{equation}
\nabla^\mu \nabla_\mu \Phi(x) + V^{'}[\Phi(x);x] = 0 \; ,
\end{equation}
where the prime denotes differentiation with respect to $\Phi$.  For
the $\lambda \Phi^4$ theory in FRW spacetime, this becomes
\begin{equation}
\left[\nabla^\mu \nabla_\mu + m^2 + \xi {\cal R}(t) +
\frac{\lambda}{3!}\Phi^2(x) \right] \Phi(x) = 0 \; .
\end{equation}

In order to seek solutions of these field equations, it is convenient
to introduce a mode decomposition of the field $\Phi$.  Restricting
ourselves to a {\it free} scalar field in an FRW cosmology, we may write
\begin{equation}
\Phi(x) = \int d\tilde{\mu}(\vec{k}) \left[ a_{\vec{k}} u_{\vec{k}}(x) +
a_{\vec{k}}^\dagger u_{\vec{k}}^*(x) \right] \; ,
\label{modedecomp}
\end{equation}
where $a_{\vec{k}}$ and $a_{\vec{k}}^\dagger$ are respectively the
annihilation and creation operators for the field quanta.  These
operators satisfy the relations
\begin{eqnarray}
\left[a_{\vec{k}},a_{\vec{k}^{'}}^\dagger \right] &=&
\delta_{\vec{k}\vec{k}^{'}} \; , \\
\left[a_{\vec{k}},a_{\vec{k}^{'}} \right] &=& 0 \; , \\
\left[a_{\vec{k}}^\dagger,a_{\vec{k}^{'}}^\dagger \right] &=& 0 \; .
\end{eqnarray}

We may make a separation of the modes into their space and time
dependent pieces
\begin{equation}
u_{\vec{k}}(x) =  {\cal Y}_{\vec{k}}(\vec{x}) U_{\vec{k}}(t) \; .
\end{equation}
The mode functions $U_{\vec{k}}(t)$ satisfy the mode equations
\begin{equation}
\left[\frac{d^2}{dt^2} + 3\frac{\dot{a(t)}}{a(t)} \frac{d}{dt} + k^2 
+ m^2 + \xi {\cal R}(t) \right] U_{\vec{k}}(t) = 0 \; .
\end{equation}
The spatial functions ${\cal Y}_{\vec{k}}(\vec{x})$ satisfy the relation
\begin{equation}
\nabla^i \nabla_i {\cal Y}_{\vec{k}}(\vec{x}) = - (k^2 - K) {\cal
Y}_{\vec{k}}(\vec{x}) \; ,
\label{Yeqn}
\end{equation}
where $\nabla_i$ is the covariant derivative on the three dimensional
spatial metric $h_{ij}$ of Eq.~(\ref{spatialmetric}).

The solutions to (\ref{Yeqn}) have been determined for each of the
three FRW cosmologies.  In the flat, $K=0$ case, they are simply
\begin{equation}
{\cal Y}_{\vec{k}}(\vec{x}) = e^{i \vec{k} \cdot \vec{x}} \quad (K=0) 
\; ,
\end{equation}
with $\vec{k} = (k_x,k_y,k_z)$.  For the closed and open cases
respectively, these functions may be written as
\begin{equation}
{\cal Y}_{\vec{k}}(\vec{x}) = \Pi^{\pm}_{kJ}(\chi) Y_J^M(\theta,\phi)
\quad (K = \pm 1) \; .
\end{equation}
Here, $\vec{k} = (k,J,M)$ and the $Y_J^M$ are the usual spherical
harmonics.  For $K=+1$, $k$ takes on positive integer values with $J =
0,1,\dots,k-1$, while for $K=-1$, $k$ takes on any positive indefinite
value while $J$ is a positive indefinite integer.  In either case, $M$
takes on the values $M=-J,-J+1,\dots,J$.

The function $\Pi^{-}_{kJ}(\chi)$ is
\begin{equation}
\Pi^{-}_{kJ}(\chi) = \left\{ \frac{\pi}{2} k^2(k^2+1)\dots(k^2+J^2)
\right\}^{-1/2} \sinh^J\chi \left(\frac{d}{d \cosh \chi}\right)^{J+1}
\cos \left(k\chi\right) \; ,
\end{equation}
and $\Pi^{+}_{kJ}(\chi)$ is given by the same expression taking $\chi
\to -i \chi$.

Finally, we return to the measure $d\tilde{\mu}(\vec{k})$ appearing in the
mode decomposition (\ref{modedecomp}).  We have,
\begin{equation}
\int d\tilde{\mu}(\vec{k}) = \left\{ \begin{array}{c}
\int \frac{d^3k}{(2\pi)^3} \quad  (K=0) \\ \\
\sum_{k,J,M} \quad (K=+1) \\ \\
\int dk \sum_{J,M} \quad (K=-1)
\end{array}
\right.
\end{equation}

In what follows and throughout, we restrict ourselves to the case of a
spatially flat ($K=0$) FRW cosmological model.  This is done for convenience
and because an initially curved space takes on the properties of a flat
universe soon after inflation begins.

\subsection*{Conformal Coordinates}
Quantum field theory may be formulated using a number of different
coordinate systems in which the field equations take on different
forms, while the physics described is the same.  Of these, few have
particular physical significance.  However, the system of conformal
coordinates in flat FRW spacetime is of importance due to its
relationship with Minkowski spacetime.  

A conformal transformation is one in which the spacetime metric is
multiplied by an overall factor.  That is $g_{\mu\nu} \to C(x)
g_{\mu\nu}$.  The new spacetime metric $C(x) g_{\mu\nu}$ is said to be
conformal to $g_{\mu\nu}$.  If we examine the flat FRW metric
$$
ds^2 = dt^2 - a^2(t)\left[dx^2+dy^2+dz^2\right] \; ,
$$
we see that if we define the conformal time variable
$$
\eta \equiv \int^t \frac{dt^{'}}{a(t^{'})} \; ,
$$
then the metric becomes
\begin{equation}
ds^2 = a^2(\eta)\left[d\eta^2-dx^2-dy^2-dz^2\right] \; ,
\label{metrica}
\end{equation}
which is conformal to the Minkowski metric.  This is of particular
importance because we may define a vacuum in conformal coordinates in
which the adiabatic limit $\dot{a}/a \to 0$ is equivalent to the
Minkowski vacuum.  This presents us with a foundation on which we can
build our vacuum state.

In addition to working in conformal time, it is convenient to make a 
conformal rescaling of the field by writing 
\begin{equation}
\Phi(\vec{x},t) = \frac{\chi(\vec{x},\eta)}{a(\eta)} \; .
\label{confres}
\end{equation}
The action in terms of the new field $\chi(\vec{x},\eta)$ becomes,
after integration by parts and dropping the surface term
\begin{equation}
S = \int d^3x d\eta \left\{ \frac12 \chi^{'2} - \frac12 \left(
\vec{\nabla} \chi \right)^2 - {\cal V}(\chi) \right\} \; ,
\label{confaction}
\end{equation}
where primes denote differentiation with respect to conformal time
$\eta$ and 
$$
{\cal V}(\chi) = a^4(\eta) V\left(\frac{\chi}{a(\eta)}\right) -
a^2(\eta) \frac{{\cal R}}{12} \chi^2 \; ,
$$
with the Ricci scalar ${\cal R}=6a^{''}(\eta)/a^3(\eta)$.  Here we see that
the action (\ref{confaction}) takes on the same form as that of a
scalar field in Minkowski spacetime with a potential ${\cal V}(\chi)$.

If $\Phi(\vec{x},t)$ is a field with $\lambda \Phi^4$ potential
(\ref{lphi4pot}), then the conformal potential is
$$
{\cal V}(\chi) = \frac{a^2(\eta)}{2}\left[m^2 + \left(\xi-\frac16
\right) {\cal R}(\eta) \right] \chi^2 + \frac{\lambda}{4!} \chi^4 \; .
$$
From this, we recognize an important property of the action
(\ref{confaction}).  If $\Phi(\vec{x},t)$ is a massless ($m^2=0$),
conformally coupled ($\xi=1/6$) field, then the action for the
conformal field $\chi(\vec{x},\eta)$ (\ref{confaction}) is invariant
under conformal transformations.

\subsection*{Renormalization in Curved Spacetime}

Renormalization is a delicate issue in curved spacetime for a number
of reasons\cite{birrell}.  First, whereas in Minkowski spacetime the zero point
energy has no physical meaning, such an energy becomes important in
the gravitational field equations.  A divergent zero point energy
therefore cannot be removed in a more or less arbitrary manner as in
Minkowski spacetime, rather it must be subtracted in such a way to
yield a sensible finite result.  Second, one must be concerned about
preserving the property of covariant conservation of the
energy momentum tensor.  

We will require the following properties of our renormalization
scheme:
\begin{enumerate}
\item It must result in numerically amenable renormalized equations of
motion.
\item It must preserve covariant conservation of the energy momentum
in the limit that the regulator is removed.
\item All subtractions must be removed through {\it time-independent}
renormalizations of the parameters of the theory.
\item The zero point energy must vanish in the true vacuum of the
theory. 
\end{enumerate}

The standard methods of dimensional regularization and covariant point
splitting are inherently covariant schemes. However, they are not well
suited to numerical analysis.  An exception to this rule is provided
by the recently constructed renormalization scheme of Baacke,
Heitmann, and P\"atzold\cite{baacke} which uses dimensional
regularization combined with a clever separation of the divergences
from the finite terms to provide a numerically amenable and inherently
covariant regularization scheme.  

We choose to turn to a non-covariant scheme for which we have to
verify covariant conservation after the fact.  We introduce a large
momentum cutoff to regularize divergent integrals, and show that the
resulting cutoff dependent terms may be used to renormalize the
parameters of the theory.  The scheme satisfies all of the required
properties and in particular allows for efficient numerical analysis
of the resulting equations of motion.  

\section{Metric Perturbations}

One of the most important aspects of inflationary theory is its
ability to produce primordial metric perturbations on scales relevant
to the formation of large scale structure and to the temperature
inhomogeneities of the CBR.  Here, we review the elements of the
formalism of cosmological perturbations that will be important for our
study of inflation.  We utilize the gauge invariant formalism of
Mukhanov, Feldman, and Brandenberger\cite{mukhanov}, to which we refer the
reader for a more complete and detailed exposition of the subject.

Obviously the treatment of cosmological perturbations requires a
departure from the purely homogeneous FRW cosmology.  We therefore
write the metric in terms of a background FRW spacetime denoted by
$^{(0)}g_{\mu\nu}$ and a perturbation metric $\delta g_{\mu\nu}$ which
will be considered to be small:
$$
ds^2 = \, ^{(0)} \! g_{\mu\nu} dx^{\mu}dx^{\nu} + \delta g_{\mu\nu}
dx^{\mu}dx^{\nu} \; .
$$
Of course, there are 10 independent components making up $\delta
g_{\mu\nu}$.  Those may be categorized in terms of their
transformation properties under spatial coordinate transformations.
In particular, the perturbations may transform as spatial scalars,
vectors, or tensors.  In conformal time, we may write\footnote{Here,
our notation differs somewhat from \cite{mukhanov} since we continue to
represent the spatial metric by $h_{ij}$ ($\gamma_{ij}$ in
\cite{mukhanov}).  We use $k_{ij}$ for the tensor perturbation ($h_{ij}$ in
\cite{mukhanov}).} 
\begin{equation}
\delta g_{\mu\nu} = - a^2(\eta) \left( 
\begin{array}{cc} 
-2\phi \quad & B_{|i}-S_i \\
B_{|i}-S_i \quad & -2\left(\psi h_{ij} - E_{|ij}\right) + F_{i|j} + F_{j|i} 
+ k_{ij} 
\end{array}
\right) \; ,
\end{equation}
where $h_{ij}$ is the spatial background metric (\ref{spatialmetric})
and $f_{|i}$ is the covariant derivative with respect to $h_{ij}$.
The vector components satisfy the constraints 
$$
S_i^{|i} = F_i^{|i} = 0 \; ,
$$
while the tensor component satisfies
$$
k^i_i = h_{ij}^{|j} = 0 \; ,
$$
to yield a total of 10 independent functions.

There exists a gauge freedom in general relativity which corresponds
to a choice of coordinates on the curved spacetime, and it is
important to note that the metric perturbation is not invariant under
these gauge transformations.  A general, infinitesimal coordinate
transformation is given by
\begin{equation}
x^\alpha \to \tilde{x}^\alpha = x^\alpha + \xi^\alpha \; .
\label{trans}
\end{equation}
The transverse piece of the 3-vector $\xi^i$ satisfying $\xi^i_{T \,
|i} = 0$ contributes only to vector perturbations.  We will be
interested in the remaining pieces $\xi^0$ and $\xi$ where the latter
quantity is defined by the expression 
$$
\xi^{|i}_{|i} \equiv \xi^i_{|i} \; .
$$
Given a gauge transformation which preserves the scalar nature of
perturbations, with $\xi^i_T=0$, the four scalar functions of $\delta
g_{\mu\nu}$ transform as
\begin{eqnarray}
\tilde{\phi} &=& \phi - \frac{a^{'}}{a}\xi^0 - \xi^{0'} \; , \\
\tilde{\psi} &=& \psi + \frac{a^{'}}{a}\xi^0 \; , \\
\tilde{B} &=& B + \xi^0 - \xi^{'} \; , \\
\tilde{E} &=& E - \xi \; .
\end{eqnarray}
From these, it is possible to construct two gauge invariant
combinations.  The simplest are
\begin{eqnarray}
\Phi &=& \phi + \frac{1}{a} \left[ (B-E^{'}) a \right]^{'} \; , \\
\Psi &=& \psi - \frac{a^{'}}{a}(B-E^{'}) \; .
\end{eqnarray}

There exists, of course, a gauge freedom corresponding to the
transformation (\ref{trans}).  Of particular note is the longitudinal
gauge which is defined by $B=E=0$.  As a result, $\phi_l = \Phi$ and
$\psi_l = \Psi$ where the subscript denotes that the quantity is
evaluated in the longitudinal gauge.
 
We will be interested in metric perturbations due to fluctuations in a
scalar field $\varphi(\vec{x},t)$, which we separate into a
homogeneous, classical part $\varphi_0(t)$ and an inhomogeneous
fluctuation $\delta \varphi(\vec{x},t)$.  We write
$$
\varphi(\vec{x},t) = \varphi_0(t) + \delta \varphi(\vec{x},t) \; ,
$$
where it is assumed that $\delta \varphi \ll \varphi_0$.  For metric
perturbations which result from the fluctuations of a scalar field, it
turns out that $\Psi = \Phi$ and we may reduce the perturbation
equations to include just one of these gauge invariant quantities.  We
therefore focus on the dynamics of $\Phi$ which is referred to as the
Bardeen potential\cite{bardeen}.  In the longitudinal gauge, $\Phi$
may be identified with the Newtonian potential, the gradient of which
provides the gravitational acceleration of a test particle relative to
the global inertial coordinates.  The importance of this quantity is
due to its close relationship with the power spectrum per logarithmic
$k$ interval $|\delta_k(t)|^2$ which gives the amplitude of scalar
metric perturbations.  For modes which have physical wavelength larger
than the horizon scale, $k < a(t)d_h(t)$, this relation is simply
$$
|\delta_k(t)|^2 = k^3|\Phi_k(t)|^2 \; ,
$$
where $\Phi_k$ are the Fourier components of $\Phi$.

The power spectrum $|\delta_k(t)|^2$ is directly related to the
temperature variation of the cosmic microwave background radiation on
large angular scales\cite{mukhanov} (greater than one degree).
Assuming that the last scattering surface at recombination is
infinitely thin, this relation is simply
\begin{equation}
\frac{\Delta T(\theta)}{T} \simeq \frac13 
|\delta_{k_\theta}(t_{\mbox{rec}})|
\label{dTT}
\end{equation}
where $\theta$ is the angular separation corresponding to the
wavenumber $k_\theta$ at recombination.

The equation of motion for these Fourier modes of the Bardeen potential
is\cite{mukhanov}
\begin{equation}
\ddot{\Phi}_k(t) + \left[H(t) 
- 2\frac{\ddot{\varphi}_0(t)}{\dot{\varphi}_0(t)} \right]
\dot{\Phi}_k(t) + \left[\frac{k^2}{a^2(t)} + 2\left(\dot{H}(t) - H(t)
\frac{\ddot{\varphi}_0(t)}{\dot{\varphi}_0(t)} \right) \right]
\Phi_k(t) = 0 \; . 
\label{bardeen}
\end{equation}
Here, $H \equiv \dot{a}/a$ is the Hubble expansion rate.  The Bardeen
potential may also be directly related to the field fluctuations
$\delta \varphi$, or rather the gauge invariant combination $\delta
\varphi^{gi} = \delta \varphi + \varphi_0(B-E^{'})$, by the constraint
equation\cite{mukhanov}
\begin{equation}
\frac{d}{dt}(a\Phi_k) = \frac{4\pi}{M^2_{Pl}} a \delta \varphi_k^{gi}
\dot{\varphi}_0 \; .
\label{constraint}
\end{equation}
This gauge invariant scalar field fluctuation obeys the equation of
motion 
\begin{equation}
\left[ \frac{d^2}{dt^2} + 3H\frac{d}{dt} + \frac{k^2}{a^2} + {\cal
M}^2 \right] \delta \varphi_k^{gi} - 4 \dot{\varphi} \dot{\Phi}_k 
+ 2 V^{'}(\varphi_0) \Phi_k = 0 \; ,
\label{gauginv}
\end{equation}
where ${\cal M}$ is the (generally time dependent) mass and
$V(\varphi_0)$ is the potential for the classical field $\varphi$.

While the above equations are useful in determining the power spectrum
for modes of physical wavelengths greater than the horizon size, it is
not until these fluctuations re-enter the horizon long after
inflation has ended that they will be relevant to structure formation
or observations of the CBR.  A convenient way of progressing from the
spectrum of fluctuations during inflation to the spectrum of density
perturbations at re-entry is the conservation law\cite{mukhanov}
\begin{equation}
\xi_k = \frac{2}{3} \frac{\dot{\Phi}_k/H + \Phi_k}{1+p/\rho} +
\Phi_k \; ; \quad \dot{\xi}_k = 0 \; ,
\label{xivar}
\end{equation}
which is valid after the first horizon crossing of modes with
wavevector $k<a d_h$.

\chapter{Non-equilibrium Field Theory}

\section{Introduction to the Closed Time Path Formalism}

The out of equilibrium dynamics of a quantum field is obtained
from the Closed Time Path formalism of Keldysh
and Schwinger\cite{neq}.  The starting point for determining
the time evolution of a quantum system is the Liouville equation.
This equation determines the evolution of the quantum density 
matrix, 
$\hat{\rho}[\phi,\tilde{\phi};t]$, in terms of the Hamiltonian operator 
$\hat{H}(t)$ of the system.  In the Schr\"odinger picture,
in which operators are constant while the states change with time,
the Liouville equation reads
\begin{equation}
i\frac{\partial \hat{\rho} (t)}{\partial t} = [\hat{H}(t),\hat{\rho}(t)].
\label{liouville}
\end{equation}
Here, we allow for an explicitly time
dependent Hamiltonian $\hat{H}(t)$, as will be necessary to treat quantum 
fields in the time dependent background of an expanding universe.  

A formal solution of this equation may be written in terms of the
time evolution operator
$$
\hat{U}(t,t^{'}) \equiv \exp \left[-i \int_{t^{'}}^t
dt^{''}\hat{H}(t^{''}) \right] \; ,
$$
in the form
\begin{equation}
\hat{\rho}(t) = \hat{U}(t,t_0)\hat{\rho}(t_0)\hat{U}^{-1}(t,t_0).
\label{rhoevol}
\end{equation}
The quantity $\hat{\rho}(t_0)$ determines the initial state of the system.
One possible choice is to let the initial state be one of local 
thermodynamic equilibrium at a temperature $T=1/\beta$.
The unnormalized initial density matrix then takes the form
\begin{equation}
\hat{\rho}(t_0) = \exp\left(-\beta \hat{H}(t_0)\right) \; .
\label{rhoinit}
\end{equation}
While it is not obvious that this choice is the most appropriate initial
state for the highly non-equilibrium processes of early universe, it is
a very convenient and simple choice which includes both the case of
an initial finite temperature as well as that of an initial state at
zero temperature (the vacuum state) in the limit $\beta \to \infty$.
The latter state may be particularly appropriate for the dynamics after
an initial stage of inflation.  Of course, there exist other possible 
initial states.  

\begin{figure}
\epsfig{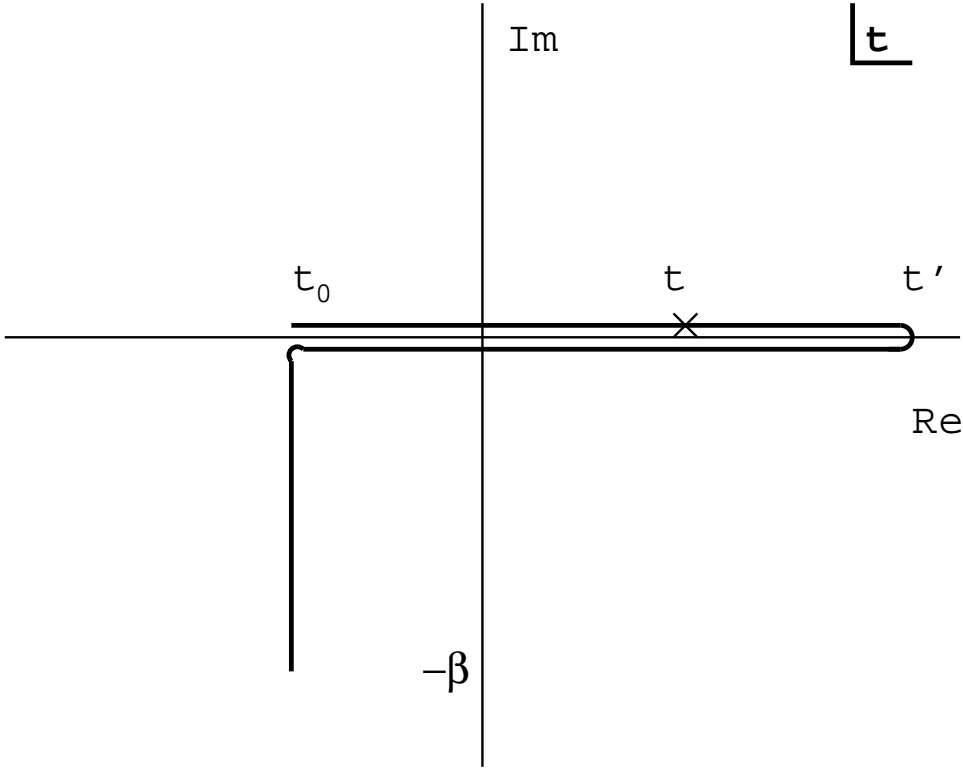}
\caption{The Closed Time Path contour in complex time.}
\label{contour}
\end{figure}

Given the evolution of the density matrix (\ref{rhoevol}), ensemble averages
of operators are given by the expression\footnote{From here on, we
drop the hats from the operators.} (again in the Schr\"odinger picture)
\begin{equation}
\langle{\cal O}(t)\rangle = 
\frac{Tr[\rho(t){\cal O}]}{Tr\rho(t)} =
\frac{Tr[\rho(t_0)U(t_0,t^{'})U(t^{'},t){\cal O}U(t,t_0)]}{Tr\rho(t_0)},
\label{expect}
\end{equation}
where we have inserted the identity, $U(t,t^{'})U(t^{'},t)$ with $t^{'}$ an
arbitrary time which may be taken to infinity.  The form of the initial
density matrix (\ref{rhoinit}) allows us to extend this expression into 
the imaginary time plane by writing
\begin{equation}
\langle{\cal O}(t)\rangle = 
\frac{Tr[U(t_0-i\beta,t_0)U(t_0,t^{'})U(t^{'},t){\cal O}U(t,t_0)]}{TrU(t_0-i\beta,t_0)},
\label{expectT}
\end{equation}
We give the following interpretation to this expression, reading from right 
to left.  The state is first evolved forward from the initial time $t_0$ to 
$t$ when the operator ${\cal O}$ is inserted.  We then evolve this state forward to time 
$t^{'}$, back again to the initial time, and down the imaginary axis to 
$t_0 - i\beta$.  This is normalized by the evolution in the imaginary time
direction from $t_0$ to $t_0 - i\beta$.  This process is depicted by the
contour in Fig.~\ref{contour}.  This is the Closed Time Path (CTP) 
formalism.  Notice that the expectation values
defined in this way are {\it in-in} expectation values as both the initial
and final (real) times appearing in the trace (\ref{expectT}) are given by $t_0$, 
and all observable quantities are determined with respect to the Hamiltonian 
at the initial time.

The actual evolution of various quantities in the theory are evaluated
by either constructing the appropriate Green functions \cite{us1}, or by
choosing an explicit ansatz for the functional form of the time dependent
density matrix so that the trace in (\ref{expect}) may be explicitly 
evaluated as a functional integral\cite{frw1}.  Within the approximations
we will use, the methods are equivalent.  For completeness, we will
describe both methods, beginning with the Green function approach.
We focus on the case of a real scalar field with no derivative couplings 
appearing in the Lagrangian.  The analysis will be generalized to the case 
of a vector $O(N)$ multiplet in a later section.

\subsection*{Computation of the Green Functions}

We introduce the generating functional with sources, $J^+$, $J^-$, and 
$J^{\beta}$, defined on each of the three legs of the contour in 
Fig.~\ref{contour}
\begin{equation}
Z[J^+,J^-,J^\beta] = Tr \left[ U(t_0-i\beta,t_0;J^\beta) U(t_0,t^{'};J^-)
U(t^{'},t_0;J^+) \right] \; ,
\end{equation}
where we have defined the sourced time evolution operators which for a
scalar field $\Phi$ have the form
\begin{equation}
U(t,t^{'};J) = \exp \left\{ -i \int_{t^{'}}^t dt^{''}\left\{ H(t^{''}) 
- \int d^3x J(x) \Phi(x) \right] \right\} \; .
\end{equation}
By inserting a complete set of field eigenstates, which we shall denote
as $\Phi$, between the time evolution operators, this may be written as
\begin{eqnarray}
Z[J^+,J^-,J^\beta] &=& \int {\cal D}\Phi^+ {\cal D}\Phi^- {\cal D}\Phi^\beta
\exp \left\{i \int d^3x \left[ \int_{t_0}^{t^{'}} dt \{{\cal L}[\Phi^+,J^+]
-{\cal L}[\Phi^-,J^-]\} \right. \right. \nonumber \\
&+& \left. \left. \int_{t_0}^{t_0-i\beta} dt {\cal L}[\Phi^\beta,J^\beta]
\right] \right\}
\label{genfunctional}
\end{eqnarray}
with the boundary conditions $\Phi^+(t_0) = \Phi^\beta(t_0-i\beta)$,
$\Phi^+(t^{'}) = \Phi^-(t^{'})$, and $\Phi^-(t_0) = \Phi^\beta(t_0)$.
Here, ${\cal L}[\Phi,J]$ represents the Lagrangian density for the scalar
field $\Phi$ with source $J$ and is defined in terms of the Lagrangian
$L$ as
$$
{\cal L}[\Phi,J] \equiv \sqrt{-g}L[\Phi] + J\Phi \; .
$$
It is clear that this generating functional corresponds to a path 
integral along the complex time contour of Fig.~\ref{contour}.

The Lagrangian density may now be broken up into pieces quadratic in
the field and into interaction terms, ${\cal L}_{\mbox{{\tiny int}}}$, 
which are of higher order in the
field variables.  The quadratic integrals may be evaluated explicitly,
leaving the following expression in terms of the currents and a set
of Green's functions\cite{niemi}:
\begin{eqnarray}
Z[J^+,J^-,J^\beta] &=& \exp \left\{i \int d^3x \left[ \int_{t_0}^{t^{'}} dt 
\{{\cal L}_{\mbox{{\tiny int}}}(-i\delta/\delta J^+) 
- {\cal L}_{\mbox{{\tiny int}}}(-i\delta/\delta J^-) \} \right. \right.
\nonumber \\
&+& \left. \left. \int_{t_0}^{t_0-i\beta} dt 
{\cal L}_{\mbox{{\tiny int}}}(-i\delta/\delta J^\beta)\right] \right\}
\nonumber \\
&\times & \; \exp \left\{\frac{i}{2} \int_c d^4x_1 d^4x_2 
J^a(x_1)G^{ab}(x_1,x_2)J^b(x_2)\right\}
\; . \label{partition}
\end{eqnarray}
Here, the subscript $c$ in the last integrals indicates that the time
portion of the integrals are to be taken over the contour in Fig.~\ref{contour}
and the superscripts $a$ and $b$ have values given by $+$, $-$, and $\beta$
corresponding to the location on this contour of the times $t_1$ and $t_2$.

This expression simplifies somewhat for the computation of the real time 
correlation functions in which we will be interested.  For such computations,
the contribution from terms in which $a=b=\beta$ in (\ref{partition}) cancels
between the numerator and denominator of Eq.~(\ref{expectT}).  In addition,
in the limit $t_0 \to -\infty$ the contributions from terms in which either
$a=\beta$ or $b=\beta$ vanish for real time correlation functions as a 
consequence of the Riemann-Lebesgue lemma.  We arrive at the generating
functional\cite{calzetta}
\begin{eqnarray}
Z[J^+,J^-,J^\beta] &=& \exp \left\{i \int d^3x \left[ \int_{t_0}^{t^{'}} dt 
\{{\cal L}_{\mbox{{\tiny int}}}(-i\delta/\delta J^+) 
- {\cal L}_{\mbox{{\tiny int}}}(-i\delta/\delta J^-) \} \right] \right\}
\nonumber \\
&& \times \; \exp \left\{\frac{i}{2} \int_c d^4x_1 d^4x_2 
J^a(x_1)G^{ab}(x_1,x_2)J^b(x_2)\right\}
\; ,
\end{eqnarray}
where $a$ and $b$ now take on the values $+$ and $-$.  It should be noted
that the finite temperature initial condition on $\hat{\rho}$ will enter 
through the boundary conditions on 
the Green's functions as we will see below.  

The Green's functions which appear along the contour may be written in 
terms of the Wightman function 
$G(x_1,x_2) = \langle \Phi(x_1) \Phi(x_2)\rangle$. 
They are
\begin{eqnarray}
G^{++}(x_1,x_2) &=& G(x_1,x_2)\Theta(t_1-t_2)+G(x_2,x_1)\Theta(t_2-t_1) \; , 
\label{G++} \\
G^{--}(x_1,x_2) &=& G(x_2,x_1)\Theta(t_1-t_2)+G(x_1,x_2)\Theta(t_2-t_1) \; , 
\label{G--} \\
G^{+-}(x_1,x_2) &=& G^{-+}(x_1,x_2) = -G(x_1,x_2) 
\label{G+-} \;.
\end{eqnarray}
These Green's functions are subject to the periodicity condition in imaginary
time:
\begin{equation}
G(\vec{x}_2,t_2;\vec{x}_1,t_0) = G(\vec{x}_1,t_0-i\beta;\vec{x}_2,t_2) \; .
\label{periodicity}
\end{equation}
The Wightman function $G(x_1,x_2)$ is constructed from the homogeneous solutions
of the quadratic form, subject to the periodicity conditions of 
Eq.~(\ref{periodicity}) and appropriate initial conditions.  We will explicitly
construct these solutions within the the Hartree and large $N$ approximations
for a scalar field below.  In addition, we will derive the equations of
motion for the field and its expectation value.  An important feature of the
Green's functions $G^{ab}$ of Eqs.~(\ref{G++})--(\ref{G+-}) is that at 
equal times, $t_1=t_2$, they are all equivalent and equal to $G(x_1,x_2)$.

\section{Field Equations in FRW}
We now derive the equations of motion for both a real scalar field 
and an $O(N)$ invariant scalar field in the vector representation.
As was discussed in the introductory chapter, it is often necessary to 
study inflationary dynamics using a non-perturbative framework.
To this end, we derive the equations of motion for the real and $O(N)$
fields within the self-consistent Hartree\cite{hartree} and the large 
$N$ limit\cite{largeN} respectively.  

These approximations have a variety of important features which make them
useful in the treatment of inflation.  
\begin{enumerate}
\item Within the realm of validity
of perturbation theory, the approximations reproduce the results of a 
one loop perturbative expansion.  
\item The approximations remain self-consistent even when amplitudes
become non-perturbatively large (order $1/\lambda$).
\item As we shall see, the approximations
are renormalizable for potentials containing terms up to fourth order
in the scalar field.  
\item The equations of motion preserve the
property of covariant conservation of the energy-momentum tensor.
\item The approximations result in local integro-differential evolution
equations which are quite amenable to integration on a computer.  
\end{enumerate}
In addition to these properties, the large
$N$ approximation preserves the Ward identities corresponding to the
$O(N)$ symmetry.  As a consequence, in the case of spontaneous symmetry
breaking, the late time dynamics will satisfy Goldstone's theorem.

We will assume, here and throughout, the spatially flat, 
homogeneous and isotropic expanding Robertson-Walker
cosmology described in Ch.~2.

\subsection{Real Scalar Field: Hartree Dynamics}

We now derive the equations of motion for a real scalar field with
Lagrangian (\ref{genlagrange}) within
the self-consistent Hartree approximation\cite{hartree}.
We will find it convenient to break the field $\Phi$ into its expectation
value plus a fluctuation about this value:
\begin{eqnarray}
\Phi(\vec{x},t) &=& \phi(t) + \psi(\vec{x},t) \; , \\
\phi(t) & \equiv & \langle \Phi(\vec{x},t) \rangle \; .
\label{separate}
\end{eqnarray}
Here, $\phi$ depends only on time due to space translation invariance
as is consistent with the metric (\ref{metric}).  By definition
$\langle \psi(\vec{x},t) \rangle = 0$; the expectation values
in these expressions are defined by Eq.~(\ref{expect}).

The Hartree approximation consists of replacing $\psi^{2n}$ by
$c_1\langle \psi^2 \rangle^{n-1}\psi^2 + c_2\langle \psi^2 \rangle^n$
and $\psi^{2n+1}$ by $c_3 \langle \psi^2 \rangle^n \psi$, where the
constant factors $c_i$ will be determined below.  In this
way, the Lagrangian (\ref{genlagrange}) is rendered quadratic in the
field variable $\psi$, and the functional integrals in (\ref{genfunctional})
become purely Gaussian.  

To determine the coefficient $c_1$, we first make the replacement
$$
\psi^{2n} \to {2n \choose 2} \psi^2 \langle \psi^{2n-2} \rangle
+ c_2 \langle \psi^2 \rangle^n \; ,
$$
where the numerical factor appearing before the first term on the
right hand side is determined by matching the n-point functions of
$\psi$.  
Next, we use Wick's theorem to write
$$
\langle \psi^{2n-2} \rangle = \frac{(2n-2)!}{2^{n-1}(n-1)!}
\langle \psi^2 \rangle^{n-1},
$$
from which we find 
$$
c_1 = \frac{(2n)!}{2^n (n-1)!} \; .
$$
Using Wick's theorem for the quantity $\langle \psi^{2n} \rangle$ then
gives us the value of $c_2$:
$$
c_2 = -\frac{(2n)!(n-1)}{2^n n!} \; .
$$
The coefficient $c_3$ is determined in the same way and is found to 
be given by
$$
c_3 = \frac{(2n+1)!}{2^n n!} \; .
$$
The Hartree factorization may be summarized as follows:
\begin{eqnarray}
\psi^{2n} &\to& \frac{(2n)!}{2^n (n-1)!} \langle \psi^2 \rangle^{n-1}\psi^2 
- \frac{(2n)!(n-1)}{2^n n!} \langle \psi^2 \rangle^n \; , \nonumber \\
\psi^{2n+1} &\to& \frac{(2n+1)!}{2^n n!} \langle \psi^2 \rangle^n \psi \; .
\label{factorization}
\end{eqnarray}

Given this factorization, any function $F(\Phi)$ becomes
\begin{equation}
F(\phi + \psi) = \sum_{n=0}^{\infty} \frac{1}{n!} \left(
\frac{\langle \psi^2 \rangle}{2} \right)^n \left\{
F^{(2n)}(\phi) + \psi F^{(2n+1)}(\phi) +\frac12 \left[ \psi^2 
- \langle \psi^2 \rangle \right] F^{(2n+2)}(\phi) \right\} ,
\label{factorF}
\end{equation}
where we use the notation
\begin{equation}
F^{(n)}(\phi) \equiv \frac{\delta^n}{\delta \phi^n} F(\phi) \; .
\end{equation}

The equations of motion for the scalar field are given by the tadpole
condition $\langle \psi \rangle = 0$ \cite{tadpole}.  
We find the equation for the mean field
\begin{equation}
\nabla^{\mu}\nabla_{\mu}\phi(t) + \left\langle V^{(1)}\left(\phi(t)
+\psi(\vec{x},t)\right) \right\rangle = 0 \; .
\label{phieqn}
\end{equation}
Defining the Fourier transform of the Wightman function by the expression
$$
G(\vec{x},t;\vec{x}^{'},t^{'}) = \int \frac{d^3k}{(2\pi)^3}
e^{i\vec{k}\cdot(\vec{x}-\vec{x}^{'})} G_k(t,t^{'}) \; ,
$$
where we have again used the property of space translation invariance.
The quantity $\langle \psi^2(t) \rangle \equiv -i G^{++}(x,x)$ is constructed
from the mode functions obeying the equation
\begin{equation}
[G_k(t,t)]^{-1} f_k(t) = 0 \; ,
\end{equation}
together with the boundary conditions (\ref{periodicity}), where 
the operator $[G_k]^{-1}$
is given by the quadratic form appearing in the generating functional 
(\ref{genfunctional}).
Explicitly, the $f_k(t)$ obey
\begin{equation}
\left[\frac{d^2}{dt^2} + 3\frac{\dot{a}^2(t)}{a^2(t)}\frac{d}{dt} + 
\frac{k^2}{a^2(t)} + \left\langle V^{(2)}\left(\phi(t) + 
\psi(\vec{x},t)\right)\right\rangle \right] f_k(t) = 0 \; .
\label{fkeqn}
\end{equation}
To this equation we will have to append an appropriately chosen set of
initial conditions for the mode functions $f_k(t)$.

Implementing the periodicity condition (\ref{periodicity}), we find
\begin{equation}
G_k(t,t^{'}) = \frac{i}{2\omega_k} \frac{1}{1-e^{-\beta \omega_k}}
\left[ f_k(t) f_k^*(t^{'}) + e^{-\beta \omega_k}f_k^*(t) f_k(t^{'})
\right] \; ,
\label{wightman}
\end{equation}
and we finally arrive at the expression
\begin{equation}
\langle \psi^2(t) \rangle = \int \frac{d^3k}{2(2\pi)^3} |f_k(t)|^2 
\coth \left(\frac{\omega_k}{2T}\right) \; ,
\label{psipsi}
\end{equation}
where $T$ is the initial temperature.  Note that in the $T \to 0$
limit, the hyperbolic cotangent has 
the value $1$.  The frequency $\omega_k$ is given by
\begin{equation}
\omega_k^2 = k^2 + a^2(t_0) \left\langle V^{(2)}\left(\phi(t_0) + 
\psi(\vec{x},t_0)\right)\right\rangle 
- \frac{a^2(t_0) R(t_0)}{6} \; .
\label{omega_k}
\end{equation}
We have included in this definition of the initial frequency a term proportional
to the Ricci scalar.
This term arises when one considers initial conditions corresponding to the adiabatic
vacuum state in conformal time\cite{frwbig} as was discussed in
Sec.~2.4.  Using such a vacuum state for the mode
functions leads to the following set of initial conditions on the $f_k(t)$:
\begin{eqnarray}
f_k(t_0) &=& \frac{1}{\sqrt{\omega_k}} \; , \nonumber \\
\dot{f}_k(t_0) &=& \left(-\frac{\dot{a}(t_0)}{a(t_0)} - i\omega_k \right) f_k(t_0) \; .
\label{initcond1}
\end{eqnarray}

This completes our set of equations of motion of the matter fields within the
Hartree approximation.  We will return to the specific case of the
$\lambda \Phi^4$ theory in Sec.~\ref{seclphi4}.

\subsection{$O(N)$ Scalar Field: Large $N$ Dynamics}

We now consider $\vec{\Phi}$ to be an $N$ component scalar field in the vector
representation of the $O(N)$ symmetry group.  The Lagrangian is
\begin{equation}
L[\vec{\Phi}] = \frac12 \nabla^{\mu}\vec{\Phi}(x)
\cdot\nabla_{\mu}\vec{\Phi}(x)-V\left(\vec{\Phi}(x)\cdot\vec{\Phi}(x)\right) \; ,
\label{lagrangianN}
\end{equation}
Again, we will separate the mean field value from the fluctuation.  Without
loss of generality, we choose the expectation value of the $i^{th}$ component 
of the field $\langle \Phi_i \rangle \equiv \sqrt{N} \bar{\sigma} \delta_{i,1}$ 
such that the mean value of all components of $\vec{\Phi}$ except the first vanish.
We therefore decompose the field as follows:
\begin{equation}
\vec{\Phi}(\vec{x},t) = \left(\sqrt{N}\bar{\sigma}(t)+\hat{\sigma}(\vec{x},t), 
\, \vec{\pi}(\vec{x},t) \right) \; .
\label{separateN}
\end{equation}
The quantity $\hat{\sigma}$ is the fluctuation in the ``sigma'' direction, whereas 
$\vec{\pi}$ is the fluctuation in the $N-1$ ``pion'' directions.  Since each of
the pion directions are equivalent, we may write
$$
\vec{\pi}(\vec{x},t) = \pi(\vec{x},t) (1,1,\cdots,1) \; .
$$

For simplicity, we will write the $O(N)$ generating functional keeping
only terms on the $+$ time contour in Eq.~(\ref{genfunctional}) and we
will drop the contour superscripts.  When we explicitly construct the
Green's functions below, it will be important to recall that these are
closed time path Green's functions.   

Defining the currents $J$ and $\vec{J}$ corresponding to the $\hat{\sigma}$
and $\vec{\pi}$ fields respectively, the generating functional becomes
\begin{eqnarray}
Z[J,\vec{J}] &=& \int {\cal D}\hat{\sigma} {\cal D}\vec{\pi}
\exp \left\{-i \int d^4x \sqrt{-g} \left[\frac12 
(\sqrt{N}\bar{\sigma}+\hat{\sigma}) 
\Box (\sqrt{N}\bar{\sigma}+\hat{\sigma}) - J \hat{\sigma} \right. \right. 
\nonumber \\
&+&  \left. \left. \frac12 \vec{\pi} \cdot \Box \vec{\pi} - \vec{J} \vec{\pi} 
+ N V\left(\frac{(\sqrt{N}\bar{\sigma} + \hat{\sigma})^2 + 
\vec{\pi}^2}{N} \right) \right] \right\} \; ,
\label{genfuncN}
\end{eqnarray}
with $\Box \equiv \nabla_\mu \nabla^\mu$.
We now make the functional pion integral Gaussian by introducing an auxiliary 
field $\chi \equiv \vec{\pi}^2/N$.  The potential term in (\ref{genfuncN}) 
may be written as
\begin{equation}
e^{-i \int d^4x N V\left(\frac{(\sqrt{N}\bar{\sigma} + \hat{\sigma})^2 
+ \vec{\pi}^2}{N}\right)} = \int {\cal D}\chi \, \delta 
\left(\chi - \frac{\vec{\pi}^2}{N}\right) e^{-i \int d^4x N 
V\left(\frac{(\sqrt{N}\bar{\sigma} + \hat{\sigma})^2}{N} + \chi\right) }\; .
\label{constraint2}
\end{equation}
It will be convenient to impose the delta function constraint of 
(\ref{constraint2}) by introducing a Lagrange multiplier field $\lambda$.
The generating functional now reads
\begin{eqnarray}
Z[J,\vec{J}] &=& \int {\cal D}\hat{\sigma} {\cal D}\vec{\pi} {\cal D}\chi 
{\cal D}\lambda
\exp \left\{-\frac{i}{2} \int d^4x \left[(\sqrt{N}\bar{\sigma}+\hat{\sigma}) 
\Box (\sqrt{N}\bar{\sigma}+\hat{\sigma}) \right. \right. \nonumber \\
&-& \left. \left. 2J \hat{\sigma} 
+ \vec{\pi} \cdot \Box \vec{\pi} - 2\vec{J} \cdot \vec{\pi} 
\right] \right\} \nonumber \\
& \times & \exp \left\{i \int d^4x \lambda\left(\chi-
\frac{\vec{\pi}^2}{N}\right)\right. \nonumber \\
&-& \left. iN \int d^4x V\left(\frac{(\sqrt{N}\bar{\sigma} 
+ \hat{\sigma})^2}{N} + \chi\right) \right\} \; .
\end{eqnarray}

Since we are interested in the dynamics to leading order in the $1/N$
expansion, we now Taylor expand the potential $V$ about the
quantity $\hat{\sigma}/\sqrt{N}$, keeping terms to order $1/N$.  
The resulting generating functional is quadratic in the $\vec{\pi}$
and $\hat{\sigma}$ fields.  Letting $\beta = \lambda/N$, defining 
the functions
\begin{eqnarray}
G^{-1}[\beta] &\equiv& \Box + 2\beta, \label{greensN}\\
G_\sigma^{-1}[\chi] &\equiv& \Box + 2V^{'}(\bar{\sigma}^2+\chi)
+ 2\bar{\sigma}^2 V^{''}(\bar{\sigma}^2+\chi), \\
F_J[\chi] &\equiv & J - \sqrt{N}\bar{\sigma}\left(\Box 
+ 2V^{'}(\bar{\sigma}^2+\chi)\right),
\end{eqnarray}
and performing the Gaussian integrals, we arrive at the functional
\begin{eqnarray}
Z[J,J_{\pi}] &=& \int {\cal D}\chi {\cal D}\beta \exp \left\{i(N-1)
\left[\frac{i}{2} Tr \log G^{-1}[\beta] 
+ \frac12 J_{\pi} G[\beta] J_{\pi} \right] \right\} \nonumber \\
&& \times \exp\left\{i\left[\frac{i}{2}Tr \log G_{\sigma}^{-1}[\chi]
+ \frac12 F_J[\chi] G_{\sigma}[\chi] F_J[\chi] \right] \right\}
\nonumber \\
&& \times \exp \left\{iN \int d^4x \left[\beta \chi 
- V(\bar{\sigma}^2+\chi) \right] \right\}.
\label{genN2}
\end{eqnarray}
Since each of the pion directions is equivalent, we have set each of the 
$N-1$ components of the current $\vec{J}$ equal to $J_{\pi}$.  

Taking $N$ to be large, we can now evaluate the $\chi$ and $\beta$
integrals in (\ref{genN2}) via the saddle point method.  Varying
the exponentials with respect to $\beta$ and $\chi$, we find the 
respective stationary point values $\bar{\chi}$ and $\bar{\beta}$:
\begin{eqnarray}
\bar{\chi} &=& -iG[\beta] - J_{\pi}G[\beta]G[\beta]J_{\pi}, 
\label{stationarychi} \\
\bar{\beta} &=& V^{'}(\bar{\sigma}^2+\chi) - \frac{\delta}{\delta \chi}
\left[\frac{1}{2N} F_J[\chi]G_{\sigma}[\chi]F_J[\chi] \right],
\label{stationarybeta}
\end{eqnarray}
where the ${\cal O}(1/N)$ contribution is dropped in the
latter variation in determining $\bar{\beta}$.

For consistency, we
drop the ${\cal O}(1/N)$ terms to arrive at the generating functional
for normalized $n$-point functions
\begin{eqnarray}
W[J,J_{\pi}] &=& \frac{1}{N} \ln Z[J,J_{\pi}] \nonumber \\
&=& i\left[\frac{1}{2} Tr \log G^{-1}[\bar{\beta}] 
+ \frac12 J_{\pi}G[\bar{\beta}]J_{\pi} + \int d^4x \bar{\beta}\bar{\chi}
- \int d^4x V(\bar{\sigma}^2 + \bar{\chi}) \right] \nonumber \\
&&+ \frac{i}{2N} F_J[\bar{\chi}] G_{\sigma}[\bar{\chi}] F_J[\bar{\chi}] \; .
\label{WJJpi}
\end{eqnarray}
We now impose the trivial tadpole relation $\langle \hat{\sigma} \rangle = 0$,
which yields the equation of motion for the zero mode $\bar{\sigma}$:
\begin{equation}
0 = \langle \hat{\sigma} \rangle \equiv \left.
-i \frac{\delta W[J,J_{\pi}]}{\delta J} \right|_{J=0} =
i \sqrt{N}\left[\Box + 2V^{'}(\bar{\sigma}^2+\bar{\chi})\right]\bar{\sigma} \; .
\label{tadpole}
\end{equation}
From this, we see that we may replace $F_J[\bar{\chi}] = J$ and, because
the corresponding contribution to $W[J,J_{\pi}]$ and to $\bar{\beta}$
is of ${\cal}(1/N)$ with respect to the leading order contributions, 
the current $J$ decouples completely from the leading order dynamics.

To leading order, the generating functional $W[J,J_{\pi}]$ becomes
\begin{equation}
W[J_{\pi}] = iN\left[\frac{1}{2} Tr \log G^{-1}[\bar{\beta}] 
+ \frac12 J_{\pi}G[\bar{\beta}]J_{\pi} + \int d^4x \bar{\beta}\bar{\chi}
- \int d^4x V(\bar{\sigma}^2 + \bar{\chi}) \right] \; .
\label{WJpi}
\end{equation}
The two point function is given by
\begin{equation}
\langle \pi^2 \rangle = (-i)^2 \left.
\frac{\delta^2 W[J_{\pi}]}{\delta J_{\pi} \delta J_{\pi}}\right|_{J_{\pi}=0} = 
-i G[\bar{\beta}],
\end{equation}
where, setting $J_{\pi}=0$,
\begin{eqnarray}
G^{-1}[\bar{\beta}] &=& \Box + 2\bar{\beta} \;, \\
\bar{\beta} &=& V^{'}(\bar{\sigma}^2+\bar{\chi}) \;, \\
\bar{\chi} &=& -i G[\bar{\beta}] \; .
\end{eqnarray}
The field $\sigma$ obeys the equation of motion given by Eq.~(\ref{tadpole}).

We now recall that the Green's function, (\ref{greensN}), is defined on
the CTP contour of fig. (\ref{contour}) and that we must impose the
periodicity condition (\ref{periodicity}).  The results of this
procedure [see Eq.~(\ref{wightman})] 
are as follows
\begin{equation}
\left[ \Box + 2 V'\left(\bar{\sigma}^2(t)+\langle \pi^2(t) \rangle\right)\right]
\bar{\sigma}(t) = 0 \;,
\label{zeromodeN}
\end{equation}
\begin{equation}
\langle \pi^2(t) \rangle = \int \frac{d^3k}{2(2\pi)^2}|f_k(t)|^2
\coth \left(\frac{\omega_k}{2T}\right) \;,
\label{pisq}
\end{equation}
where $\omega_k$ is defined by 
\begin{equation}
\omega_k^2 = k^2 + 2 a^2(t_0) V^{'}\left(\bar{\sigma}^2(t_0) 
+ \langle \pi^2(t_0)\rangle \right) 
- \frac{a^2(t_0) {\cal R}(t_0)}{6} \; ,
\label{omega_kN}
\end{equation}
and the mode functions obey the equation
\begin{equation}
\left[ \Box + 2 V'(\bar{\sigma}^2(t)+\langle \pi^2(t) \rangle)\right]
f_k(t) = 0 \;.
\label{modesN}
\end{equation}
To this set of equations of motion we add the initial conditions 
on the mode functions, which again take the form (\ref{initcond1}) with
the frequencies $\omega_k$ now defined by Eq.~(\ref{omega_kN}).

\subsection{Density Matrix Formalism}

An alternative method of computing the equations of motion without explicit
computation of the Green's functions is to make an {\it ansatz} for the 
form of the density matrix and compute its time evolution via the 
Liouville equation (\ref{liouville}).  Expectation values of operators
may then be evaluated explicitly using Eq.~(\ref{expect}).  To ensure
that such a computation is tractable, it is necessary to choose a 
Gaussian form for the density matrix.  We will find that this choice
is consistent with the approximations we will use to study the dynamics.

We restrict our analysis to a real scalar field.  The case of an
$N$-component field is completely analogous.
With the metric (\ref{metric}) and Lagrangian 
density (\ref{genlagrange}), the momentum operator conjugate to $\Phi(x)$ 
is defined as
\begin{equation}
\Pi(x) \equiv \sqrt{-g}\dot{\Phi}(x) \; .
\label{conjmomentum}
\end{equation}
The Hamiltonian may therefore be written as 
\begin{equation}
\hat{H}(t) = \int d^3x \left[\frac{\Pi^2(x)}{2 a^3(t)} 
+ \frac{a(t)}{2}\left(\vec{\nabla}\Phi(x)\right)^2
+ a^3(t) V\left(\Phi(x)\right)\right] \; .
\label{hamiltonian}
\end{equation} 
In the Schr\"odinger picture, the canonical momentum operator at an
arbitrary time $t_0$ has the representation 
\begin{equation}
\Pi(\vec{x},t_0) = -i \frac{\delta}{\delta \Phi(\vec{x},t_o)} \; .
\end{equation}

As before, we decompose the field $\Phi(x)$ into its expectation
value $\phi(t)$ and a fluctuation term $\psi(x)$ as in Eq.~(\ref{separate}).  
In terms of the density matrix $\hat{\rho}(t)$,
\begin{equation}
\phi(t) = \langle \Phi(\vec{x},t) \rangle \equiv 
Tr \left[\hat{\rho}(t) \Phi(\vec{x}) \right] \; .
\end{equation}
Again, we have by definition
\begin{equation}
\langle \psi ( \vec{x},t) \rangle = 0 \; .
\end{equation}
In terms of these new variables, the Hamiltonian becomes
\begin{eqnarray}
\hat{H}(t) &=& \int d^3x \left\{
- \frac{1}{2 a^3(t)} \frac{\delta^2}{\delta \psi^2(x)} 
+ \frac{a(t)}{2} \left( \vec{\nabla} \psi(x)\right)^2 \right. \nonumber \\
&+& \left. a^3(t) \left[ V\left(\phi(t)\right) + V^{'}\left(\phi(t)\right)
\psi(x) + \frac12 V^{''}\left(\phi(t)\right)\psi^2(x) + \cdots \right] 
\right\} \; . \nonumber \\
& & \;
\label{ham}
\end{eqnarray}
Keeping terms only to quadratic order in $\psi(x)$ corresponds to the
first order term in a loop expansion.  Alternatively, the higher order
terms represented by the ellipses of (\ref{ham}) may be approximated
as in Eqs.~(\ref{factorization}), leading to the Hartree approximation.  
We will return to this point shortly.

It will be convenient to introduce a Fourier representation for the 
Hamiltonian.  Letting $\Omega$ represent the spatial volume, we define
\begin{equation}
\psi(\vec{x},t) \equiv \frac{1}{\sqrt{\Omega}} \int \; d^3x \,
\psi_{\vec{k}}(t) e^{-i \vec{k} \cdot \vec{x}} \; .
\end{equation}
Defining the effective frequencies
\begin{equation}
W_k^2(t) \equiv \frac{\vec{k}^2}{a^2(t)} + V^{''}\left(\phi(t)\right) \; ,
\end{equation}
the Hamiltonian (\ref{ham}) becomes, to quadratic order ($1$-loop), 
\begin{equation}
\hat{H}(t) = \Omega a^3(t) V\left(\phi(t)\right) 
+ \sum_{\vec{k}} \left\{ - \frac{1}{2a^3(t)} 
\frac{\delta^2}{\delta\psi_{\vec{k}}(t)\delta\psi_{-\vec{k}}(t)}
+ \frac{a^3(t)}{2} W_k^2(t) \psi_{\vec{k}}(t) \psi_{-\vec{k}}(t) \right\}
\; .
\end{equation}

We now introduce an explicit Gaussian {\it ansatz} for the form of the 
density matrix:
\begin{eqnarray}
\hat{\rho}\left[\Phi,\tilde{\Phi};t\right] & = & \prod_{k=0}^{\infty} 
\rho_k(t), \nonumber \\
\rho_k(t) & = & {\cal N}_k(t) \exp \left\{- \frac12 A_k(t) \Phi_k \Phi_{-k}
- \frac12 A_k^*(t) \tilde{\Phi}_k \tilde{\Phi}_{-k} 
- B_k(t) \Phi_k \tilde{\Phi}_{-k}\right. \nonumber \\
&+&\left. i \pi_k(t) \left[\Phi_{-k} - \tilde{\Phi}_{-k}\right] \right\}.
\label{ansatz}
\end{eqnarray}
This particular form is dictated by the Hermiticity of the density matrix,
$$
\rho^{\dagger}[\Phi,\tilde{\Phi};t] = \rho^*[\tilde{\Phi},\Phi;t] \; ,
$$
which also leads to the result that the mixing parameter $B_k(t)$ is real.
We choose the factorized form since there is no mode-mode mixing in
the Gaussian approximation.

The non-equilibrium evolution of the system is now determined by the 
Liouville equation (\ref{liouville}), which reads
\begin{eqnarray}
i \frac{\partial \hat{\rho}\left[\psi,\tilde{\psi};t\right]}{\partial t}
&=& \sum_{\vec{k}} \left\{ -\frac{1}{2a^3(t)} \left[ 
\frac{\delta^2}{\delta\psi_{\vec{k}}\delta\psi_{-\vec{k}}}
- \frac{\delta^2}{\delta\tilde{\psi}_{\vec{k}}\delta\tilde{\psi}_{-\vec{k}}}
\right] \right. \nonumber \\
& + & \left. a^3(t) V^{'}\left(\phi(t)\right)\sqrt{\Omega}\delta_{\vec{k},0}
\left(\psi_{\vec{k}}-\psi_{-\vec{k}}\right)\right. \nonumber \\
& + & \left. \frac{a^3(t)}{2}W_k^2(t)
\left(\psi_{\vec{k}}\psi_{-\vec{k}} 
- \tilde{\psi}_{\vec{k}}\tilde{\psi}_{-\vec{k}}\right)\right\} 
\hat{\rho}\left[\psi,\tilde{\psi};t\right] \; .
\end{eqnarray}
Inserting our explicit ansatz for the density matrix yields the equations 
of motion for the parameters:
\begin{eqnarray}
\frac{\dot{\cal N}_k(t)}{{\cal N}_k(t)} & = & \frac{-i}{2a^3(t)} 
\left(A_k(t)-A_k^*(t)\right) \equiv \frac{A_{kI}(t)}{a^3(t)} \; , \\
\dot{A}_k(t) & = & -i \left[ \frac{A_k^2-B_k^2}{a^3(t)} - a^3(t) W_k^2(t)\right]
\; , \\
\dot{B}_k(t) & = & -i \frac{B_k(t)}{a^3(t)}\left(A_k(t)-A_k^*(t)\right)
\equiv 2 \frac{B_k(t)A_{kI}(t)}{a^3(t)} \; .
\end{eqnarray}
Here, we have introduced the notations $A_{kR}$ and $A_{kI}$ for the real
and imaginary parts of $A_k$ respectively.

There are two important notes to make regarding these equations:
\begin{itemize}
\item An initial pure state with $B_k(t_0)=0$ remains pure.
\item The time evolution is unitary as indicated by the relations
\begin{equation}
\frac{d}{dt}\left(\frac{B_k(t)}{A_{kR}(t)}\right)  =  0 \; ,
\label{AkRBk} 
\end{equation}
\begin{equation}
\frac{d}{dt}\left(\frac{{\cal N}(t)}{\sqrt{A_{kR}(t)+B_k(t)}}\right) = 0\; .
\end{equation}
\end{itemize}

The relation (\ref{AkRBk}) suggests that we introduce a new function 
\begin{eqnarray}
{\cal A}_{kR} & \equiv & C_k A_{kR}(t) = D_k B_k(t) \; , \\
{\cal A}_{kI} & \equiv & A_{kI}(t) \; ,
\end{eqnarray}
where the constants $C_k$ and $D_k$ depend on the choice of initial
conditions.  These definitions yield the following Ricatti-type
equation of motion for the quantity ${\cal A}_k \equiv {\cal A}_{kR}
+ i{\cal A}_{kI}$:
\begin{equation}
\dot{\cal A}_k(t) = -i\left[\frac{{\cal A}_k^2(t)}{a^3(t)}
-a^3(t)W_k^2(t)\right] \; .
\end{equation}
This equation may be linearized by a change of variables.  We define
\begin{equation}
{\cal A}_k(t) \equiv -i a^3(t) \frac{\dot{f}_k(t)}{f_k(t)} \; .
\end{equation}

The mode functions $f_k(t)$ then obey the equation:
\begin{equation}
\ddot{f}_k(t) + 3 \frac{\dot{a}(t)}{a(t)}\dot{f}_k(t) + W_k^2(t) f_k(t) = 0 
\; .
\end{equation}
To again impose the Hartree approximation, we make a Hartree factorization
of the effective frequencies by writing
$$
W_k^2(t) = \frac{k^2}{a^2(t)} + \langle V^{''}\left(\phi(t) +
\psi(\vec{x},t)\right) \rangle \; ,
$$
where the latter term on the right hand side is factorized as in 
Eq.~(\ref{factorF}). 
In terms of these mode functions, the original parameters of the density
matrix are given by:
\begin{eqnarray}
A_{kR}(t) &=& \frac{1}{C_k |f_k(t)|^2} \; , \\
A_{kI}(t) &=& -\frac{a^3(t)}{2|f_k(t)|^2}
\left(\dot{f}_k(t)f_k^*(t)+\dot{f}_k^*(t)f_k(t) \right) \; , \\
B_k(t) &=& \frac{1}{D_k |f_k(t)|^2} \; .
\end{eqnarray}
The two-point fluctuation becomes
\begin{equation}
\langle \psi^2(t) \rangle = \int \frac{d^3k}{(2\pi)^3}\frac{1}{2(A_{kR}+B_k)}
= \int \frac{d^3k}{(2\pi)^3} \frac{|f_k(t)|^2}{2} \frac{C_k D_k}{C_k+D_k} \; ,
\end{equation}
and the equation of motion of the expectation value of $\Phi$ becomes
\begin{equation}
\ddot{\phi}(t) + 3 \frac{\dot{a}(t)}{a(t)}\dot{\phi}(t) 
+ V^{'}\left( \phi(t) \right) + \frac{V^{'''}\left(\phi(t)\right)}{2}
\langle \psi^2(t) \rangle = 0 \; .
\end{equation}
It is clear from these expressions that for a particular choice of the
constants $C_k$ and $D_k$ and the initial conditions on the $f_k$, we 
have reproduced the set of equations of motion determined in the last
section using a Green's function approach.  In particular, the choice
\begin{eqnarray}
C_k &=& \tanh(\beta \omega_k) \; , \\
D_k &=& - \sinh(\beta \omega_k) \; , \\
f_k(t_0) &=& \frac{1}{\sqrt{\omega_k}} \; , \\
\dot{f}_k(t_0) &=& - \left(\frac{\dot{a}(t_0)}{a(t_0)} + i\omega_k\right)f_k(t_0)
\; ,
\end{eqnarray}
along with the appropriate choice of initial normalization
\begin{equation}
{\cal N}_k(t_0) = \left[ \frac{a^3(t_0)\omega_k}{\pi}
\tanh\left(\frac{\beta\omega_k}{2}\right)\right]^{1/2} \; ,
\end{equation}
yields a density matrix (\ref{ansatz}) which is in local thermodynamic 
equilibrium at the initial time at a temperature $T=1/\beta$.
The two-point fluctuation becomes:
\begin{equation}
\langle \psi^2(t) \rangle = \int \frac{d^3k}{(2\pi)^3}\frac{1}{2(A_{kR}+B_k)}
= \int \frac{d^3k}{(2\pi)^3} \frac{|f_k(t)|^2}{2} 
\coth \left(\frac{\beta\omega_k}{2}\right) \; .
\end{equation}

\subsection{Gravitational Dynamics}

The gravitational sector includes the usual Einstein term in addition
to a higher order curvature term and a cosmological constant term 
which are necessary to renormalize the theory. The action for
the gravitational sector is therefore:
\begin{equation}
S_g  =  \int d^4x\; {\cal L}_g = \int d^4x \, a^3(t) \left[\frac{{\cal
R}(t)}{16\pi G}  
+ \frac{\alpha}{2}\; {\cal R}^2(t) - K\right].
\end{equation}
with $K$ being the cosmological constant (we use $K$ rather than the conventional
$\Lambda/8\pi G$ to distinguish the cosmological constant from the 
ultraviolet cutoff $\Lambda$ we introduce to regularize the theory).
In principle, we also need to include the terms $R^{\mu\nu}R_{\mu\nu}$
and $R^{\alpha\beta\mu\nu}R_{\alpha\beta\mu\nu}$ as they are also terms
of fourth order in derivatives of the metric (fourth adiabatic order),
but the variations resulting from these terms turn out not to be 
independent of that of ${\cal R}^2$ in the flat
FRW cosmology we are considering.

The variation of the action $S = S_g + S_m$ with respect to the 
metric $g_{\mu\nu}$ gives us Einstein's equation
\begin{equation}
\frac{G_{\mu\nu}}{8\pi G} + \alpha H_{\mu\nu} + K g_{\mu\nu}
= - T_{\mu\nu} \; ,
\label{extendEinstein}
\end{equation}
where $G_{\mu\nu}$ is the Einstein tensor given by the variation of
$\sqrt{-g}{\cal R}$, $H_{\mu\nu}$ is the higher order curvature term given
by the variation of $\sqrt{-g}{\cal R}^2$, and $T_{\mu\nu}$ is the contribution 
from the matter Lagrangian. 
 With the metric (\ref{metric}), the various components
of the curvature tensors in terms of the scale factor are:
\begin{eqnarray}
G^{0}_{0} & = & -3(\dot{a}/a)^2, \label{G00} \\
G^{\mu}_{\mu} & = & -{\cal R} = -6\left(\frac{\ddot{a}}{a}
+\frac{\dot{a}^2}{a^2}\right), \label{Gmumu} \\ 
H^{0}_{0} & = & -6\left(\frac{\dot{a}}{a}\dot{{\cal R}} + 
\frac{\dot{a}^2}{a^2}{\cal R} - \frac{1}{12}{\cal R}^2\right), 
\label{hzerozero} \\
H^{\mu}_{\mu} & = & -6\left(\ddot{{\cal R}} + 
3\frac{\dot{a}}{a}\dot{{\cal R}}\right).
\label{htrace}
\end{eqnarray}

However, it should be clear that even this extended Einstein's equation 
(\ref{extendEinstein}) is not self-consistent in the present case,
since the energy momentum tensor on the right hand side is a quantum
quantity, while the gravitational tensors on the left hand side are
classical.  This leads us to one of the major problems of the physics
of the early universe: the reconciliation of quantum field theory
with the cosmological theory based on Einstein's equation of general
relativity.  In particular, there has yet to be constructed a full
quantum theory of gravity.  One must therefore rely on
approximations.  A promising candidate is semi-classical gravity,
based on the equation
\begin{equation}
\frac{G_{\mu\nu}}{8 \pi G_N} + \alpha H_{\mu\nu} + K g_{\mu\nu} = 
- \langle T_{\mu\nu} \rangle \; ,
\label{semiclass}
\end{equation}
where the expectation value of the quantum energy momentum tensor acts
as a classical source to the gravitational field.  This equation may
be considered the first approximation to a more complete expansion
which includes expectation values of terms of higher order in
$T_{\mu\nu}$.  The hope is that, since each additional factor of
$T_{\mu\nu}$ requires a coefficient which by dimensional analysis is
of order $1/M_{Pl}^4$, for energy densities sufficiently removed from
the Planck scale, Eq.~(\ref{semiclass}) is a good approximation to the
complete theory.  Of course, without a complete quantum theory of
gravity with which this approximation could be compared, there is no
definitive way to judge its validity (see the discussion in \cite{birrell}).

Eventually, when we have fully renormalized the theory, we will
simplify the theory by setting $\alpha_R$ to zero, and we will keep as our
only contribution to $K_R$ a piece related to the matter fields which
we shall incorporate into $T_{\mu\nu}$.  Note that the dimensional
analysis indicates that $\alpha$ is of order $1/M_{Pl}^2$ so that
there may be some justification to neglecting this contribution in
determining the gravitational equations of motion for scales
significantly removed from the Planck scale.

\section{The $\lambda \Phi^4$ Theory} \label{seclphi4}
We now specialize the equations of motion we have derived 
above for very general potentials to those appropriate
to the $\lambda \Phi^4$ theory.  

It is convenient to treat the Hartree and large $N$ approximations
simultaneously.  We may do so by writing the potential as
\begin{equation}
V(\Phi(x)) = K + \frac12 \left(m^2 + \xi {\cal R}(t)\right)\Phi^2(x)
+ \frac{\gamma \lambda}{8N} \Phi^4(x) \; ,
\label{phi4potential}
\end{equation}
where in the Hartree approximation, the quantity $\gamma$ has the value $1/3$ 
and we take $N=1$, while in the large $N$ limit, $\gamma=1$ and it is to be
understood that $\Phi^2 \equiv \vec{\Phi} \cdot \vec{\Phi}$.  The constant
quantity $K$ is a cosmological constant contribution which does not
directly affect the equations of motion of $\Phi(x)$ but will be 
necessary in the treatment of spontaneous symmetry breaking.

Given this potential and using a unified notation by letting 
$\bar{\sigma} \to \phi$ and $\langle \pi^2 \rangle \to \langle \psi^2 \rangle$
in the equations of motion for the $O(N)$ scalar field, we arrive at the
following set of equations of motion in the $\lambda \Phi^4$ theory:
\begin{eqnarray}
\ddot{\phi}(t)+3\frac{\dot{a}(t)}{a(t)} \dot{\phi}(t)+\left(m^2+\xi{\cal
R}(t)\right)\phi(t) 
+\frac{\gamma \lambda}{2}\phi^3(t)+\frac{\lambda}{2}\phi(t)
\langle\psi^2(t)\rangle 
& = & 0\; ,
 \nonumber \\  \label{hartphieq} \\
\left[\frac{d^2}{dt^2}+3\frac{\dot{a}(t)}{a(t)}\frac{d}{dt}+
\frac{k^2}{a^2(t)}
+m^2+\xi{\cal R}(t)+\frac{\lambda}{2} \phi^2(t)+
\frac{\lambda}{2}\langle\psi^2(t)\rangle\right]
f_k(t) & = & 0\; , \nonumber \\
\label{hartukeq}
\end{eqnarray}
\begin{eqnarray}
\langle\psi^2(t)\rangle &=& \int \frac{d^3k}{2(2\pi)^3}\, |f_k(t)|^2
\coth \left(\frac{\beta \omega_k}{2} \right) \; .
\label{fluct} \\ \nonumber
\end{eqnarray}
The initial conditions on the mode functions are 
\begin{equation}
f_k(t_0) = \frac{1}{\sqrt{\omega_k(t_0)}}, \quad \dot{f}_k(t_0) =\left[
-\frac{\dot{a}(t_0)}{a(t_0)}-i\omega_k(t_0)\right]f_k(t_0),
\label{initcond}
\end{equation}
with the frequencies $\omega_k(t_0)$ given by
\begin{eqnarray}
\omega_k(t_0) &=& \left[k^2+{\cal{M}}^2(t_0)\right]^{1/2} \; , \nonumber \\
{\cal{M}}^2(t) &=& \left[m^2+(\xi-1/6){\cal R}(t)+
\frac{\lambda}{2}\phi^2(t)+\frac{\lambda}{2}\langle\psi^2(t)\rangle \right].
\label{freq}
\end{eqnarray}
 
Notice that the only difference between the expressions for the case
of a real scalar field in the Hartree approximation and the case of an
$N$-component field in the large $N$ limit
is a factor of three appearing in the self interaction term in
the equations for the zero mode.  However, as we will see, the two 
approximations
are describing theories with distinct symmetries and there will be 
qualitative differences in the results.  
An important point to note in the large $N$ equations of motion is that the
form of the equation for the zero mode (\ref{hartphieq}) is the same as for the
$k=0$ mode function (\ref{hartukeq}). It will be this identity that allows
solutions of these equations in a symmetry broken scenario to satisfy
Goldstone's theorem.

\section{Renormalization and Energy Momentum}

Upon examination of the equal time correlator (\ref{fluct}), one finds that 
the
integral is divergent and thus must be regulated.  This can be done by a 
number
of methods, but we require a scheme which is amenable to numerical 
calculation.
We therefore introduce a large momentum cutoff which renders the integral
finite, and one finds that it is possible to remove the terms depending both
quadratically and logarithmically on the cutoff by a renormalization of the
parameters of the theory\cite{frw1}. After these terms are subtracted, the
cutoff may be taken to infinity, and the remaining quantity is both physical
and finite.  A similar process is required to regulate the expressions for 
the
energy density and the pressure as will be described in more detail below
\cite{desitter}.

In terms of the variables introduced in section II above, the renormalization
of (\ref{fluct}) proceeds almost identically in the Hartree and large $N$
approximations.  We begin by undertaking a WKB analysis of the mode
functions $f_k(t)$ to reveal their large $k$ behavior.  We introduce
the WKB modes
$$
D_k = \exp \left(\int^t_{t_0} R_k(t^{'}) dt^{'} \right) \; ; \quad
D_k(t_0) = \frac{1}{\sqrt{\omega_k}} \; .
$$
We then have 
$$
f_k = (1-\gamma_k)D_k^* + \gamma_k D_k \; .
$$
The $D_k$ satisfy the same equation of motion as the $f_k$, whereas
the function $R_k$ obeys the Riccati equation
\begin{equation}
\dot{R}_k(t) + R_k^2(t) + 3 \frac{\dot{a}(t)}{a(t)} R_k(t) +
\frac{k^2}{a^2(t)} + {\cal M}^2(t) = 0 \; .
\label{Riccati}
\end{equation}
Writing $R_k(t)$ in the form
$$
R_k(t) = -i \frac{k}{a(t)} + R_0(t) -i\frac{a(t)}{k}R_1(t) +
\frac{a^2(t)}{k^2}R_2(t) -i\frac{a^3(t)}{k^3}R_3(t) +
\frac{a^4(t)}{k^4}R_4(t) + \cdots \; ,
$$
we may determine $R_k(t)$ term by term by matching coefficients of
different powers of $k$ in the Riccati equation (\ref{Riccati}).  In
this way, we find
\begin{eqnarray}
R_0(t) &=& -\frac{\dot{a}(t)}{a(t)} \; , \\
R_1(t) &=& \frac12 \left[ {\cal M}^2(t) - \frac{R(t)}{6} \right] \; ,
\\ 
R_2(t) &=& -\frac12 \left[ \dot{R}_1(t) 
+ 2\frac{\dot{a}(t)}{a(t)}R_1(t) \right] \; , \\
R_3(t) &=& \frac12 \left[ \dot{R}_2(t) +
3\frac{\dot{a}(t)}{a(t)}R_2(t) - R_1^2(t) \right] \; , \\
R_4(t) &=& -\frac12 \left[ \dot{R}_3(t) + 4\frac{\dot{a}(t)}{a(t)}R_3(t)
+ 2R_1(t)R_(t) \right]  \; . 
\end{eqnarray}

The coefficient $\gamma_k$ is determined from the initial conditions
on the mode functions (\ref{initcond}). We find for the real and
imaginary parts respectively
$$
\gamma_{k,R} = 1 + {\cal O}(1/k^4) \; , \quad \gamma_{k,I} = {\cal
O}(1/k^3) \; .
$$

Using these results, we find the high $k$ behavior of the mode
functions and their derivatives.  
In terms of the effective mass term for the large $ N $ limit
and defining the quantities
\begin{eqnarray}
B(t) &\equiv& a^2(t)\left({\cal M}^2(t)-{\cal{R}}/6 \right),\label{boft}\\
{\cal M}^2(t) &=& -m^2_B+\xi_B {\cal R}(t)+\frac{\lambda_B}{2}\phi^2(t)
+\frac{\lambda_B}{2}\langle\psi^2(t)\rangle_B \; , \label{masso}
\end{eqnarray}
where the subscript $B$ stands for bare quantities,  
we find the following large $k$ behavior for the case of an {\em arbitrary}
scale factor $a(t)$ (with $a(0)=1$):
\begin{eqnarray}
|f_k(t)|^2 &=& \frac{1}{ka^2(t)}- \frac{1}{2k^3
a^2(t)}\;B(t)  \nonumber \\ &+&
{1 \over {8 k^5 \; a^2(t) }}\left\{  3 B(t)^2 + a(t)
\frac{d}{dt} \left[ a(t) {\dot B}(t) \right]  \right\} +
{\cal{O}}(1/k^7)  \nonumber \\
& = & {\cal S}^{(2)}+ {\cal{O}}(1/k^5) \; ,
\label{sub1}\\ 
|\dot{f}_k(t)|^2 &=&
\frac{k}{a^4(t)}+\frac{1}{2ka^4(t)}\left[B(t)+2\dot{a}^2 \right] 
\nonumber \\
& + & {1 \over {8 k^3 \; a^4(t) }}\left\{ - B(t)^2 - a(t)^2 {\ddot
B}(t) + 3 a(t) 
{\dot a}(t)
{\dot B}(t) - 4 {\dot a}^2(t) B(t) \right\} \nonumber \\
& + &  {\cal{O}}(1/k^5) \nonumber \\
& = & {\cal S}^{(1)}+ {\cal{O}}(1/k^5) \; ,
\label{sub2}  
\end{eqnarray}
\begin{eqnarray}
\frac12 \left[f_k(t)\dot{f}_k^*(t)+\dot{f}_k(t)f_k^*(t)\right] &=&
-\frac{1}{k \; a^2(t)}\frac{\dot{a}(t)}{a(t)} - \frac{1}{4 k^3 a^2(t)}
\left[\dot{B}(t) - 2\frac{\dot{a}(t)}{a(t)}\;B(t)\right] \nonumber \\
&+&{\cal{O}}(1/k^5) \; .
\label{sub3}
\end{eqnarray}

 Because there are no counterterms to
utilize in the renormalization of the theory,
the renormalization conditions on the mass, coupling to the Ricci
scalar and coupling constant are obtained from the requirement that the
frequencies that appear in the mode equations are finite\cite{frw1}, i.e:
\begin{equation}
-m^2_B+\xi_B {\cal R}(t)+\frac{ \lambda_B}{2}\phi^2(t)
+\frac{\lambda_B}{2}\langle\psi^2(t)\rangle_B=
-m^2_R+\xi_R {\cal R}(t)+\frac{ \lambda_R}{2}\phi^2(t)
+\frac{\lambda_R}{2}\langle\psi^2(t)\rangle_R, \label{rencond}
\end{equation}
We find the following renormalization scheme:
\begin{eqnarray}
m_B^2 & + & \frac{\lambda_B}{16\pi^2}\frac{\Lambda^2}{a^2(t)} = 
m_R^2\left[1+\frac{\lambda_B}{16\pi^2}\ln(\Lambda/\kappa)\right],
\label{massrenorm} \\
\lambda_B & = & \frac{\lambda_R}
{1-\gamma\lambda_R\ln(\Lambda/\kappa)/16\pi^2},
\label{couprenorm} \\
\xi_B & = & \xi_R + \frac{\lambda_B}{16\pi^2}(\xi_R-1/6)\ln(\Lambda/\kappa),
\label{xirenorm}
\end{eqnarray}
where $\kappa$ is the renormalization point.
The subtracted equal time correlator is now given by
\begin{eqnarray}
\langle\psi^2(t)\rangle_R & = & \int^{\Lambda}\frac{d^3k}{(2\pi)^3}\left\{
\frac{|f_k(t)|^2}{2} - \frac{1}{2ka^2(t)}
 + \frac{\theta(k-\kappa)}{4k^3}\left[(\xi_R-1/6){\cal R}(t)
\right. \right. \nonumber \\
&+& \left. \left. m_R^2+
\frac{\lambda_R}{2}(\phi^2(t)+\langle\psi^2(t)\rangle_R)\right]\right\}.
\label{fluctrenorm}
\end{eqnarray}
Notice in 
particular
that in order for the renormalization of the mass to be time independent, we
must require that the cutoff $\Lambda$ be fixed in {\it physical} coordinates
and therefore have the form $\Lambda \propto a(t)$.  The ability to make
a renormalization which is independent of time is non-trivial.  In
particular, choosing initial conditions in with respect to the comoving
Hamiltonian instead of the initial conditions in conformal time (\ref{initcond})
we use here, results in renormalizations which depend explicitly on the
initial state\cite{frw1}.  Further discussions of the significance of the
choice of initial state may be found in \cite{initcond}.  It is worth noting
that for $a(t) = 1$ the renormalizations of the mass, quartic coupling, 
and correlator are the same as those in Minkowski space. 

Our treatment of the renormalization of the energy momentum tensor is similar
to the approach of \cite{anderson}, extended to the non-perturbative Hartree
and large $N$ approximations.  The expressions for the expectation values of
the energy density, $\varepsilon$, and the trace of the stress energy,
$\varepsilon-3p$, where $p$ is the pressure density are\cite{frwbig}:
\begin{eqnarray}
\frac{\varepsilon}{N} & = & \frac12\dot{\phi}^2 + \frac12m^2\phi^2 +
\frac{\gamma\lambda}{8}\phi^4 + \frac{m^4}{2\gamma\lambda} - \xi
G^{0}_{0}\phi^2 +  6\xi\frac{\dot{a}}{a}\phi\dot{\phi}
\nonumber \\
& + & \frac12\langle\dot{\psi}^2\rangle + 
\frac{1}{2a^2}\langle(\nabla\psi)^2\rangle + \frac12 m^2 
\langle\psi^2\rangle 
+ \frac{\lambda}{8}[2\phi^2\langle\psi^2\rangle + \langle\psi^2\rangle^2] 
\nonumber \\
& - & \xi G^{0}_{0}\langle\psi^2\rangle + 
6\xi\frac{\dot{a}}{a}\langle\psi\dot{\psi}\rangle, \label{energy} \\
\frac{\varepsilon-3p}{N} & = & -\dot{\phi}^2 + 2m^2\phi^2 + 
\frac{\gamma\lambda}{2}\phi^4 + \frac{2m^4}{\gamma\lambda} - \xi
G^{\mu}_{\mu}\phi^2 + 
6\xi\left(\phi\ddot{\phi} + \dot{\phi}^2 +
3\frac{\dot{a}}{a}\phi\dot{\phi}\right) 
\nonumber \\
& - & (1-6\xi)\langle\dot{\psi}^2\rangle +
\frac{1-6\xi}{a^2}\langle(\nabla\psi)^2\rangle + 
(2-6\xi)m^2\langle\psi^2\rangle - \xi
G^{\mu}_{\mu}(1-6\xi)\langle\psi^2\rangle  
\nonumber \\
& + & \frac{\lambda}{2}[(2-6\xi)\phi^2\langle\psi^2\rangle +
\langle\psi^2\rangle^2 - 
6\xi\langle\psi^2\rangle\langle\psi^2\rangle_R],
\label{trace}
\end{eqnarray}
where we have used the equations of motion in deriving this expression for 
the
trace (\ref{trace}). The quantities $G^{\mu}_{\mu}$ and $
G^{0}_{0}$ are given by Eqs.~(\ref{Gmumu}) and (\ref{G00}), 
$\langle\psi^2\rangle$ is given by Eq.~(\ref{fluct}), 
$\langle\psi^2\rangle_R$ by (\ref{fluctrenorm}), and we have
defined the following integrals:
\begin{eqnarray}
\langle(\nabla\psi)^2\rangle & = & \int \frac{d^3k}{2 (2\pi)^3}k^2
|f_k(t)|^2 \; , \label{delpsi} \\
\langle\dot{\psi}^2\rangle & = & \int \frac{d^3k}{2(2\pi)^3}
|\dot{f}_k(t)|^2 \;. \label{dotpsi}
\end{eqnarray}
The composite operator $\langle \psi \dot{\psi} \rangle$ is symmetrized by
removing a normal ordering constant to yield
\begin{equation}
\frac{1}{2}(\langle\psi\dot{\psi}\rangle + \langle\dot{\psi}\psi\rangle)
 =  \frac{1}{4} \int \frac{d^3k}{(2\pi)^3}
\frac{d |f_k(t)|^2}{dt}. 
\label{psidotpsi} 
\end{equation}

Each of these integrals is divergent and must be regularized.  We proceed in
the same manner as above, imposing an ultraviolet cutoff, $\Lambda$, and
computing the high $k$ expansions of the $f_k$, this time to fourth order in
$1/k$.  We find the following divergences in $\varepsilon$ and
$\varepsilon-3p$:
\begin{eqnarray}
\left(\frac{\varepsilon}{N}\right)_{div} & = & \frac{\Lambda^4}{16\pi^2a^4} +
\frac{\Lambda^2}{16\pi^2a^2}\left[2(\xi_R-1/6)G^{0}_{0}+m_R^2+
\frac{\lambda_R}{2}(\phi^2+\langle\psi^2\rangle_R)\right] \nonumber \\
& + & \frac{\ln(\Lambda/\kappa)}{16\pi^2}\left[-\frac{m_R^4}{2} -
m_R^2\frac{\lambda_R}{2}(\phi^2+\langle\psi^2\rangle_R) -
\frac{\lambda_R^2}{8}(\phi^2+\langle\psi^2\rangle_R)^2\right. \nonumber \\
& + & \left. 2(\xi_R-1/6)G^{0}_{0}\left(m_R^2+
\frac{\lambda_R}{2}(\phi^2+\langle\psi^2\rangle_R)\right) 
+(\xi_R-1/6)^2H^{0}_{0}\right. \nonumber \\
& - & \left.6(\xi_R-1/6)\frac{\dot{a}}{a}\frac{\lambda_R}{2}
\frac{d}{dt}(\phi^2+\langle\psi^2\rangle_R)\right], 
\label{energydiv} \\
\left(\frac{\varepsilon-3p}{N}\right)_{div} & = & 
\frac{\Lambda^2}{16\pi^2a^2}
\left[2(\xi_R-1/6)G^{\mu}_{\mu} 
+ 12(\xi_R-1/6)\frac{\dot{a}^2}{a^2}\right. \nonumber \\
&+& \left. 2m_R^2 + 
\lambda_R(\phi^2+\langle\psi^2\rangle_R)\right] \nonumber \\
& + & \frac{\ln(\Lambda/\kappa)}{16\pi^2}\left[-2m_R^4 - 
2m_R^2\lambda_R(\phi^2+\langle\psi^2\rangle_R) - \frac{\lambda_R^2}{2}
(\phi^2+\langle\psi^2\rangle_R)^2\right. \nonumber \\
& + & \left. 2(\xi_R-1/6)G^{\mu}_{\mu}
\left(m_R^2+\frac{\lambda_R}{2}(\phi^2+\langle\psi^2\rangle_R)\right)\right. 
+ (\xi_R-1/6)^2H^{\mu}_{\mu}
\nonumber \\ 
& - & \left. 6(\xi_R-1/6)\left[\frac{\lambda_R}{2}\frac{d^2}{dt^2}
(\phi^2+\langle\psi^2\rangle_R) +
3\frac{\dot{a}}{a}\frac{\lambda_R}{2}\frac{d}{dt}
(\phi^2+\langle\psi^2\rangle_R)\right]\right]. \nonumber \\
\label{tracediv}
\end{eqnarray}

The energy momentum is made finite by subtraction of the divergent pieces
(\ref{energydiv}) and (\ref{tracediv}) from the expressions for the energy
density (\ref{energy}) and the trace (\ref{trace}).  Within the context of
covariant regularization schemes such as dimensional regularization and
covariant point splitting\cite{birrell}, such a procedure has been shown to
adequately renormalize couplings appearing in the semi-classical Einstein's
equation, Eq.~(\ref{semiclass}).

The renormalizations of Newton's constant, the higher order curvature
coupling, and the cosmological constant are given by the condition of
finiteness of the semi-classical Einstein equation:
\begin{equation}
\frac{G^0_0}{8\pi G_B} + \alpha_B H^0_0 + K_B g^0_0 + 
\langle T^0_0 \rangle_B = \frac{G^0_0}{8\pi G_R} + 
\alpha_R H^0_0 + K_R g^0_0 + \langle T^0_0 \rangle_R \; .
\end{equation}

Finally we arrive at the following set of additional
renormalizations\cite{grav}:
\begin{eqnarray}
\frac{1}{8\pi N G_R} &=& \frac{1}{8\pi N G_B} 
- 2\left(\xi_R-\frac16\right)\frac{\Lambda^2}{16\pi^2}
+ 2\left(\xi_R-\frac16\right)m_R^2\;
\frac{\ln(\Lambda/\kappa)}{16\pi^2} \; , \nonumber \\ \\
\frac{\alpha_R}{N} &=& \frac{\alpha_B}{N} 
- \left(\xi_R-\frac16\right)^2\frac{\ln(\Lambda/\kappa)}{16\pi^2} \; , \\
\frac{K_R}{N} &=& \frac{K_B}{N} - \frac{\Lambda^4}{16\pi^2}
+ m_R^2\;\frac{\Lambda^2}{16\pi^2}
+ \frac{m_R^4}{2}\;\frac{\ln(\Lambda/\kappa)}{16\pi^2} \; .
\end{eqnarray}
As expected, the logarithmic 
terms are consistent with the renormalizations found using dimensional
regularization\cite{baacke,ramseyhu1}.  Again, we set $\alpha_R=0$ and
choose the renormalized cosmological constant such that the vacuum
energy is zero in the true vacuum.  We emphasize that while 
the regulator we have chosen does not respect the covariance of 
the theory, the renormalized energy momentum tensor defined in this 
way nevertheless retains the property of covariant conservation in the
limit when the cutoff is taken to infinity.

In the numerical analysis of inflation, the logarithmic subtractions
can be neglected because of the coupling
$\lambda \leq 10^{-12}$.  Using the Planck scale as the cutoff and the
inflaton mass $m_R$ as a renormalization point, these terms are of order
$\lambda \ln[M_{pl}/m_R] \leq 10^{-10}$, for $m \geq 10^9 \mbox{ GeV }$. An
equivalent statement is that for these values of the coupling and inflaton
masses, the Landau pole is well beyond the physical cutoff $M_{pl}$.
Our relative 
error in the numerical analysis is of order $10^{-8}$, so that our numerical
study is insensitive to the logarithmic corrections. Though these corrections
are fundamentally important, they are small enough to be negligible in our
numerical analyses.

\chapter{New Inflation}

\section{Introduction and Motivation}

With the introduction of the first model of inflation by
Guth\cite{guth} came an immediate problem.  This `old' inflation model
failed in that it did not include a mechanism to gracefully exit from
the inflationary phase.  The basic problem was that `old' inflation
relied on a first order phase transition in which the inflaton
begins in a meta-stable inflationary state and then tunnels to move to
the true minimum of the potential resulting in a radiation or matter
dominated phase, but the rate for such a phase transition to take
place was too small.

Soon, it was realized that inflation could occur in a second order
phase transition as well, and a model based on the Coleman-Weinberg
theory was built which had no exit problem since the
filed evolved continuously to the true vacuum.  This `new' inflation
model\cite{newinfl}, however, had a problem of its own.  Since the
coupling in the Coleman-Weinberg theory is a gauge coupling of order
one, the metric perturbations produced in this model were much too
large to be consistent with the observed homogeneity of the
CBR\cite{newproblem}.  Nevertheless, the basic idea of inflation
proceeding via a second order transition is sound and a number of
models, such as the $\lambda\Phi^4$ model with negative mass squared,
are successful in producing inflation with all the desired
properties.  

\begin{figure}
\epsfig{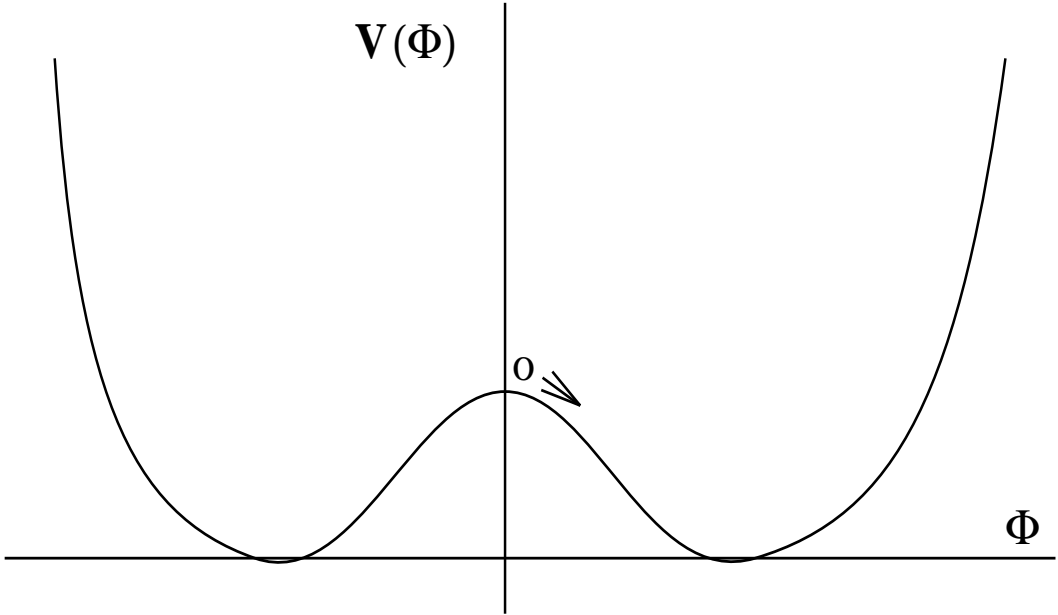}
\caption{A typical new inflation potential.  The field $\Phi$ begins
near the top of the central hill.}
\label{newpot}
\end{figure}

In this chapter, we examine the non-equilibrium dynamics of a
$\lambda\Phi^4$ new inflationary quantum field theory.  The
expectation value is initially near the symmetric point of the
potential (see Fig.~\ref{newpot}), the disordered state, and the fluctuations
about this value are assumed to be small relative to the vacuum
expectation value (vev), $v \sim \mu/\sqrt{\lambda}$ (where $\mu\equiv
|m|$). 

Because this initial state is in a region of the potential with
negative curvature, the field will grow in time due to the presence
of spinodal instabilities\cite{rolling}.  This growth continues
until the fluctuations become of the same order as the vev, 
i.e.~$\langle \Phi^2(\vec{x},t) \rangle \sim v^2 \sim \mu^2/\lambda$.
Until this time, when the field has become non-perturbatively large, 
inflation occurs.

\subsection*{On the initial state: dynamics of phase transitions} 

The issue of initial conditions on the inflaton is of particular
interest in treatments of new inflation.  In much of the 
literature\cite{rocky,lindebook},
the inflaton is treated as a classical field $\phi_{cl}(t)$ which is
taken to be initially `near' the origin.  Further, the uncertainty
relation of quantum mechanics is sometimes invoked to yield a
requirement that $\phi_{cl}(t_0) > H/2\pi$.

How this classical field relates to the quantum field
$\Phi(\vec{x},t)$, however, has been somewhat unclear.  The
straightforward identification of the classical field $\phi_{cl}(t)$
with the expectation value $\phi(t) = \langle \Phi(\vec{x},t) \rangle$
is clearly not correct, as there are no restrictions on the initial
value $\phi(t_0)$ and, in particular, a natural value in a phase
transition $\phi(t_0)=0$ would yield the incorrect result of an
unending period of inflation if $\phi(t)$ were identified with a
classical inflaton field.

One is led to the following questions:
\begin{itemize}
\item Assuming it can be done, how does one construct the classical
inflaton from the underlying quantum field theory?
\item What, if anything, does this construction tell one about the
initial state of the effective classical theory?
\end{itemize}

As we shall see, the answers to these questions lie in the quantum
fluctuations $\psi(\vec{x},t)$ of the field.  Let us take the example
of a symmetric initial condition $\phi(t_0)=0$  The zero mode remains
zero for all time $\phi(t) = 0$.  However, examination of 
Eq.~(\ref{hartukeq}) for negative mass squared indicates a spinodal 
instability for low $k$ which
results in growth of the non-zero momentum modes which make up the
quantum fluctuations $\langle \psi^2(t) \rangle$, Eq.~(\ref{fluct}).
Therefore the dynamics becomes fluctuation driven with $\langle
\psi^2(t) \rangle$ eventually becoming non-perturbatively large (that
is, $\langle \psi^2(t) \rangle \sim \mu^2/\lambda$).  One may then hope
that the classical picture of new inflation may be constructed from
the quantum evolution by the identification of the classical field
with some component of the quantum fluctuations.  

However, questions remain:
\begin{itemize}
\item How does the gravitational background respond to
non-perturbatively large quantum fluctuations?
\item How can one extract (small?) metric perturbations from
non-perturbatively large field fluctuations?
\end{itemize}
Remaining still are the details of how the classical field might be
constructed from the quantum field.  In this chapter, we address each
of these questions within a fully consistent quantum field theory
context.  

Before we move on to our analysis of new inflation, there is one more
point regarding initial conditions that deserves attention.  The
standard picture of a second order phase transition in the early
universe has the field $\Phi(\vec{x},t)$ in thermal equilibrium at
some temperature above critical, $T>T_c$, such that the initial state
is one of positive mass squared.  This localizes the field about the
origin.  Then, as the finite temperature contributions to the
effective mass are redshifted away, the phase transition takes place
and the field may evolve toward its asymptotic state.

As we shall see, however, the self-coupling of the inflaton is
constrained to be $\lambda \leq 10^{-12}$ by the magnitude of the
temperature fluctuations of the CBR.  It has therefore been argued
that the field is too weakly interacting to be in thermal equilibrium
above the critical temperature.  It is therefore not clear whether
there is any `natural' initial state for the field in new inflation.

As discussed in Sec.~3.3, we will simply assume an initial density
matrix describing either a thermal state at temperature $T=1/\beta$ or
a vacuum state of the initial Hamiltonian (an initial state in thermal 
equilibrium at a temperature $T=0$).  The assumption of an initial equilibrium 
vacuum state are essentially the same used by Linde\cite{linde2},
Vilenkin\cite{vilenkin}, as well as by Guth and Pi\cite{guthpi} in their
analyses of the quantum mechanics of inflation in a fixed de Sitter
background. 

A final note is that the initial frequencies $\omega_k$
(\ref{initcond}) are imaginary for low $k$ modes in the vacuum state and
need to be modified.  This may be done in a variety of ways with
little effect on results\cite{desitter,grav}.  Here, we choose a
smooth interpolation between low-k modes with modified frequencies and
the high $k$ modes which remain in the conformal vacuum state with
frequencies $\omega_k$ (\ref{initcond}):
\begin{equation}
\omega_k \equiv \left[k^2+{\cal{M}}^2(t_0)\tanh\left(
\frac{k^2+{\cal{M}}^2(t_0)}{|{\cal{M}}^2(t_0)|}\right)\right]^{1/2} \;
.
\label{modinit}
\end{equation}

\section{Analysis}

We will work in the large $N$ approximation throughout this chapter.
Except for the very late time dynamics with $\phi(t_0) \neq 0$, the 
evolution in the Hartree case is almost identical.  For $\phi(t_0) =0$,
there is no difference between the Hartree and large $N$ dynamics.

It is convenient to introduce the following dimensionless quantities
and definitions, with $\mu \equiv |m_R|$,
\begin{equation}
\tau = \mu t \quad ; \quad h= \frac{H}{\mu} \quad ; 
\quad q=\frac{k}{\mu} \quad \; \quad
\omega_q = \frac{W_k}{\mu} \quad ; \quad g= \frac{\lambda_R}{8\pi^2} \; ,
\label{dimvars1}
\end{equation}
\begin{equation}
\eta^2(\tau) = \frac{\lambda_R}{2m^2_R} \; \phi^2(t)
\quad ; \quad  g\Sigma(\tau) = \frac{\lambda}{2m^2_R}\; \langle \psi^2(t)
\rangle_R  \quad ; \quad f_q(\tau) \equiv \sqrt{\mu} \; f_k(t) \; .
\label{dimvars3}
\end{equation}

Choosing $\xi_R=0$ (minimal coupling)  and the renormalization
 point $\kappa = |m_R|$ and setting $a(0)=1$, 
the equations of motion become:

\vspace{2mm} 

\begin{eqnarray}
\left[\frac{d^2}{d \tau^2}+ 3h \frac{d}{d\tau}-1+\eta^2(\tau)+
g\Sigma(\tau)\right]\eta(\tau) &=& 0, 
\label{zeromode} \\
\left[\frac{d^2}{d \tau^2}+3h
\frac{d}{d\tau}+\frac{q^2}{a^2(\tau)}-1+\eta^2+g\Sigma(\tau)
\right]f_q(\tau)&=&0,
\label{modes}  
\end{eqnarray}
\begin{equation}
f_q(0) =  \frac{1}{\sqrt{\omega_q}} \; , \quad 
\dot{f}_q(0)  = \left[-h(0)-i\omega_q\right]f_q(0), \nonumber \\
\end{equation}
where
$$
\Sigma(\tau)= \int_0^{\infty} q^2 dq \left[ | f_q(\tau)|^2 - {1 \over
{ a(\tau)^2}} + {{\Theta(q - 1)}\over {2 q^3}} \left(\frac{{\cal
M}^2(\tau)}{m^2_R}-{{{\cal{R}(\tau)}}\over{6 m^2_R}}\right)\right] \; .
$$

The initial conditions for $\eta(\tau)$ will be specified later. 
An important point to notice is that the equation of
motion for the $q=0$ mode coincides with that of the zero mode
(\ref{zeromode}). Furthermore, for $\eta(\tau \rightarrow \infty) \neq
0$, a stationary (equilibrium) solution of the Eq.~(\ref{zeromode})  
is obtained when the sum rule\cite{us1,desitter,frwbig}
\begin{equation}
-1+\eta^2(\infty)+g\Sigma(\infty) = 0 \label{sumrule}
\end{equation}
is fulfilled. This sum rule is nothing but a consequence of Goldstone's
theorem and is a result of the fact that the large $ N $ approximation 
satisfies the Ward identities associated with the $ O(N) $ symmetry, since
the term  $-1+\eta^2+g\Sigma$ is seen to be the effective mass of the
modes transverse to the symmetry breaking direction, i.e. the Goldstone
modes in the broken symmetry phase.

In terms
of the zero mode $\eta(\tau)$ and the quantum mode function given
by Eq.~(\ref{modes}) we find that the Friedmann equation for the dynamics
of the scale factor in dimensionless variables is given by

\begin{equation}
 h^2(\tau)    =   4h^2_0 \; \epsilon_R(\tau) \quad  ;   \quad h^2_0 =
 \frac{4\pi N m^2_R}{3M^2_{Pl}\lambda_R} \label{hubblequation} 
 \end{equation}
and the renormalized energy and pressure are given by: 
\begin{eqnarray}
 \epsilon_R(\tau) &  =  &  
\frac{1}{2}\dot{\eta}^2+\frac{1}{4}\left(-1+\eta^2+g\Sigma \right)^2+
 \nonumber \\ 
& &  \frac{g}{2}\int q^2 dq \left[|\dot{f_q}|^2-
{\cal S}^{(1)}(q,\tau) \right. \nonumber \\
&+& \left. \frac{q^2}{a^2}\left(|f_q|^2-\Theta(q - 1)\; {\cal
S}^{(2)}(q,\tau)\right) \right]\; , 
\label{hubble} \\
 (p+\varepsilon)_R(\tau)  & = & \frac{2Nm^4_R}{\lambda_R}\left\{
\dot{\eta}^2 
 +  g \int q^2 dq \left[|\dot{f_q}|^2-
{\cal S}^{(1)}(q,\tau) \right.\right.\nonumber \\
&+&\left.\left.\frac{q^2}{3a^2}
\left(|f_q|^2-\Theta(q - 1)\;{\cal S}^{(2)}(q,\tau)\right)
\right]\right\}, \label{ppluse} 
\end{eqnarray}
where the subtractions ${\cal S}^{(1)}$ and ${\cal S}^{(2)}$ are given
by the right hand sides of Eqs.~(\ref{sub2}) and (\ref{sub1}) respectively.
 
The renormalized energy and pressure are covariantly conserved:
\begin{equation}\label{concov}
{\dot  \epsilon}_R(\tau) + 3 \, h(\tau)\, (p+\varepsilon)_R(\tau) = 0 \; .
\end{equation}

In order to provide the full
solution we now must provide the values of $\eta(0)$, $\dot{\eta}(0)$,
and $h_0$. Assuming that the 
inflationary epoch is associated with a phase transition at the GUT scale,
this requires that $ N m^4_R/\lambda_R \approx 
(10^{15}\mbox{ GeV })^4 $ and assuming the bound on the scalar
self-coupling $\lambda_R \approx 10^{-12}-10^{-14}$ (this will be seen
later 
to be a compatible requirement), we find that $h_0 \approx N^{1/4}$ which
we will take to be reasonably given by $h_0 \approx 1-10$ (for example
in popular GUT's $ N \approx 20 $ depending on particular representations). 

We will begin by studying the case of most interest from the point of view
of describing the phase transition: $\eta(0)=0$ and $\dot{\eta}(0)=0$,
which are the initial conditions that led to puzzling questions. With
these initial conditions, the evolution equation for the zero mode
Eq.~(\ref{zeromode}) determines that $\eta(\tau) = 0$ by symmetry.

\subsection{Early time dynamics}
Before engaging in the numerical study, it proves illuminating to
obtain an estimate of the relevant time scales and an intuitive idea of
the main features of the dynamics. Because the coupling is so weak 
($ g \sim 10^{-12} \ll 1 $) and because after
renormalization the contribution from the quantum fluctuations to 
the equations of motion is finite, we can neglect  all the terms
proportional to $ g $ in Eqs.~(\ref{hubble}) and (\ref{modes}). 

The evolution equations for the 
mode functions are those for an inverted oscillator in de Sitter space-time,
which have been studied by Guth and Pi\cite{guthpi}. One
obtains the approximate solution 
\begin{eqnarray}
h(t) & \approx & h_0, \nonumber \\
f_q(t) & \approx & e^{-3 h_0\tau/2} \left[A_q \;
J_{\nu}\left(\frac{q}{h_0}e^{-h_0\tau}\right)+ 
B_q  \; J_{-\nu}\left(\frac{q}{h_0}e^{-h_0\tau}\right)\right], \nonumber \\
\nu & = & \sqrt{\frac{9}{4}+\frac{1}{h^2_0}}, \label{earlytime} 
\end{eqnarray}
where $ J_{\pm \nu}(z) $ are Bessel functions, and $A_q$ and $B_q$ are
determined by the initial conditions on the mode functions:
\begin{eqnarray}
B_q = - \frac{1}{\sqrt{\omega_q}}{{\pi \, q}\over { 2 h_0 \, \sin{\nu \pi} }} 
\; \left[ {{i \omega_q -
\frac12 \, h_0 }\over q} \; J_{\nu}\left(\frac{q}{h_0}\right) - 
J'_{\nu}\left(\frac{q}{h_0}\right) \right],
\label{coefb} \\
A_q =   \frac{1}{\sqrt{\omega_q}}{{\pi \, q}\over { 2 h_0 \, \sin{\nu \pi} }} 
\; \left[ {{i \omega_q -
\frac12 \, h_0 }\over q} \;  J_{-\nu}\left(\frac{q}{h_0}\right) -  
J'_{-\nu}\left(\frac{q}{h_0}\right)\right] \;. \label{coefa}
\end{eqnarray}

After the physical wavevectors cross the horizon, i.e. when $qe^{-h_0
\tau}/h_0 \ll 1$ we find that the mode functions factorize: 
\begin{equation}
f_q(\tau) \approx  {{B_q} \over {\Gamma(1-\nu)}} 
\; \left( {{2h_0\, }\over q}\right)^{\nu}e^{(\nu-3/2)h_0 \tau}. \label{factor}
\end{equation}
 This  result reveals a very
important feature: because of the negative mass squared
term in the matter Lagrangian leading to symmetry breaking (and $\nu > 3/2$), we see
that all of the mode functions {\em grow exponentially} after horizon
crossing (for positive mass squared $ \nu < 3/2 $, and  
they would {\em decrease
exponentially} after horizon crossing). This exponential growth is a 
consequence of the spinodal instabilities which 
is a hallmark of the process of phase separation that
occurs to complete the phase transition. 
 We note, in addition that the time 
dependence is exactly given by that of the $ q=0 $ mode, i.e. the zero 
mode, which is a consequence of the redshifting of the wavevectors and 
the fact that after horizon crossing the contribution of the term
$q^2/a^2(\tau)$ in the equations of motion become negligible.
 We clearly  see that the quantum fluctuations grow exponentially and
they will begin to be of the order of the tree level terms in the
equations of motion when $g\Sigma(\tau) \approx 1$. At large
times 
$$
\Sigma(\tau) \approx {\cal F}^2(h_0) h_0^2
e^{(2\nu-3)h_0 \tau} \; ,  
$$
with
${\cal F}(h_0)$ a finite constant that depends on the initial conditions and
is found numerically to be of ${\cal O}(1)$ [see Fig.~\ref{fofh}].  

In terms of the initial dimensionful variables, the
 condition  $ g\Sigma(\tau) \approx 1$ translates
to $ \langle \psi^2(\vec x,t) \rangle_R \approx 2m^2_R/\lambda_R $, i.e. the quantum
fluctuations sample the minima of the (renormalized) tree level potential.
We find that the
time at which the contribution of the 
quantum fluctuations becomes of the same order as the tree level terms is
estimated to be\cite{desitter}
\begin{equation}
\tau_s \approx \frac{1}{(2\nu-3)h_0}
\ln\left[\frac{1}{g\, h_0^2 {\cal F}^2(h_0)}\right] 
= \frac32 h_0 \ln\left[\frac{1}{g\, h_0^2 {\cal F}^2(h_0)}\right]
+ {\cal O}(1/h_0).
\label{spinodaltime}
\end{equation}
At this time, the contribution of the quantum fluctuations makes the
back reaction very important and, as will be seen numerically, this
translates into the fact that $\tau_s$ also determines the end of the
De Sitter era and the end of inflation. The total number of e-folds during
the stage of exponential expansion of the scale factor (constant
$h_0$) is  given by   
\begin{equation}
N_e \approx \frac{1}{2\nu-3}\;
\ln\left[\frac{1}{g\; h_0^2\; {\cal F}^2(h_0)}\right] 
= \frac32\; h_0^2\; \ln\left[\frac{1}{g\;  h_0^2 \;{\cal
F}^2(h_0)}\right]  
+ {\cal O}(1)\label{efolds}
\end{equation}
For large $h_0$ we see that the number of e-folds scales as $h^2_0$ as well
as with the logarithm of the inverse coupling. 
These results (\ref{factor}) -- (\ref{efolds}) will be
confirmed numerically below and will be of paramount importance for the
interpretation of the main consequences of the dynamical evolution. 

\subsection{Classical or quantum behavior?}

Above we have analyzed the situation when $\eta(0) =0$ (or in dimensionful
variables $\phi(0)=0$). The typical analysis of inflaton dynamics in
the literature involves the {\em classical} evolution of $ \phi(t) $ 
with an initial condition in which $ \phi(0) $ is very close to zero (i.e.
the top of the potential hill) in the `slow-roll' regime, for which
$ \ddot{\phi} \ll 3H\dot{\phi}$. Thus, it is important
to quantify the initial conditions on $ \phi(t) $ for which the
dynamics will be 
determined by the classical evolution of $ \phi(t) $ and those for
which the quantum 
fluctuations dominate the dynamics. We can provide a criterion to
separate classical from quantum dynamics by analyzing the relevant time
scales, estimated by neglecting
non-linearities and backreaction effects. We consider the evolution 
of the zero mode in terms of dimensionless variables, and choose
$ \eta(0) \neq 0 $ and $\dot{\eta}(0) = 0 $.
($\dot{\eta}(0) \neq 0$ simply corresponds
to a shift in origin of time). We  assume  $ \eta(0)^2 << 1
$ which is the relevant case where  spinodal instabilities are important. 
We find
\begin{equation}
\eta(\tau) \approx \eta(0) \; e^{(\nu - \frac{3}{2})h_0\tau}\; .
\end{equation}

The non-linearities will become important and eventually terminate
inflation when $\eta(\tau) \approx 1$. This corresponds to a time scale
given by
\begin{equation}
\tau_c \approx \frac{\ln\left[1/ \eta(0)\right]}{(\nu - \frac{3}{2})\;
h_0}\; .\label{classtime}
\end{equation}
 If $ \tau_c $ is much smaller than the spinodal time $ \tau_s $ given
by Eq.~(\ref{spinodaltime}) then the {\em classical} evolution of the
zero mode will dominate the dynamics and the quantum fluctuations will
not 
become very large, although they will still undergo spinodal growth. On 
 the other hand, if $\tau_c \gg \tau_s$ the quantum fluctuations will
grow to be very large well before the zero mode reaches the non-linear
regime. In this case the dynamics will be determined completely by
the quantum fluctuations. Then the criterion for the classical or quantum
dynamics is given by
\begin{eqnarray} 
\eta(0) & \gg & \sqrt{g}\;h_0 \Longrightarrow \mbox{ classical dynamics }
\nonumber \\
\eta(0) & \ll & \sqrt{g}\;h_0 \Longrightarrow \mbox{ quantum dynamics }
\label{classquandyn} 
\end{eqnarray}
or in terms of dimensionful variables $\phi(0) \gg H_0$ leads to 
{\em classical dynamics} and $\phi(0) \ll H_0$ leads to 
{\em quantum dynamics}. 

However, even when the classical evolution of the
zero mode dominates the dynamics, the quantum fluctuations grow
exponentially after horizon crossing unless the value of $\phi(t)$ is
very close to the minimum of the tree level potential. In the large $
N $ approximation the spinodal line, that is the values of $\phi(t)$ for  
which there are spinodal instabilities, reaches all the way to the minimum
of the tree level potential as can be seen from the equations of motion for
the mode functions. 
Therefore even in the
classical case one must understand how to deal with quantum fluctuations
that grow after horizon crossing.  

\subsection{Numerics}
\begin{figure}
\epsfig{file=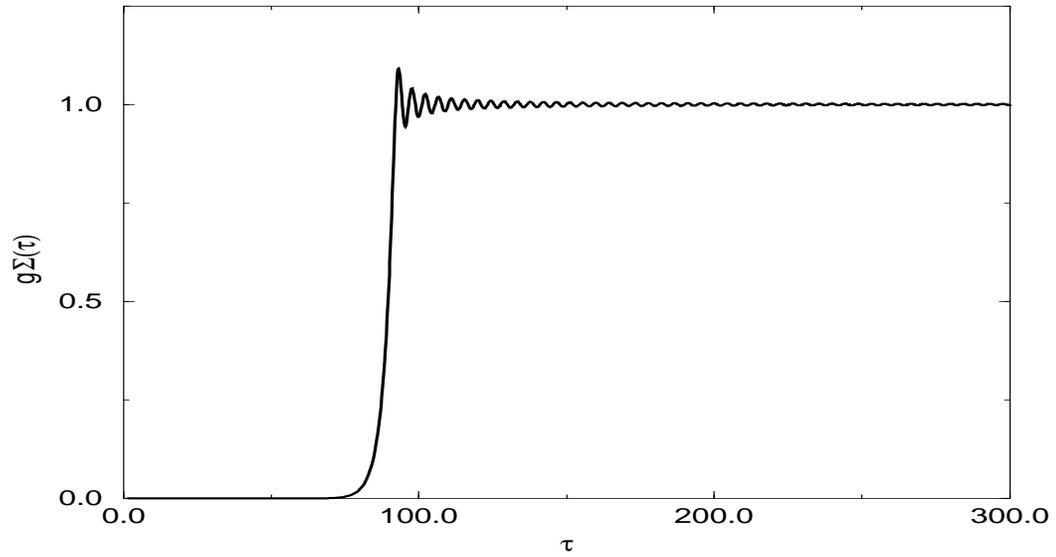,width=5.5in,height=3.2in}
\caption{ $g\Sigma$ vs. $\tau$, for $\eta(0)=0, \dot{\eta}(0)=0,
g = 10^{-14}, h_0 = 2.0$. }
\label{gsigma}
\end{figure}

\begin{figure}
\epsfig{file=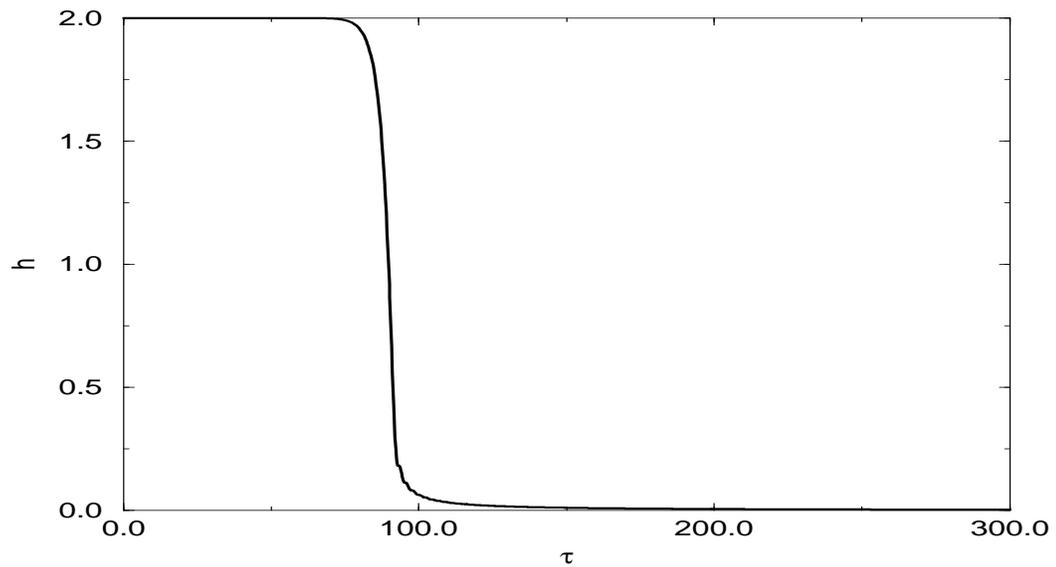,width=5.5in,height=3.2in}
\caption{$H(\tau)$ vs. $\tau$, for $\eta(0)=0, \dot{\eta}(0)=0, 
g = 10^{-14}, h_0 = 2.0 $. }
\label{hubblefig}
\end{figure}

\begin{figure}
\epsfig{file=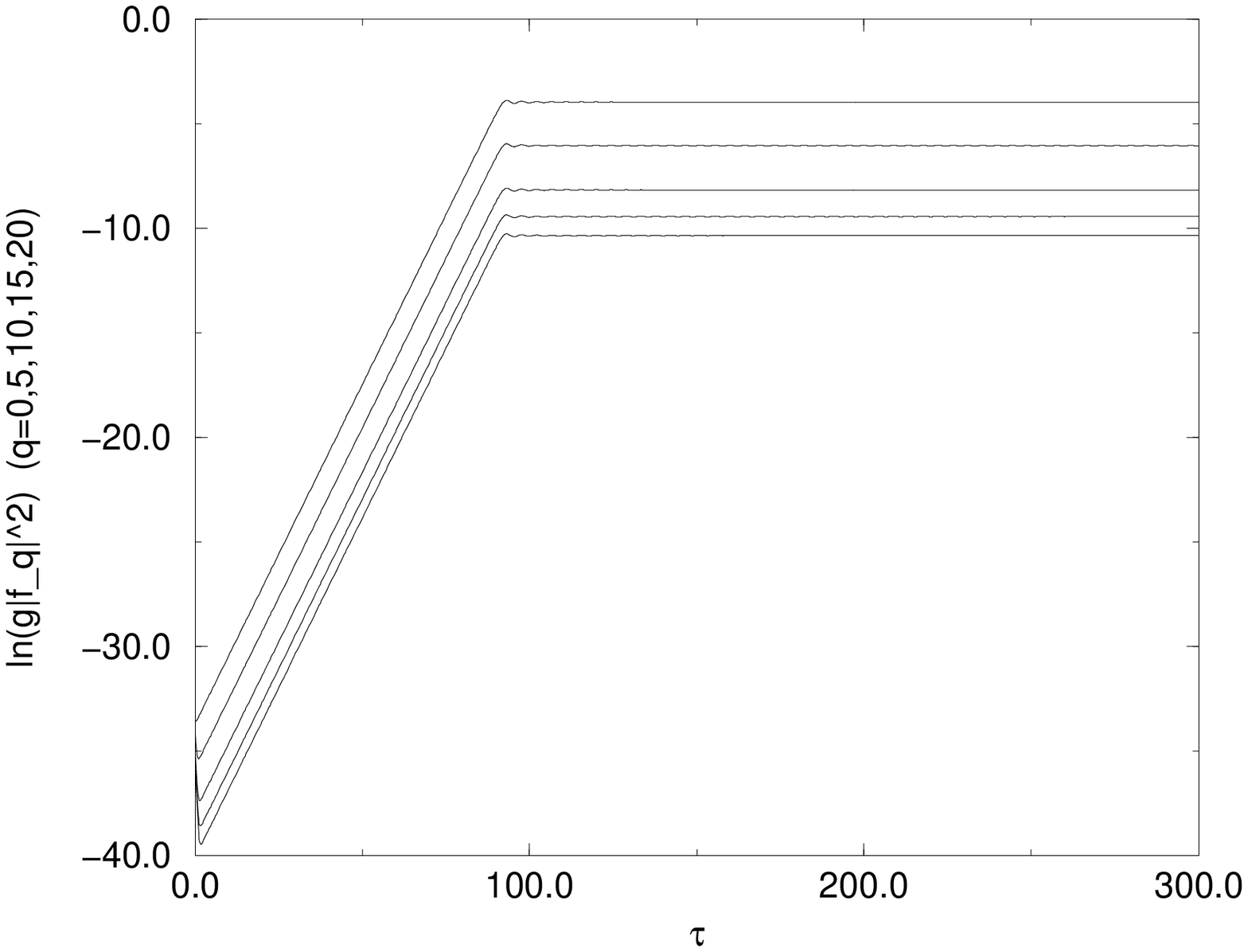,width=5.5in,height=3.2in}
\caption{$\ln(|f_q(\tau)|^2)$ vs. $\tau$, for $\eta(0)=0,
\dot{\eta}(0)=0,  g = 10^{-14}, h_0=2.0$ for
$q=0.0,5,10,15,20$ with smaller 
$q$ corresponding to larger values of $\ln(|f_q(\tau)|^2)$.}
\label{modu}
\end{figure}

\begin{figure}
\epsfig{file=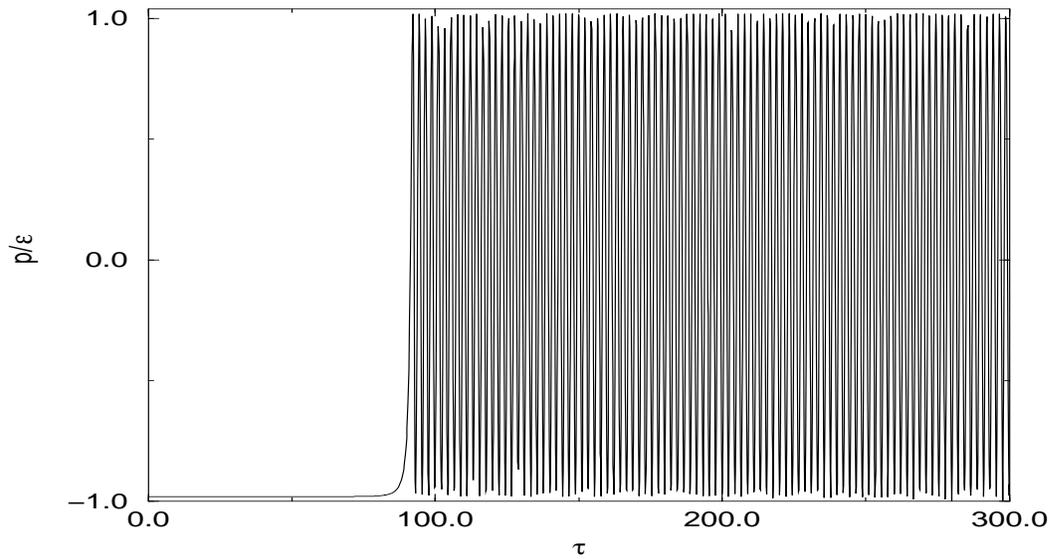,width=5.5in,height=3.2in}
\caption{$p/\varepsilon$ vs. $\tau$, for $\eta(0)=0, \dot{\eta}(0)=0,
g = 10^{-14}, h_0=2.0$.}
\label{povere}
\end{figure}

\begin{figure}
\epsfig{file=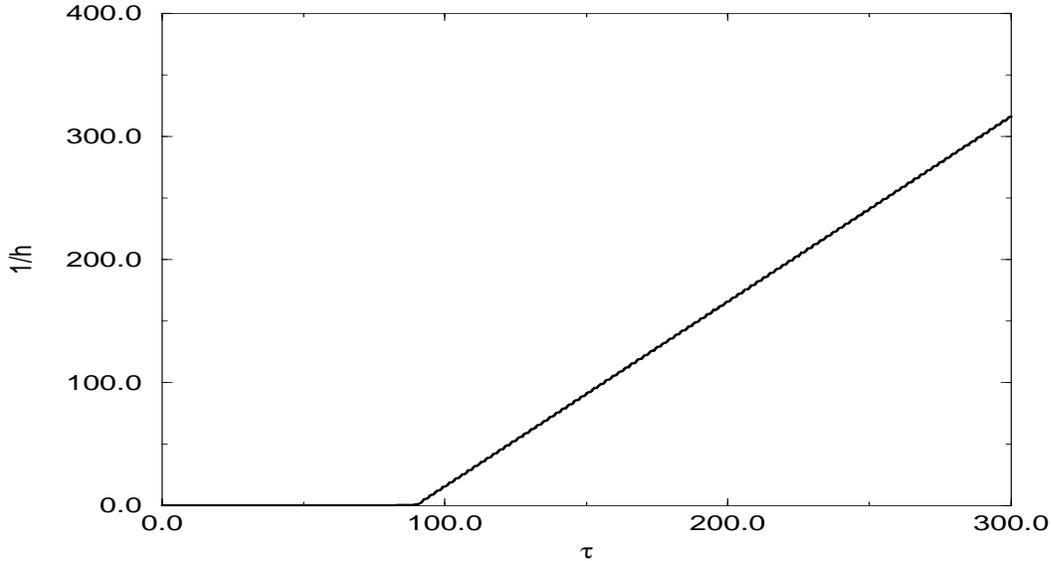,width=5.5in,height=3.2in}
\caption{$1/h(\tau)$ vs. $\tau$, for $\eta(0)=0, \dot{\eta}(0)=0,
g = 10^{-14}, h_0=2.0$. }
\label{hinverse}
\end{figure}

The time evolution is carried out by means of a fourth order Runge-Kutta
routine with adaptive step sizing while the momentum 
integrals are carried out using an 11-point Newton-Cotes integrator.  
The relative errors in both
the differential equation and the integration are of order $10^{-8}$.
We find that the energy is covariantly conserved throughout the evolution
to better than a part in a thousand.  Figures \ref{gsigma}--\ref{modu} 
show $g\Sigma(\tau)$ vs. $\tau$,
$h(\tau)$ vs. $\tau$ and $\ln(|f_q(\tau)|^2)$ vs. $\tau$ for several values
of $q$ with larger $q's$ corresponding to successively lower curves. 
Figures \ref{povere} and \ref{hinverse} show $p(\tau)/\varepsilon(\tau)$ 
and the horizon size $h^{-1}(\tau)$ for 
$g = 10^{-14} \; ; \; \eta(0)=0 \; ; \; \dot{\eta}(0)=0$
and we have chosen the representative value $h_0=2.0$.

Figures \ref{gsigma} and \ref{hubblefig} show clearly that 
when the contribution of the quantum
fluctuations $ g\Sigma(\tau) $ becomes of order 1 inflation ends,
and the time scale for $ g\Sigma(\tau) $ to reach ${\cal O}(1)$ is very well
described by  the estimate (\ref{spinodaltime}). From Fig.~\ref{gsigma} we see
that this happens for $\tau =\tau_s\approx 90$, leading to a number of
$e$-folds  
$N_e \approx 180$ which is correctly estimated by  (\ref{spinodaltime},
\ref{efolds}). 

Figure \ref{modu} shows clearly the factorization of the modes after they
cross the horizon as described by Eq.~(\ref{factor}).
 The slopes of all the curves after they become
straight lines in Fig.~\ref{modu} is given exactly by $(2\nu-3)$, whereas the
intercept depends on the initial condition on the mode function and
the larger the value of $ q $ the smaller the intercept because the
amplitude of the mode function is smaller initially. Although the
intercept depends on the initial conditions on the long-wavelength
modes, the slope is independent of the value of $q$ and is the same as
what would be obtained in the linear approximation for the {\em
square} of the zero mode at times 
long enough that the decaying solution can be neglected but short enough
that the effect of the non-linearities is very small.
 Notice from the figure that when inflation ends and
the non-linearities become important all of the modes effectively saturate.
This is also what one would expect from the solution of the zero mode:
exponential growth in early-intermediate times (neglecting the
decaying solution), with a growth exponent
given by $(\nu - 3/2)$ and an asymptotic behavior of small oscillations
around the equilibrium position, which for the zero mode is $\eta =1$, but
for the $q \neq 0$ modes depends on the initial conditions. 
All of the mode functions have this behavior once they cross the horizon.
We have also studied the phases of the mode functions and we found that 
they freeze after horizon crossing in the sense that they become independent
of time. This is natural since both the
real and imaginary parts of $ f_q(\tau) $ obey the same equation but
with different 
boundary conditions. After the physical wavelength crosses the horizon, the
dynamics is insensitive to the value of $q$ for real and imaginary parts and
the phases become independent of time. Again, this is a consequence of the
factorization of the modes.  

The growth of the quantum fluctuations is sufficient to end inflation
at a time given by $\tau_s$ in Eq.~(\ref{spinodaltime}). Furthermore 
Fig.~\ref{povere} shows that during the inflationary epoch
$p(\tau)/\varepsilon(\tau)  
\approx -1$ and the end of inflation is rather sharp at $\tau_s$ with
$p(\tau)/\varepsilon(\tau)$ oscillating between $\pm 1$ with zero average
over the cycles, resulting in matter domination. Figure
\ref{hinverse} shows this 
feature very clearly; $h(\tau)$ is constant during the De Sitter epoch and
becomes matter dominated after the end of inflation with $h^{-1}(\tau) 
\approx \frac32 (\tau -\tau_s)$. There are small oscillations around
this value because both $p(\tau)$ and $\varepsilon(\tau)$
oscillate. These oscillations 
are a result of small oscillations of the mode functions after they 
saturate, and are also a
feature of the solution for a zero mode. 

\begin{figure}
\epsfig{file=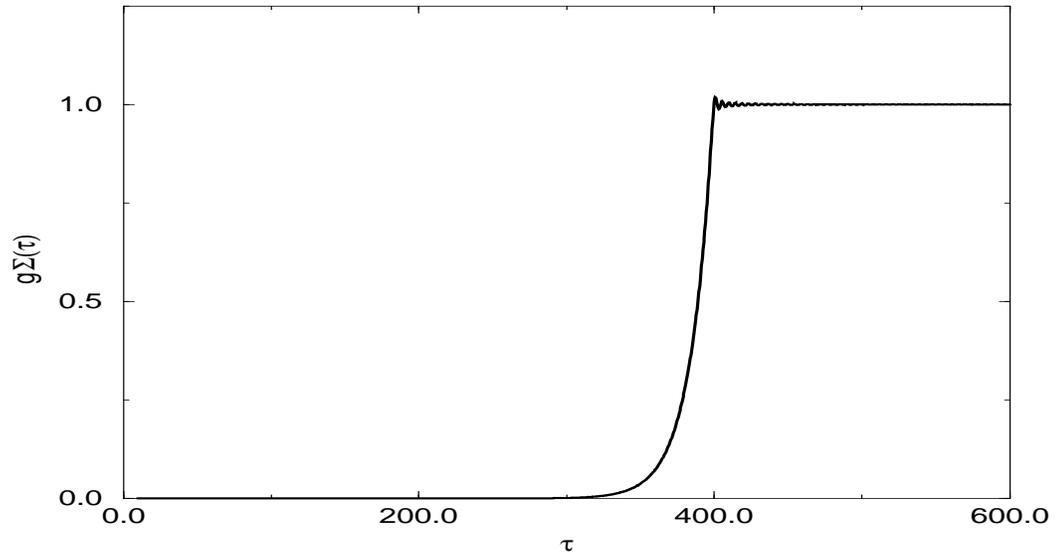,width=5.5in,height=3.2in}
\caption{ $g\Sigma$ vs. $\tau$, for $\eta(0)=0, \dot{\eta}(0)=0,
g = 10^{-14}, h_0 = 10.0$. }
\label{gsigma10}
\end{figure}

\begin{figure}
\epsfig{file=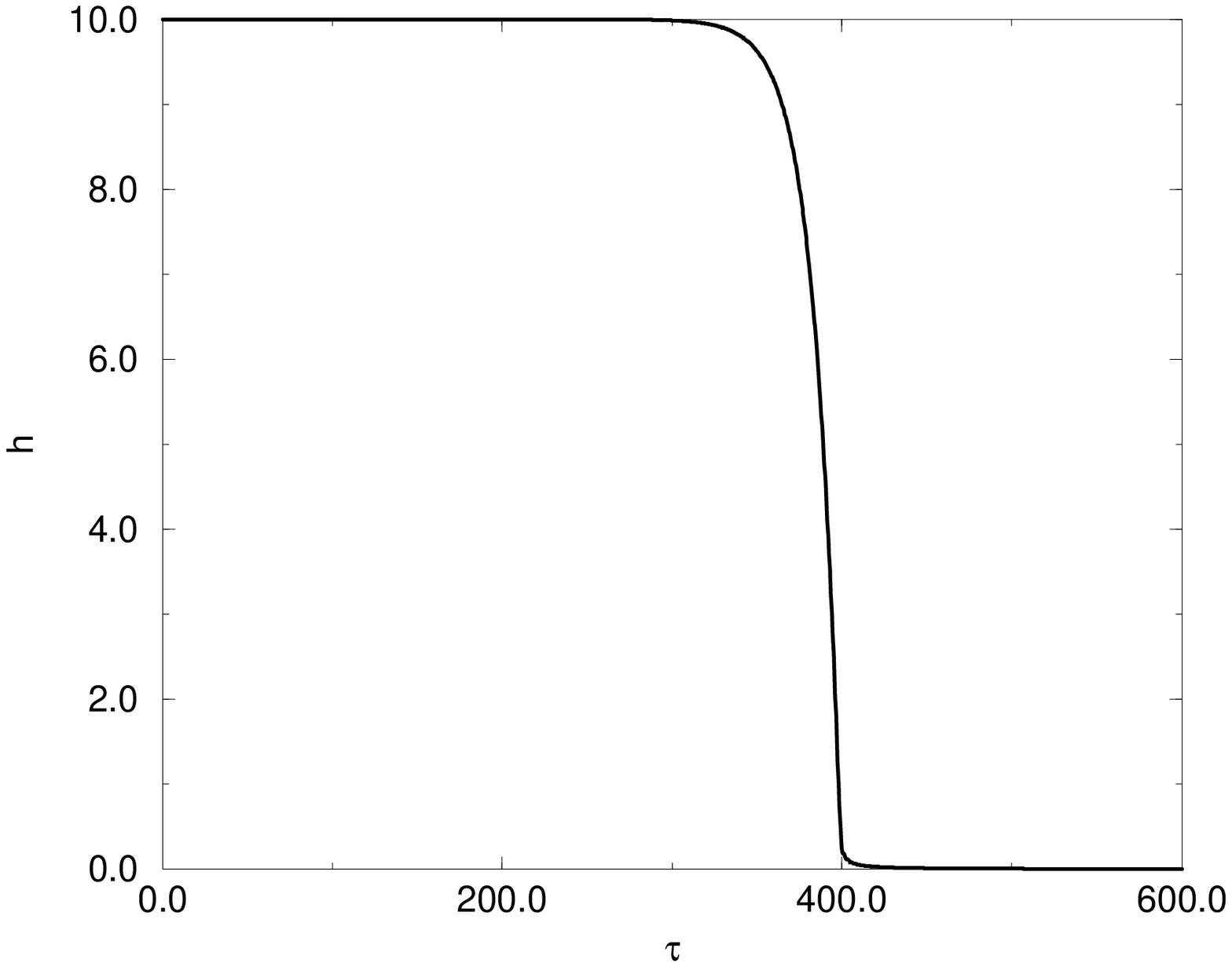,width=5.5in,height=3.2in}
\caption{$H(\tau)$ vs. $\tau$, for $\eta(0)=0, \dot{\eta}(0)=0, 
g = 10^{-14}, h_0 = 10.0 $. }
\label{hubble10}
\end{figure}

\begin{figure}
\epsfig{file=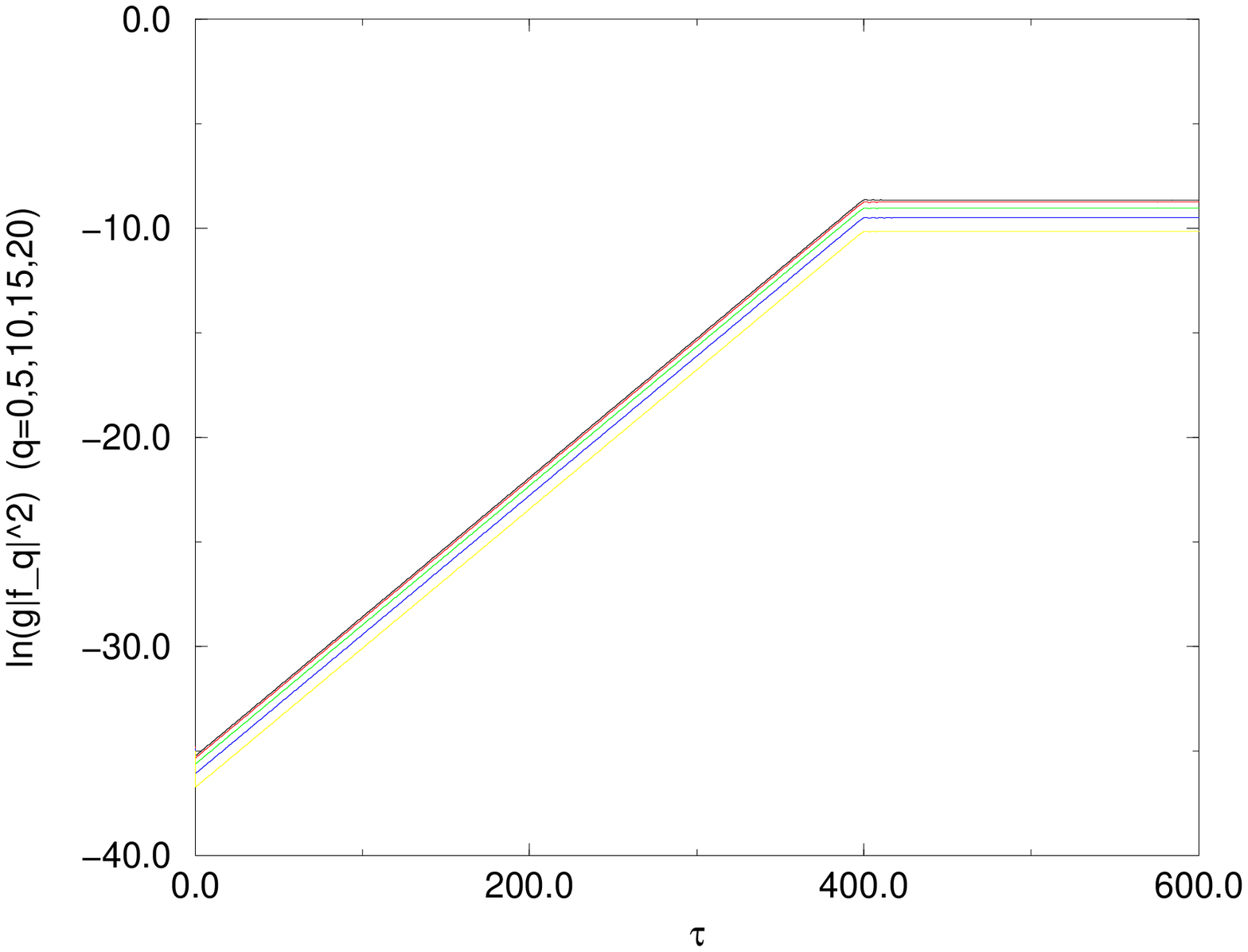,width=5.5in,height=3.2in}
\caption{$\ln(|f_q(\tau)|^2)$ vs. $\tau$, for $\eta(0)=0,
\dot{\eta}(0)=0,  g = 10^{-14}, h_0=10.0$ for
$q=0.0,5,10,15,20$ with smaller 
$q$ corresponding to larger values of $\ln(|f_q(\tau)|^2)$.}
\label{modu10}
\end{figure}

\begin{figure}
\epsfig{file=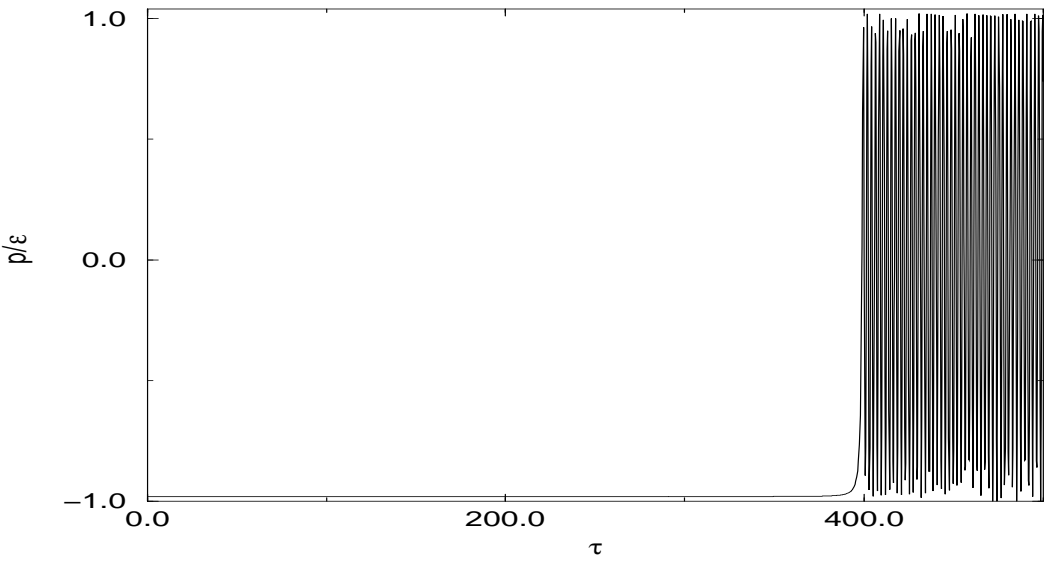,width=5.5in,height=3.2in}
\caption{$p/\varepsilon$ vs. $\tau$, for $\eta(0)=0, \dot{\eta}(0)=0,
g = 10^{-14}, h_0=10.0$.}
\label{povere10}
\end{figure}

All of these features hold for a variety of initial conditions.  As an
example, we show in Figs.~\ref{gsigma10}--\ref{povere10} the
plots corresponding to Figs.~\ref{gsigma}--\ref{povere} for the
case of an initial Hubble parameter of $h_0=10$.

\section{Zero Mode Assembly}
This remarkable feature of factorization of the mode functions after
horizon crossing can be elegantly summarized as
\begin{equation}
f_k(t)|_{k_{ph}(t) \ll H_0} = g(q,h_0)f_0(\tau),\label{factor2}
\end{equation}
with $k_{ph}(t) = k\,e^{-H_0t}$ being the physical momentum, 
$ g(q,h_0)$ a complex constant, and $f_0(\tau)$ a {\em real} function
of time that satisfies the mode equation with $q=0$ and real initial
conditions which will be inferred later.
 Since the factor $g(q,h_0)$ depends solely on the initial
conditions on the mode functions, it turns out that for two mode
functions corresponding to momenta $k_1,k_2$ that have crossed the
horizon at times $t_1 > t_2$, the ratio of the two mode functions  at
time $t$, ($t_s>t >t_1 >t_2$) is 
$$
{{f_{k_1}(t)}\over {f_{k_2}(t)}} \propto
e^{(\nu-\frac{3}{2})h_0 (\tau_1 - \tau_2)} > 1 \; . 
$$
Then if we consider the 
contribution of these modes to the  {\em renormalized} quantum
fluctuations a long time after the beginning of inflation (so as to
neglect the decaying solutions), we find that 
$$g\Sigma(\tau) \approx {\cal C}e^{(2\nu-3)h_0 \tau} + \mbox{ small} \; , 
$$
where 
`small' stands for
the contribution of mode functions associated with momenta that have not
yet crossed the horizon at time $\tau$, which give a perturbatively
small (of order $\lambda$) contribution.  
 We find that several e-folds
after the beginning of inflation but well before inflation ends, this
factorization of superhorizon modes implies the following:

\begin{eqnarray}
g\int q^2 dq \; |f^2_q(\tau)| & \approx & 
 |C_0|^2 f^2_0(\tau), \label{int1} \\ 
g\int q^2 dq \; |\dot{f}^2_q(\tau)| & \approx & |C_0|^2 \dot{f}^2_0(\tau),
\label{int2} \\
g\int \frac{q^4}{a^2(\tau)} dq \; |f^2_q(\tau)| & \approx & 
 \frac{|C_1|^2 }{a^2(\tau)}f^2_0(\tau), \label{int3}
\end{eqnarray}
where we have neglected the weak time dependence arising from the perturbatively small
contributions of the short-wavelength modes that have not yet crossed the
horizon, and the integrals above are to be understood as the fully
renormalized (subtracted), finite integrals. For $\eta = 0$, we note that (\ref{int1}) and the fact that $f_0(\tau)$ obeys the equation of motion for the mode with $q=0$ leads at once to the conclusion that 
in this regime $\left[g\Sigma(\tau)\right]^{\frac{1}{2}} = |C_0|f_0(\tau)$ obeys the zero mode equation of motion
\begin{equation}
\left[\frac{d^2}{d \tau^2}+ 3h \frac{d}{d\tau}-1+
(|C_0|f_0(\tau))^2\right]|C_0|f_0(\tau) = 0 \; . 
\label{zeromodeeff}
\end{equation}

\begin{figure}
\epsfig{file=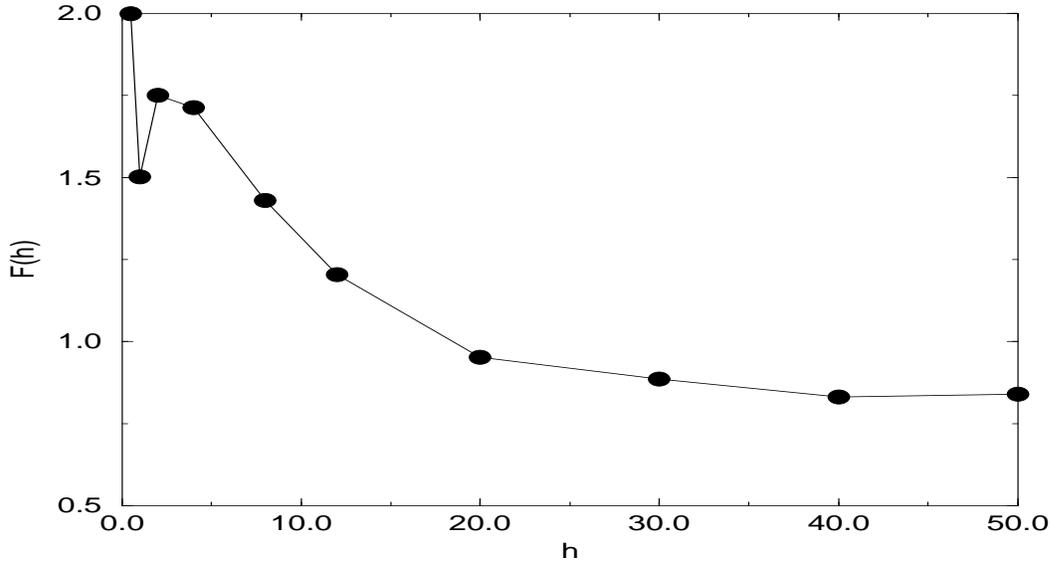,width=5.5in,height=3.2in}
\caption{${\cal F}(H/m)$ vs. $H$, where ${\cal F}(H/m)$
is defined by the relation
$\phi_{eff}(0) = (H/2\pi) {\cal F}(H/m)$ 
[see Eqs.~(\ref{effectivezeromode}) and (\ref{effzeromodein})].}
\label{fofh}
\end{figure}

It is clear that 
several e-folds after the beginning of inflation, we can define an
  effective zero mode   as 
\begin{equation}
\eta^2_{eff}(\tau) \equiv g\Sigma(\tau), \mbox{ or in dimensionful
variables, } \phi_{eff}(t) \equiv \left[\langle \psi^2(\vec x, t)
\rangle_R \right]^{\frac{1}{2}} 
\label{effectivezeromode}
\end{equation}
Although this identification seems natural, we emphasize that it
is by no means a trivial or ad-hoc statement. There are several
important features that allow an {\em unambiguous} identification:
i) $\left[\langle \psi^2(\vec x, t) \rangle_R \right]$ is a fully 
renormalized operator product and hence finite, ii) because of  the
factorization 
 of the superhorizon modes that enter in the evaluation of 
$\left[\langle \psi^2(\vec x, t) \rangle_R \right]$,  
$\phi_{eff}(t)$ (\ref{effectivezeromode}) 
{\em obeys the equation of motion for the zero mode}, iii) this identification 
is valid several e-folds after the beginning of inflation,
after the transient decaying solutions have died away and the integral
in $\langle \psi^2(\vec x,t) \rangle$
is dominated by the modes with wavevector $k$ that have crossed the horizon at 
$t(k) \ll t$.
Numerically we see that this identification holds throughout the
dynamics except for a very few e-folds at the beginning of inflation. This
factorization determines at once the initial conditions of the effective
zero mode that can be extracted numerically: after the first few e-folds and
long before the end of inflation we find
\begin{equation}
\phi_{eff}(t) \equiv \phi_{eff}(0)\; e^{(\nu-
\frac{3}{2})H_0t} \; \; , 
\label{effzeromodein} 
\end{equation}
where we parameterized 
$$
\phi_{eff}(0) \equiv \frac{H_0}{2\pi} \; {\cal F}(H_0/m)
$$
to make contact with the literature. 
As is shown in Fig.~\ref{fofh}, we find numerically that $ {\cal F}(H_0/m)  \approx
{\cal O}(1) $ for a large range of $ 0.1 \leq H_0/m \leq 50 $ and that this quantity
depends on the initial conditions of the long wavelength modes.  

Therefore, in summary, the effective composite zero mode obeys
\begin{equation}
\left[\frac{d^2}{d \tau^2}+ 3h \frac{d}{d\tau}-1+
\eta^2_{eff}(\tau)\right]\eta_{eff}(\tau) = 0 
\; ; \; \dot{\eta}_{eff}(\tau = 0) = (\nu -
\frac{3}{2}) \; \eta_{eff}(0) \; , \label{effzeromode}
\end{equation}
where $ \eta_{eff}(0)
 \equiv {{\sqrt{\lambda_R/2}}\over {m_R}} \; \phi_{eff}(0) $
 is obtained numerically for a given $ h_0 $ by fitting
the intermediate time behavior of  $ g\Sigma(\tau) $ with the growing zero
mode solution.

 Furthermore, this analysis
shows that in the case $\eta = 0$,  the renormalized energy and
pressure in this regime in which the renormalized integrals are
dominated by the superhorizon modes are given by  
\begin{eqnarray}
 \varepsilon_R(\tau) &  \approx   & \frac{2Nm^4_R}{\lambda_R} \left\{
\frac{1}{2}\dot{\eta}^2_{eff}+\frac{1}{4}\left(-1+\eta^2_{eff}\right)^2
\right\},
\label{effenergy} \\
 (p+\varepsilon)_R  & \approx & \frac{2Nm^4_R}{\lambda_R}\left\{
\dot{\eta}^2_{eff}\right\} \label{ppluseeff},
\end{eqnarray}
where we have neglected the contribution proportional to $1/a^2(\tau)$ 
because it is effectively red-shifted away after just a few e-folds.
We found numerically that this term is negligible after the interval
of time necessary for the superhorizon modes to dominate the contribution
to the integrals. 
Then the dynamics of the scale factor is given by 
\begin{equation}
h^2(\tau) = 4 h^2_0 \left\{
\frac{1}{2}\dot{\eta}^2_{eff}+\frac{1}{4}\left(-1+\eta^2_{eff}\right)^2
\right\}.
\label{effscalefactor}
\end{equation}

\begin{figure}
\epsfig{file=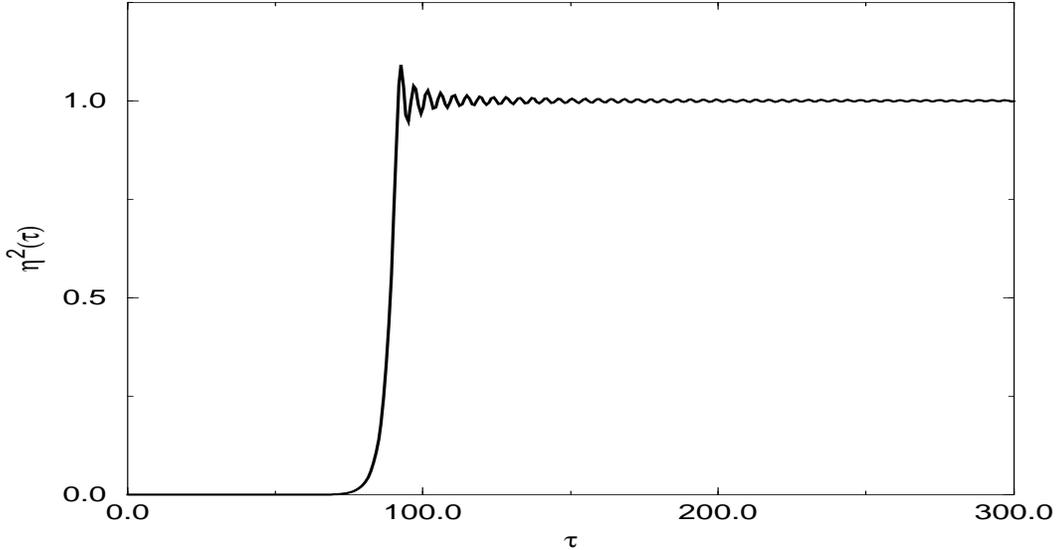,width=5.5in,height=3.2in}
\caption{ $\eta_{eff}^2(\tau)$ vs. $\tau$, for $\eta_{eff}(0)=3.94 \times
10^{-7}, \dot{\eta}_{eff}(0)=0.317\eta_{eff}(0),
g = 10^{-14}, h_0 = 2.0$. The initial conditions were obtained by
fitting the intermediate time regime of $g\Sigma(\tau)$ in
fig. (\ref{gsigma}). $\eta_{eff}(\tau)$ is the solution of
Eq.~(\ref{effzeromode}) 
with these initial conditions.}
\label{etaclas}
\end{figure}

\begin{figure}
\epsfig{file=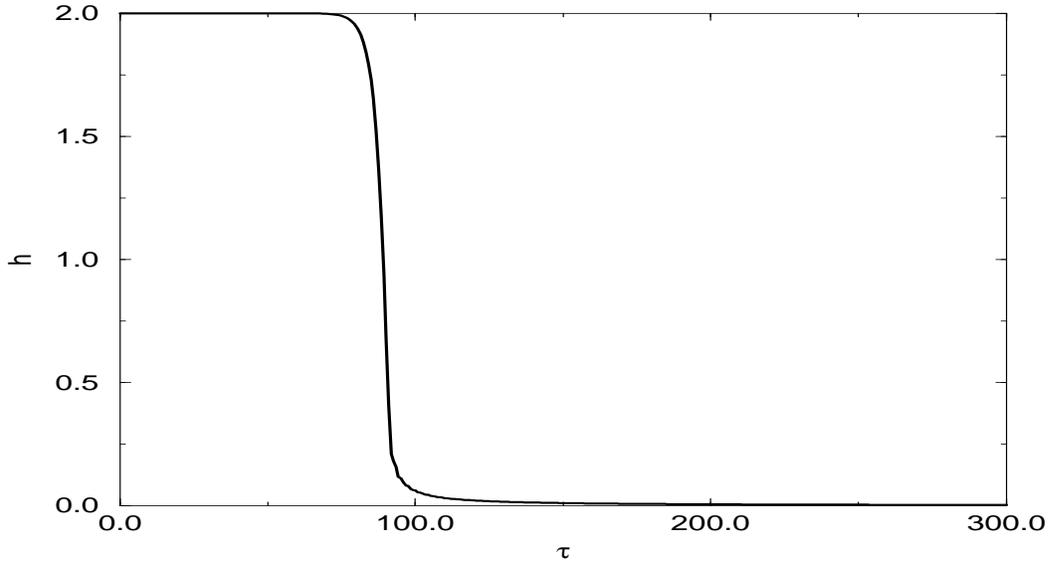,width=5.5in,height=3.2in}
\caption{$h(\tau)$ vs. $\tau$, obtained from the solution of
eqns. (\ref{effzeromode}) and (\ref{effscalefactor}) 
with the conditions of fig. (\ref{etaclas})).}
\label{hubclas}
\end{figure}

We have numerically evolved the set of effective equations (\ref{effzeromode}, \ref{effscalefactor}) by extracting the initial
condition for the effective zero mode from the intermediate time behavior
of $g\Sigma(\tau)$, and found a remarkable agreement  between the
evolution of $\eta^2_{eff}$ and $g\Sigma(\tau)$ and between 
the dynamics of the scale factor in terms of the evolution of
$\eta_{eff}(\tau)$, and the {\em full}
dynamics of the scale factor and quantum fluctuations within our numerical
accuracy. Figs.~\ref{etaclas} and \ref{hubclas} show the evolution
of $\eta^2_{eff}(\tau)$ and $h(\tau)$ respectively from the {\em
classical} evolution 
equations (\ref{effzeromode}) and (\ref{effscalefactor}) using the initial
condition  $ \eta_{eff}(0) $ extracted from the exponential fit of
$ g\Sigma(\tau) $ in the intermediate regime. These figures should be 
compared to Figs.~\ref{gsigma} and \ref{hubblefig}. We have also
numerically compared  
$p/\varepsilon$ given solely by the dynamics of the effective zero mode
and it is again numerically indistinguishable from that obtained with the
full evolution of the mode functions. 

This is one of the main results of our work. In summary: the modes that
become superhorizon sized and grow through the spinodal instabilities assemble
themselves into an effective composite zero mode a few e-folds after
the beginning of inflation. This effective zero mode drives the dynamics
of the FRW scale factor, terminating inflation when the non-linearities
become important. In terms of the underlying fluctuations, the spinodal
growth of superhorizon modes gives a non-perturbatively large contribution
to the energy momentum tensor that drives the dynamics of the scale factor.
Inflation terminates when the root mean square fluctuation probes the
equilibrium minima of the tree level potential. 

This phenomenon of zero mode assembly, i.e. the `classicalization'
of quantum mechanical fluctuations that grow after horizon crossing is
very similar to the interpretation of `decoherence without decoherence'
of Starobinsky and Polarski\cite{polarski}.

The extension of this analysis to the case for which $\eta(0) \neq 0$
is straightforward.  Since both $\eta(\tau)$ and 
$\sqrt{g\Sigma(\tau)} = |C_0|f_0(\tau)$ obey the equation for the
zero mode, Eq.~(\ref{zeromode}), it is clear that we can generalize
our definition of the effective zero mode to be
\begin{equation}
\eta_{eff}(\tau) \equiv \sqrt{\eta^2(\tau)+g\Sigma(\tau)}\; .
\label{effeta}
\end{equation}
which obeys the equation of motion of a {\em classical} zero mode:
\begin{equation}
\left[\frac{d^2}{d\tau^2}+3h\frac{d}{d\tau}-1+\eta_{eff}(\tau)^2\right]
\eta_{eff}(\tau) = 0 \; . \label{effzeroeqn}
\end{equation}
If this effective zero mode is to drive the FRW expansion, then the
additional condition
\begin{equation}
\dot{\eta}^2 f_0^2 - 2\eta\dot{\eta}f_0\dot{f_0} + \eta^2 \dot{f_0}^2 = 0
\end{equation}
must also be satisfied, which is simply the condition that the contribution 
of the kinetic term in the effective theory is the same as that in
the full theory.  One can easily show that this relation is indeed
satisfied if the mode functions factorize as in (\ref{factor2}) and if
the integrals (\ref{int1}) -- (\ref{int3}) are dominated by the 
contributions of the superhorizon mode functions.  This leads to the
conclusion that the gravitational dynamics is given by eqns. 
(\ref{effenergy}) -- (\ref{effscalefactor}) with $ \eta_{eff}(\tau) $ defined
by (\ref{effeta}).

We see that in {\em all} cases, the full large $N$ quantum dynamics in these 
models of inflationary phase transitions is well approximated by the
equivalent dynamics of a homogeneous, classical scalar field with initial
conditions on the effective field 
$\eta_{eff}(0) \geq \sqrt{g} h_0 {\cal F}(h_0)$.  
We have verified these
results numerically for the field and scale factor dynamics, finding that
the effective classical dynamics reproduces the results of the full
dynamics to within our numerical accuracy.  
We have also checked numerically
that the estimate for the classical to quantum crossover given by 
Eq.~(\ref{classquandyn}) is quantitatively correct. Thus in the classical case in
which $\eta(0) \gg \sqrt{\lambda}\; h_0$ we find that $\eta_{eff}(\tau) = \eta(\tau)$,
 whereas in the opposite, quantum case $\eta_{eff}(\tau) =
\sqrt{g\Sigma(\tau)}$. 

This remarkable feature of zero mode assembly of long-wavelength,
spinodally unstable modes is a consequence of the presence of the horizon.
It also explains why, despite the fact that asymptotically the
fluctuations sample the broken symmetry state, the equation of state is
that of matter. 
Since the excitations in the broken symmetry state are massless Goldstone
bosons one would expect radiation domination. However, the assembly
phenomenon, i.e. the redshifting of the wave vectors, makes these modes behave
exactly like zero momentum modes that give an equation of state of matter
(upon averaging over the small oscillations around the minimum).  

Sub-horizon modes at the end of inflation with $ q > h_0 \, e^{h_0
\tau_s} $ do not participate in the zero mode assembly. The behavior of such
modes do depend on $ q $ after the end of inflation. Notice that these
modes have extremely large comoving $ q $ since $  h_0 \, e^{h_0
\tau_s} \geq 10^{26} $. As discussed in Ref.~\cite{frwbig} such modes decrease
with time after inflation as $ \sim 1/a(\tau) $. 

\subsection*{Making sense of the fluctuations}
Having recognized the effective classical variable that can be interpreted
as the component of the field that drives the FRW background and rolls
down the classical potential hill, we want to recognize unambiguously
the small fluctuations. We have argued above that after horizon crossing,
all of the mode functions evolve proportionally to the zero mode,
and the question arises: which modes are assembled into the effective
zero mode whose dynamics drives the evolution of the FRW scale factor
and which modes are treated as perturbations? In principle every
$k\neq 0$ mode provides some spatial inhomogeneity, and assembling these
into an effective homogeneous zero mode seems in principle to do away with
the very inhomogeneities that one wants to study. However, scales of cosmological importance today first crossed the horizon during the
last 60 or so e-folds of inflation. Recently Grishchuk\cite{grishchuk}
 has argued that the
sensitivity of the measurements of $ \Delta T/T $ probe inhomogeneities on
scales $\approx 500$ times the size of the present horizon. Therefore scales
that are larger than these and that have first crossed the horizon much earlier than the last 60 e-folds of inflation are unobservable today and
can be treated as an effective homogeneous component, whereas the scales that
can be probed experimentally via the CMB inhomogeneities today must be treated 
separately as part of the inhomogeneous perturbations of the CMB. 

Thus a consistent description of the dynamics in terms of an effective
zero mode plus `small' quantum fluctuations can be given provided
the following requirements are met:
a) the total number of e-folds $N_e \gg 60$, b) all the modes that have
crossed the horizon {\em before} the last 60-65 e-folds are assembled into
an effective {\em classical} zero mode via $\phi_{eff}(t) = 
\left[\phi^2_0(t)+ \langle \psi^2(\vec x,t) \rangle_R
\right]^{\frac{1}{2}}$, c) the modes that cross the horizon during the
last 60--65 e-folds are accounted as `small' perturbations. The reason
for the requirement a) is that in the separation 
$\phi(\vec x, t) = \phi_{eff}(t)+\delta \phi(\vec x,t)$ one requires that
$\delta \phi(\vec x,t)/\phi_{eff}(t) \ll 1$. As argued above, after the 
modes cross the horizon, the ratio of amplitudes of the mode functions remains
constant and given by $e^{(\nu - \frac{3}{2})\Delta N}$ with $\Delta N$ 
being the number of e-folds between the crossing of the smaller $ k $ and the
crossing of the larger $ k $. Then for $\delta \phi(\vec x, t)$ to be much
smaller than the effective zero mode, it must be that the Fourier components
of $\delta \phi$ correspond to very large $k$'s at the beginning of inflation,
so that the effective zero mode can grow for a long time before the components
of $\delta \phi$ begin to grow under the spinodal instabilities. 
In fact requirement a) is not very severe; in Figs.~\ref{gsigma}--\ref{hinverse} 
we have taken
$h_0 = 2.0$ which is a very moderate value and yet for $\lambda = 10^{-12}$
the inflationary stage lasts for well over 100 e-folds, and as argued above, the
larger $h_0$ for fixed $\lambda$, the longer is the inflationary stage. 
Therefore under this set of conditions, the classical dynamics of the effective zero mode $\phi_{eff}(t)$  drives the FRW background, whereas
the inhomogeneous fluctuations $\delta \phi(\vec x,t)$, which are made up
of Fourier components with wavelengths that are much smaller than the
horizon at the beginning of inflation and that cross the horizon during
the last 60 e-folds, provide the inhomogeneities that seed density
perturbations.

\section{Scalar Metric Perturbations}
Having identified the effective zero mode and the `small perturbations',
we are now in position to provide an estimate for the amplitude and spectrum
of scalar metric perturbations. We use the formulation by Mukhanov,
Feldman and Branden\-berger\cite{mukhanov}, a short review of which was provided
in Sec.~2.5.

We are interested in determining the dynamics of $\Phi_k$ for those
wavevectors that cross the horizon during the last 60 e-folds before the
end of inflation. During the inflationary stage the numerical analysis
yields, to a very good approximation
\begin{equation}
H(t) \approx H_0  \; ; \; \phi_{eff}(t) = \phi_{eff}(0)\; e^{(\nu-
\frac{3}{2})H_0t}, \label{infla}
\end{equation}
where $H_0$ is the value of the Hubble constant during inflation, leading to 
\begin{equation}
\Phi_k(t) = e^{(\nu -2)H_0t}\left[a_k\;
H^{(1)}_{\beta}\left(\frac{ke^{-H_0t}}{H_0}\right) 
+b_k \; H^{(2)}_{\beta}\left(\frac{ke^{-H_0t}}{H_0}\right)\right]
\; ; \; \beta= \nu-1 \; .
\label{solbardeen}
\end{equation}
The coefficients $a_k$ and $b_k$ are determined by the initial conditions.

Since we are interested in the wavevectors that cross the horizon during
the last 60 e-folds, the consistency for the zero mode assembly and
the interpretation of `small perturbations' requires that there must be
many e-folds before the {\em last} 60. We are then considering wavevectors
that were deep inside the horizon at the onset of inflation. 
 Mukhanov et. al.\cite{mukhanov} show that $\Phi_k(t)$ is related to
the canonical `velocity field' that determines scalar  perturbations
of the metric and which is quantized with Bunch-Davies initial
conditions for the large $k$-mode functions. The relation between
$\Phi_k$ and $v$ and the  
initial conditions on $v$ lead at once to a determination of the
coefficients $a_k$ and $b_k$ for $k >> H_0$\cite{mukhanov} 
\begin{equation}
a_k = -\frac{3}{2} \left[\frac{8\pi}{3M^2_{Pl}}\right] \dot{\phi}_{eff}(0)
\sqrt{\frac{\pi}{2H_0}} \frac{1}{k} \quad ; \quad b_k = 0\; .
\label{coeffs}
\end{equation} 

Thus we find that the amplitude of scalar metric perturbations after 
horizon crossing is given by
\begin{equation}
|\delta_k(t)| = k^{\frac{3}{2}}|\Phi_k(t)| \approx
\frac{3}{2} \left[\frac{8\sqrt{\pi}}{3M^2_{Pl}}\right] \dot{\phi}_{eff}(0)
\left(\frac{2H_0}{k}\right)^{\nu -\frac{3}{2}}
e^{(2\nu-3)H_0t}\; .
\label{perts}
\end{equation}
The time dependence of $ |\delta_k(t)| $ displays the
unstable growth associated with the spinodal instabilities of
superhorizon modes and 
is a hallmark of the phase transition.
This time dependence can be understood from the constraint 
equation (\ref{constraint})
that relates the Bardeen potential to the gauge invariant field 
fluctuations\cite{mukhanov}, which in longitudinal gauge are identified
with $\delta \phi(\vec x,t)$.

Since the right hand side of (\ref{constraint}) is proportional to
$\dot{\phi}_{eff}/M^2_{Pl} \ll 1 $ during the inflationary epoch in this
model,  we can neglect the terms proportional
to $\dot{\Phi}_k $ and $\Phi_k$ on the left hand side of (\ref{gauginv}),
in which case the equation for the gauge invariant scalar field
fluctuation is the same as for the mode functions. In fact, since $ 
\delta \phi^{gi}_k$ is gauge invariant we can evaluate it in the 
longitudinal gauge wherein it is identified with the mode functions
$f_k(t)$. Then, absorbing a constant of integration in the initial
conditions for the Bardeen variable, we find
\begin{equation}
\Phi_k(t) \approx \frac{4\pi}{M^2_{Pl}a(t)}\int_{t_o}^t
a(t')\;\phi_{eff}(t')\; f_k(t')\; dt' + {\cal O}\left(\frac{1}{M^4_{Pl}}\right) ,
\label{bard}
\end{equation}
and using that $\phi(t) \propto e^{(\nu-3/2)H_0t} $ and that
 after horizon crossing $f_k(t) \propto e^{(\nu-3/2)H_0t}$, one obtains
at once the time dependence of the Bardeen variable after horizon
crossing. In particular the time dependence is found to be $\propto 
e^{(2\nu-3)H_0t}$. It is then clear that the time dependence is a reflection of
the spinodal (unstable) growth of the superhorizon field fluctuations. 

 To obtain the amplitude and spectrum
of density perturbations at {\em second} horizon crossing we use the
conservation law, Eq.~(\ref{xivar}),
which is valid after horizon crossing of the mode with wavevector $ k $.
Although this conservation law is an exact statement of superhorizon mode
solutions of Eq.~(\ref{bardeen}), 
we have obtained solutions assuming that during
the inflationary stage $H$ is constant and have neglected the $\dot{H}$ term in
Eq.~(\ref{bardeen}). Since during the inflationary stage,
\begin{equation}
\dot{H}(t) = -\frac{4\pi}{M^2_{Pl}}\, \dot{\phi}^2_{eff}(t) \propto
H^2_0 \; \left(\frac{d\eta_{eff}(\tau)}{d\tau}\right)^2\ll H^2_0 \label{Hdot}
\end{equation}
and $\ddot{\phi}/\dot{\phi} \approx H_0$, the above approximation is
justified. We then see that $\phi^2_{eff}(t) \propto
e^{(2\nu-3)H_0t}$ which is the same time dependence as that of
$\Phi_k(t)$. Thus the
term proportional to $1/(1+p/\varepsilon)$ in Eq.~(\ref{xivar}) 
is indeed constant in time after horizon crossing. On the other
hand, the term that
does not have this denominator evolves in time but is of order 
$(1+p/\varepsilon) =
-2\dot{H}/3H^2 \ll 1$ with respect to the constant term and therefore can be
neglected. Thus, we confirm that the variable $\xi$ is conserved
up to the small term proportional to $(1+p/\varepsilon)\Phi_k$ which
is negligible during the inflationary stage. 
This small time dependence is consistent with the fact
that we neglected the $\dot{H}$ term in the equation of motion for $\Phi_k(t)$.
 
 However, upon second horizon crossing
it is straightforward to see that $\dot{\Phi}_k(t_f) \approx 0$. The 
reason for this assertion can be seen as follows: Eq.~(\ref{gauginv}) shows that
at long times, when the effective zero mode is oscillating around the minimum
of the potential with a very small amplitude and when the time dependence of
the fluctuations has saturated (see Fig.~\ref{modu}), $\Phi_k$ will redshift
as $\approx 1/a(t)$\cite{frwbig} and its derivative becomes extremely small. 

 Using
this conservation law,  assuming matter domination at second horizon 
crossing,  and $\dot{\Phi}_k(t_f)\approx 0$\cite{mukhanov}, we find
\begin{equation}
|\delta_k(t_f)| = \frac{12 \, \Gamma(\nu)\,\sqrt{\pi}}{5\, (\nu-\frac{3}{2})\, 
{\cal F}(H_0/m)} \left(\frac{2H_0}{k}\right)^{\nu-\frac{3}{2}},
\label{amplitude}
\end{equation}
where ${\cal F}(H_0/m)$ determines the initial amplitude of the effective
zero mode (\ref{effzeromodein}). 
We can now read the power spectrum per logarithmic $k$ interval
\begin{equation}
{\cal P}_s(k) = |\delta_k|^2 \propto k^{-2(\nu-\frac{3}{2})} 
\equiv k^{n_s-1} \; .
\end{equation}
leading to the index for scalar density perturbations
\begin{equation}
n_s = 1-2\left(\nu-\frac{3}{2}\right) \; . \label{index}
\end{equation}

For $H_0/m \gg 1$, we can expand $\nu-3/2$ as a series in $m^2/H_0^2$ in
Eq.~(\ref{amplitude}).  Given that the comoving wavenumber of the mode which 
crosses the horizon $n$ $e$-folds before the end of inflation is $k=H_0 e^{(N_e-n)}$ 
where $N_e$ is given by (\ref{efolds}), we arrive at the following expression
for the amplitude of fluctuations on the scale corresponding to $n$ 
in terms of the De Sitter Hubble constant and the coupling $\lambda$:
\begin{equation}
|\delta_n(t_f)| \simeq \frac{9 H_0^3}{5\sqrt{2} m^3} \left(2e^n\right)^{m^2/3H_0^2}
\sqrt{\lambda} \left[1+\frac{2m^2}{3H_0^2} \left(\frac76 - \ln 2 - \frac{\gamma}{2}
\right) + {\cal O}\left( \frac{m^4}{H_0^4} \right) \right] \; .
\label{amplitude_n}
\end{equation}
Here, $\gamma$ is Euler's constant.  Note the explicit dependence of the 
amplitude of density perturbations on $\sqrt{\lambda}$.  For $n \approx 60$,
the factor $\exp(nm^2/3H_0^2)$ is ${\cal O}(100)$ for $H_0/m = 2$, while
it is ${\cal O}(1)$ for $H_0/m \geq 4$.  Notice that for $H_0/m$ large,
the amplitude increases approximately as $(H_0/m)^3$, which will place 
strong restrictions on $\lambda$ in such models.

We remark that we have not included the small corrections to the dynamics
of the effective zero mode and the scale factor arising from the
non-linearities. We have found numerically that these nonlinearities 
are only significant for the
modes that cross about 60 $e$-folds before the end of inflation for
values of the Hubble parameter $H_0/m_R > 5$.  The effect of these
non-linearities in the large $N$ limit is to slow somewhat the exponential
growth of these modes, with the result of shifting the power spectrum
closer to an exact Harrison-Zeldovich spectrum with $n_s =1$.  Since
for $H_0/m_R > 5$ the power spectrum given by (\ref{index}) differs from
one by at most a few percent, the effects of the non-linearities are
expected to be observationally unimportant.
The spectrum given by (\ref{amplitude}) is
similar to that obtained in references\cite{lindebook,guthpi} although
the amplitude differs from that obtained there. In addition, we do not
assume slow roll for which $(\nu - \frac{3}{2})\ll 1$, although 
this would be the case if $N_e \gg 60$.

We emphasize an important
feature of the spectrum: it has more power at {\em long
wavelengths} because $\nu-3/2 > 0$. This is recognized to be a
consequence 
of the spinodal instabilities that result in the growth of long wavelength
modes and therefore in more power for these modes.  
This seems to be a robust prediction of new inflationary scenarios in
which the potential has negative second derivative in the region of field
space that produces inflation.  

 It is at this
stage that we recognize the consistency of our approach for separating
the composite effective zero mode from the small fluctuations. We have
argued above that many more than 60 $e$-folds are required for consistency,
and that the small fluctuations correspond to those modes that cross
the horizon during the last 60 $e$-folds of the inflationary stage. For these
modes $H_0/k = e^{-H_0 t^*(k)}$ where $t^*(k)$ is the time since the beginning 
of inflation of horizon crossing of the mode with wavevector $k$. 
The scale that  corresponds to the Hubble radius today $\lambda_0
=2\pi/k_0$ is the first to cross during the last 60 or so $e$-folds
before the end of 
inflation. Smaller scales today will correspond to $k > k_0$ at the
onset of inflation since they will cross the first horizon later and
therefore will reenter earlier. The bound on $|\delta_{k_0}| \propto  
\Delta T/ T \leq  10^{-5}$ on
these scales provides a lower bound on the number of $e$-folds required for
these type of models to be consistent:
\begin{equation}
N_e >
60+\frac{12}{\nu-\frac{3}{2}}-\frac{\ln(\nu-\frac{3}{2})}{\nu-\frac{3}{2}}\; ,
\label{numbofefolds}
\end{equation}
where we have written the total number of $e$-folds as $N_e=H_0\; t^*(k_0)+60$.
This in turn can be translated into a bound on the coupling constant using
the estimate given by Eq.~(\ref{efolds}).

Since the COBE satellite observes microwave background temperature 
variation with a resolution of a single degree or larger, which is 
approximately the size of the particle horizon at the time of 
recombination, we may directly compare the power spectrum for thermal
fluctuations measured by COBE with the spectrum computed here for 
fluctuations at second horizon crossing [see Eq.~(\ref{dTT})].
The four year COBE  DMR Sky Map\cite{gorski} gives $n \approx 1.2 \pm 0.3$
thus providing an upper bound on $\nu$
\begin{equation}
0 \leq \nu-\frac{3}{2} \leq 0.05 \label{cobebound}
\end{equation}
corresponding to $h_0 \geq 2.6$. We then find that these values of $h_0$ and
$\lambda \approx 10^{-12}-10^{-14}$ provide sufficient $e$-folds to satisfy
the constraint for scalar density perturbations.

\section{Decoherence: Quantum to Classical Transition During Inflation} 

An important aspect of cosmological perturbations is that 
they are of quantum origin but eventually they become classical as
they are responsible for the small classical metric perturbations. This
quantum to classical crossover is associated with a decoherence process and
has received much attention\cite{polarski,salman}. 

Recent work on decoherence focused on the description of the evolution of the
density matrix for a free scalar massless field that represents the ``velocity
field''\cite{mukhanov} associated with scalar density perturbations\cite{polarski}.  In this section we study
the quantum to classical transition of superhorizon modes for the Bardeen 
variable by relating these to the field mode functions and analyzing the full
time evolution of the density matrix of the matter field. This is accomplished
with the identification given by  equation (\ref{bard}) which relates the mode
functions of the Bardeen variable with those of the scalar field. This relation
establishes that  in the models under consideration the classicality of the Bardeen variable is determined by the
classicality of the scalar field modes.

In the situation under consideration, long-wavelength field modes become
spinodally unstable and grow exponentially after horizon crossing. The 
factorization (\ref{factor}) results in the phases of these modes ``freezing
out''. This feature and the growth in amplitude entail that these modes become
classical. The relation (\ref{bard}) in turn implies that these features 
also apply to the superhorizon modes of the  Bardeen potential. 

 Therefore we can address the quantum
to classical transition of the Bardeen variable (gravitational potential) by
analyzing the evolution of the density matrix for the matter field. 

To make contact with previous work\cite{polarski,salman} we choose to study
the evolution of the field density matrix in conformal time (see Sec.~2.4), 
although the same features will be displayed in comoving time.  Note
that we use $\eta$ to denote conformal time throughout this section;
it should not be confused with the dimensionless, comoving time zero
mode $\eta(t)$.

The conformal time Hamiltonian operator, which is the generator of translations
in $\eta$, is given by
\begin{equation}
H_{\eta}= \int d^3x \left\{ \frac{1}{2}\Pi^2_{\vec{\chi}}+\frac{1}{2}
(\vec{\nabla}\vec{\chi})^2+{\cal{V}}(\vec{\chi}) \right \}, \label{confham}
\end{equation}
with $\vec{\Pi}_{\chi}$ being the canonical momentum conjugate to $\vec{\chi}$,
$\vec{\Pi}_{\chi} = \vec{\chi}'$. 
Separating the zero mode of the field $\vec{\chi}$ 
\begin{equation}
\vec{\chi}(\vec x, \eta) = \chi_0(\eta)\delta_{i,1} + {\hat{\vec{\chi}}}(\vec x,\eta),
\end{equation}
and performing the large $N$ factorization on the
fluctuations we find 
that the Hamiltonian becomes linear plus quadratic in the fluctuations, and
similar to a Minkowski space-time Hamiltonian with a $\eta$ dependent mass term given by
\begin{equation}
{\cal{M}}^2(\eta) = a^2(\eta) \left[m^2+\left(\xi-\frac{1}{6}\right)\,{\cal{R}}
+ \frac{\lambda}{2}\,\chi_0^2(\eta) + \frac{\lambda}{2}\,
\langle \hat{\chi}^2 \rangle\right]. \label{masseff}
\end{equation}

We can now follow the steps and use the results of ref.\cite{frw1} for the
conformal time evolution of the density matrix by setting $a(t)=1$ in the
proper equations of that reference and replacing the frequencies by
\begin{equation}
\omega^2_k(\eta) = \vec{k}^2 + {\cal{M}}^2(\eta)\; . \label{freqs}
\end{equation}
The expectation value in Eq.~(\ref{masseff}) and that of the energy momentum 
tensor are obtained in this
$\eta$ evolved density matrix.
[As is clear, we obtain in this way the self-consistent dynamics in the curved cosmological background (\ref{metrica})].

The time evolution of the kernels in the density matrix (see \cite{frw1})
is determined by the mode functions that obey
\begin{equation}
\left[ \frac{d^2}{d\eta^2}+k^2+{\cal{M}}^2(\eta)\right] F_k(\eta)=0.
\label{fmodeqn}
\end{equation}
The Wronskian of these mode functions
\begin{equation}\label{wff}
{\cal{W}}(F,F^*)= F'_k F^*_k-F_k F'^{*}_k
\end{equation} 
is a constant. It is natural to impose initial conditions such that at the
initial $\eta$ the density matrix  describes a pure state which is the
instantaneous ground state of the  Hamiltonian at this initial time. 
 This implies that the initial conditions of the mode functions
$F_k(\eta)$ be chosen to be (see \cite{frw1})
\begin{equation}
F_k(\eta_o)= \frac{1}{\sqrt{\omega_k(\eta_o)}} \; \; ; \; 
F'_k(\eta_o)= -i\omega_k(\eta_o) \;  F_k(\eta_o). \label{inicond}
\end{equation}
With such initial conditions, the Wronskian (\ref{wff}) takes the value
\begin{equation}\label{wro}
{\cal{W}}(F,F^*)= -2i \; .
\end{equation}

The Heisenberg field operators $\hat{\chi}(\vec x, \eta)$  and 
their canonical momenta $\Pi_{\chi}(\vec x, \eta)$ can now be expanded as:
\begin{eqnarray}
&& \hat{\vec{\chi}}(\vec x, \eta) = \int {{d^3k}\over {(2\pi)^{3/2}}} 
\left[ \vec{a}_{ \vec k} \; F_k(\eta)+ \vec{a}^{\dagger}_{ -\vec k} \;F^*_k(\eta) \right]
 e^{i \vec k \cdot \vec x}, \label{heisop}\\ 
&& \vec{\Pi}_{\chi}(\vec x, \eta) =  
\int {{d^3k}\over {(2\pi)^{3/2}}} 
\left[ \vec{a}_{ \vec k}  \;F'_k(\eta)+ \vec{a}^{\dagger}_{ -\vec k}
\;F'^{*}_k(\eta) \right] 
 e^{i \vec k \cdot \vec x}, \label{canheisop}
\end{eqnarray}
with the time independent creation and
annihilation operators $ \vec{a}_{ \vec k} $ and $ \vec{a}^{\dagger}_{ \vec k} $ 
obeying canonical commutation relations. Since the fluctuation fields
in comoving and conformal time are related by a conformal rescaling
given by Eq.~(\ref{confres})
it is straightforward to see that the mode functions in comoving time
$ t $ are related to those in conformal time simply as
\begin{equation}
f_k(t) = \frac{F_k(\eta)}{a(\eta)}.
\end{equation}
Therefore the initial conditions given in Eq.~(\ref{inicond}) on the conformal time mode
functions and the choice $a(t_0)= a(\eta_0)=1$  imply the initial conditions for the mode functions in comoving time given by Eq.~(\ref{initcond}).

In conformal time, the density matrix takes the form
\begin{eqnarray}\label{matden}
\rho[\Phi,\tilde{\Phi},\eta] & = & \prod_{\vec{k}} {\cal{N}}_k(\eta)
\exp\left\{ 
- \frac12 A_k(\eta) \; \vec{\eta}_{\vec{k}}(\eta)\cdot
\vec{\eta}_{-{\vec{k}}}(\eta)- 
\frac12 A^*_k(\eta) \;
\tilde{\vec{\eta}}_{\vec{k}}(\eta)\cdot 
\tilde{\vec{\eta}}_{-{\vec{k}}}(\eta) 
\right. \nonumber \\  &   & \left.
- \;  B_k(\eta) \; \vec{\eta}_{\vec{k}}({\cal
T})\cdot\tilde{\vec{\eta}}_{-{\vec{k}}}(\eta) 
+i\, \vec{\pi}_{\vec{k}}(\eta)\cdot\left(\vec{\eta}_{-{\vec{k}}}(\eta)-
\tilde{\vec{\eta}}_{-{\vec{k}}}(\eta)\right) \right\} ;\ , \\
\vec{\eta}_{\vec{k}}(\eta)          & = &
\vec{\chi}_{\vec{k}}(\eta)-\chi_0(\eta)\,\delta_{i,1} \;\delta(\vec{k}) 
\; \; ; \;  \; 
\tilde{\vec{\eta}}_k(\eta)        = 
{\tilde{\vec{\chi}}}_k(\eta)-\chi_0(\eta)\, 
\delta_{i,1}\; \delta(\vec{k}) \; . \nonumber
\end{eqnarray}
$\vec{\pi}_{\vec{k}}(\eta) $ is the Fourier 
transform of $ \Pi_{\chi}(\eta,\vec{x}) $. 

In conformal time quantization and in the Schr\"odinger representation in which
the field $\chi$ is diagonal the conformal time evolution of the density matrix
is via the conformal time Hamiltonian (\ref{confham}). The evolution equations
for the covariances are obtained from those given in ref.\cite{frw1} by
setting $a(t) = 1$ and using the frequencies $\omega^2_k(\eta)=
k^2+{\cal{M}}^2(\eta)$. In particular, by setting the covariance of the
diagonal elements (given by equation (2.20) in \cite{frw1}; see
also equation (2.44) of \cite{frw1}),
\begin{equation}\label{FpF}
A_k(\eta) = -i \, \frac{F'^*_k(\eta)}{F^*_k(\eta)}.
\end{equation}

More explicitly \cite{frw1},
\begin{eqnarray}\label{coero}
{\cal{N}}_k(\eta) &=&{\cal{N}}_k(\eta_0)\;
\exp\left[\int_{\eta_0}^{\eta} A_{Ik}(\eta')\;d\eta' \right] = 
{{{\cal N }_k(\eta_0)} \over { \sqrt{\omega_k(\eta_o)} \; |F_k(\eta)|}} 
\; , \cr \cr
A_{Ik}(\eta) &=& - {d \over {d\eta}}\log|F_k(\eta)| = -{\dot a}(t)-
a(t)\;{d \over {dt}}\log|f_k(t)| \; , \cr \cr
A_{Rk}(\eta) &=& {1 \over { |F_k(\eta)|^2}} \, = { 1\over 
{ a(t)^2 \; |f_k(t)|^2 }} \; ,  \\ \cr
B_k(\eta) &\equiv & 0 \; , \nonumber
\end{eqnarray}
where $A_{Rk}$ and $A_{Ik}$ are respectively the real and imaginary parts
of $A_k$ and we have used the value of the Wronskian (\ref{wro}) in
evaluating (\ref{coero}).

The coefficients $ A_k(\eta) $ and $ {\cal{N}}_k(\eta) $
in the Gaussian density matrix (\ref{matden}) are
completely determined by  the conformal mode functions $ F_k(\eta) $
[or alternatively the comoving time mode functions $f_k(t)$].

Let us study the time behavior  of these coefficients. 
During inflation, $ a(t) \approx  e^{h_0t}$,
and the mode functions  factorize after horizon crossing. Further, superhorizon
modes  grow in cosmic time  as in 
Eq.~(\ref{factor}):
$$
a^2(t) |f_k(t)|^2 \approx { 1 \over  {\cal D}_k} \; e^{(2\nu -1)h_0t}
$$ 
where the coefficient $ {\cal D}_k $ can be read from Eq.~(\ref{factor}).

We emphasize that this is a {\em result of the full evolution} as displayed from the numerical solution in Fig.~\ref{modu}. These mode functions
encode all of the self-consistent and non-perturbative features of the
dynamics. This should be contrasted with other studies in which typically
free field modes in a background metric are used.  

 Inserting this expression in Eqs.~(\ref{coero}) yields 
\begin{eqnarray}\label{roasi}
A_{Ik}(\eta)&\buildrel{t \to \infty}\over =&-h_0 \;  e^{h_0t} \left(\nu -
\frac12 \right)+ {\cal O}(e^{-h_0t})    \; , \cr \cr
A_{Rk}(\eta) &\buildrel{t \to \infty}\over =& {\cal D}_k\; \; 
 e^{-(2\nu -1)h_0t}  . \nonumber
\end{eqnarray}

Since $ \nu -\frac12 > 1 $,
we see that the imaginary part of the covariance $  A_{Ik}(\eta) $
{\it grows} very fast. Hence, the off-diagonal elements of $
\rho[\Phi,\tilde{\Phi},\eta] $ oscillate wildly after a few $e$-folds of
inflation. In particular their contribution to expectation values of operators
will be washed out. That is, we quickly reach a {\it classical} regime where
only 
the diagonal part of the density matrix is relevant:
\begin{equation}\label{claden}
\rho[\Phi,\Phi,\eta] = \prod_{\vec{k}} {\cal{N}}_k(\eta)
\exp\left\{ -  A_{Rk}(\eta) 
\; \eta_{\vec{k}}(\eta)\;{\eta}_{-{\vec{k}}}(\eta) \right\}.
\end{equation}

The real part of the covariance $ A_{Rk}(\eta) $ (as well as any non-zero
mixing kernel $ B_k(\eta) $\cite{frw1}) {\it decreases} as $ e^{-(2\nu
-1)h_0t} $. Therefore, characteristic field configurations $ \eta_{\vec k} $
are very large (of order $ e^{(\nu -\frac12)h_0t} $). Therefore configurations
with field amplitudes up to ${\cal O}(e^{(\nu -\frac12)h_0t})$ will have a
substantial probability of occurring and being represented in the density
matrix.

Notice that $ \chi \sim  e^{(\nu -\frac12)h_0t} $  corresponds to
field configurations $ \Phi $ with amplitudes of order  $  e^{(\nu -\frac32)h_0t} $ 
[see Eq.~(\ref{confres})]. It is the fact that $ \nu -\frac32 > 0 $ which in this
situation is responsible for the ``classicalization'', which is  seen to
be a consequence of the spinodal growth of long-wavelength fluctuations.

The equal-time field correlator is given by 
\begin{eqnarray}
\langle \bar{\chi}(\vec x,\eta)\; \bar{\chi}({\vec x}',\eta) \rangle 
&=& \int \frac{d^3k}{2(2\pi)^3}\; |F_k(\eta)|^2\; e^{i{\vec k}.({\vec x}-
{\vec x}')} \quad   , \cr \cr
& = & a(t)^2 \; \int \frac{d^3k}{2(2\pi)^3}\; 
|f_k(t)|^2\; e^{i{\vec k}.({\vec x}-{\vec x}')} \quad .
\end{eqnarray}
and is seen to be dominated by the superhorizon mode functions and to grow as 
$  e^{(2\nu -1)h_0 t} $, whereas the
field commutators remain fixed showing the emergence of a classical behavior.
As a result we obtain
\begin{equation}
\langle \bar{\chi}(\vec x,\eta)\; \bar{\chi}({\vec x}',\eta) \rangle 
\propto a^2(t) \; \phi_{eff}(t) \;\phi_{eff}(t') \; G(|\vec x - \vec{x}'|)+
\mbox{ small }
\end{equation}
where $G(|\vec x - \vec{x}'|)$ falls off exponentially for distances larger than
the horizon\cite{desitter} and ``small'' refers to terms that are 
smaller in magnitude. This factorization of the correlation functions
is another indication of classicality.

Therefore, it is possible to  describe the physics by using
classical field theory. 
More precisely, one can use a classical statistical (or stochastic) field theory
described by the functional probability distribution (\ref{claden}).

These results generalize 
the decoherence treatment given in  ref.\cite{polarski}
for a free massless field in pure quantum states to the case of interacting
fields with broken symmetry.
Note that the formal decoherence or classicalization in the density matrix
appears after the modes with wave vector $k$ become superhorizon sized i.e. 
when the factorization of the mode functions becomes effective.

\section{Conclusions} 

It can be argued that the inflationary paradigm as currently understood is one
of the greatest applications of quantum field theory. The imprint of quantum
mechanics is everywhere, from the dynamics of the inflaton, to the generation of
metric perturbations, through to the reheating of the universe. It is clear
then that we need to understand the quantum mechanics of inflation in as deep a
manner as possible so as to be able to understand what we are actually testing
via the CBR temperature anisotropies.

What we have found in our work is that the quantum mechanics of inflation is
extremely subtle. We now understand that it involves both non-equilibrium as
well as non-perturbative dynamics and that what you start from may {\it not} be
what you wind up with at the end.

In particular, we see now that the correct interpretation of the
non-perturbative growth of quantum fluctuations via spinodal decomposition is
that the background zero mode must be redefined through the process of zero
mode reassembly. When this is done we can interpret inflation in terms of the 
usual slow-roll approach with
the now small quantum fluctuations around the redefined zero mode driving the
generation of metric perturbations. 

We have studied the non-equilibrium dynamics of a `new inflation' scenario in a
self-consistent, non-perturbative framework based on a large $ N $ expansion,
including the dynamics of the scale factor and backreaction of quantum
fluctuations. Quantum fluctuations associated with superhorizon modes grow
exponentially as a result of the spinodal instabilities and contribute to the
energy momentum tensor in such a way as to end inflation consistently.

Analytical and numerical estimates have been provided that establish the re\-gime
of validity of the classical approach.  We find that these superhorizon modes
re-assemble into an effective zero mode and unambiguously identify the
composite field that can be used as an effective expectation value of the
inflaton field whose {\em classical} dynamics drives the evolution of the scale
factor. This identification also provides the initial condition for this
effective zero mode.

A consistent criterion is provided to extract ``small'' fluctuations that will
contribute to cosmological perturbations from ``large'' non-perturbative
spinodal fluctuations. This is an important ingredient for a consistent
calculation and interpretation of cosmological perturbations.  This criterion
requires that the model must provide many more than 60 $e$-folds to identify the
`small perturbations' that give rise to scalar metric (curvature)
perturbations. We then use this criterion combined with the gauge invariant
approach to obtain the dynamics of the Bardeen variable and the spectrum for
scalar perturbations.

We find that during the inflationary epoch, superhorizon modes of the Bardeen
potential grow exponentially in time reflecting the spinodal
instabilities. These long-wavelength instabilities are manifest in the spectrum
of scalar density perturbations and result in an index that is less than
one, i.e. a ``red'' power spectrum, providing more power at long wavelength.  We
argue that this `red' spectrum is a robust feature of potentials that lead to
spinodal instabilities in the region in field space associated with inflation
and can be interpreted as an ``imprint'' of the phase transition on the
cosmological background. Tensor perturbations on the other hand, are not
modified by these features, they have much smaller amplitude and a
Harrison-Zeldovich spectrum.

We made contact with the reconstruction program and validated the results for
these type of models based on the slow-roll assumption, despite the fact that
our study does not involve such an approximation and is non-perturbative.

Finally we have studied the quantum to classical crossover and decoherence of
quantum fluctuations by studying the full evolution of the density matrix, thus
making contact with the concept of ``decoherence without
decoherence''\cite{polarski} which is generalized to the interacting case. In
the case under consideration decoherence and classicalization is a consequence
of spinodal growth of superhorizon modes and the presence of a horizon. The
phases of the mode functions ``freeze out'' and the amplitudes of the
superhorizon modes grow exponentially during the inflationary stage, again as a
result of long-wavelength instabilities.  As a result field configurations with
large amplitudes have non-vanishing probabilities to be represented in the
dynamical density matrix.  In the situation considered, the quantum to
classical crossover of cosmological perturbations is directly related to the
``classicalization'' of superhorizon matter field modes that grow exponentially
upon horizon crossing during inflation. The diagonal elements of the density
matrix in the Schroedinger representation can be interpreted as a classical
distribution function, whereas the off-diagonal elements are strongly
suppressed during inflation.

\chapter{Explosive Particle Production After Inflation}

\section{Introduction}

\begin{figure}
\epsfig{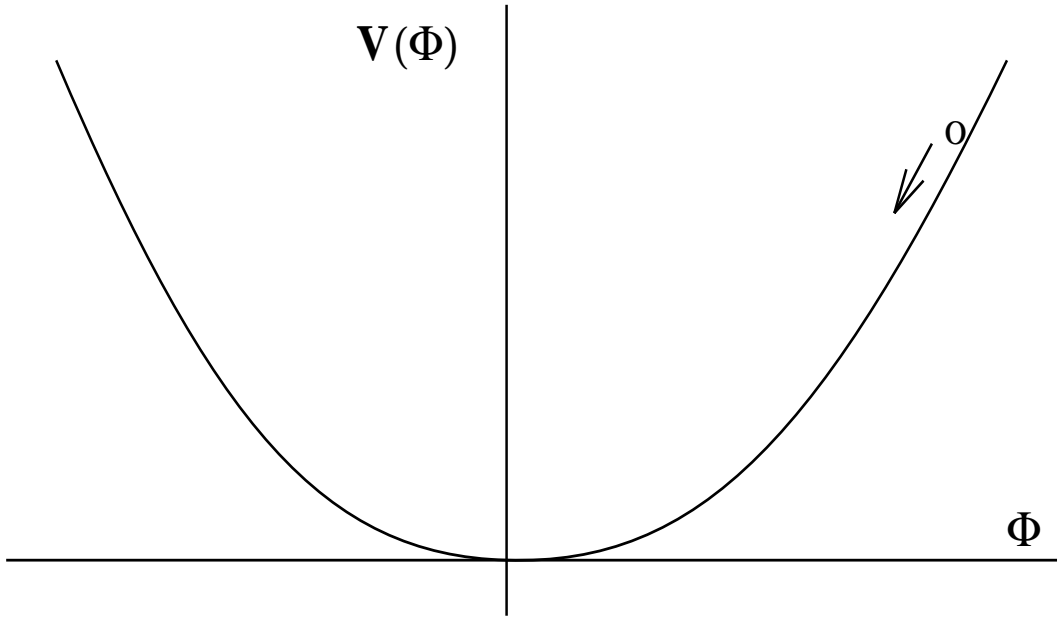}
\caption{A typical chaotic inflation potential.  The field $\Phi$ begins
at very high field values, far from the minimum.}
\label{chaospot}
\end{figure}

The most popular of current models of inflation is probably the
chaotic inflation models\cite{chaotic}.  Unlike the old or new
inflation models which rely on the inflaton being near some local
extremum of the potential, typically with $\phi(t_0) \ll
|m|/\sqrt{\lambda}$, chaotic inflation requires initial values of the
field which are large, typically with $\phi(t_0) \gg
|m|/\sqrt{\lambda}$ (see Fig.~\ref{chaospot})\footnote{One may question the
validity of such an initial state for the field.  The argument in
favor of chaotic inflation is that one requires only that there exists
a single region of the universe of size $H^{-1}$ satisfying the
condition at some time $t \sim t_{Pl}$.  Since the only (known)
restriction on the field at the Planck scale is that the energy
density satisfies $\phi \leq M_{pl}^4$, such field values exist and
chaotic inflation occurs at least in some part of the universe.
Subsequently, regions which have inflated come to dominate the
physical volume of the universe.}.  For such large field values, the
potential contribution dominates the energy momentum, leading to
inflation.  

Unlike the new inflationary scenario examined in the preceding
chapter, there are no spinodal instabilities in chaotic inflation and
the dynamics of the inflationary phase are relatively
straightforward.  In particular, during inflation, quantum
fluctuations remain small and one may simply treat the inflaton as a
classical field identified as the mean value of the quantum field.
That is, it is safe to write $\phi_{cl}(t) = \phi(t)$ and be sure that
corrections to the classical evolution remain perturbatively small.

However, as we shall see, the post inflationary evolution may not be
so straightforward.

In addition to other aspects of inflation, the question of the
reheating of the universe, the transfer of energy from the inflaton to other
particle modes, has received much attention \cite{reheat}.  The reason for 
this
is that exponential expansion during inflation causes the universe to
become very cold, as the energy in fields other than the inflaton is 
redshifted
away.  In order to achieve the standard results of nucleosynthesis, and
possibly other early processes, the universe must be reheated to temperatures
above those at which these important processes take place.  Early efforts to
account for the necessary reheating introduced {\it ad hoc} decay widths to 
the
inflaton, assuming the energy transfer occurred through single particle decay
\cite{origreheat}.  However, more recently, it has been realized that there 
are
much more efficient processes, those of either spinodal decomposition in the
case of a (new) inflationary phase transition and 
parametric amplification in the case of chaotic inflationary scenarios
\cite{branden,kls,us1,par,tkachev,son,symrest,kaiser,yoshimura,frwprl}. 
Such mechanisms, in which the rate of energy transfer grows 
exponentially, are referred to generally as preheating.

In particular, we will study the process of preheating using a wide range of
initial conditions.  We analyze the dynamics of
preheating in {\it fixed} radiation and matter dominated
Friedmann-Robertson-Walker cosmologies both analytically and 
numerically\footnote{We could easily allow the scale factor to evolve
dynamically in the semi-classical approximation to gravity, as
described in Ch.~3.  However, we find this self-consistent evolution
to be almost identical to the evolution in the appropriate fixed
gravitational background.  We therefore simplify our analysis (and
reduce our computational run time) by restricting ourselves to the
fixed background.}.

Our main results are as follows.  In the situations we analyze, we
find that the expansion of the universe allows for significant particle
production, although this production is somewhat sensitive to the exact
expansion rate and is effectively shut off for high enough rates.  
In the case of a symmetry broken potential, we
determine that in the large $N$ limit, the quantum fluctuations
decrease for  late times as $1/a^2(t)$, while, as in Ch.~4, 
these fluctuations and the zero mode 
satisfy a sum rule consistent with Goldstone's theorem 
[see Eq.~(\ref{sumrule})].  

We compute the renormalized energy density, $\varepsilon$, and the
pressure, $p$,  as a function of time. Averaging over the field oscillations,
we find immediately after preheating a cold matter equation of state 
($ p = 0 $) 
in the slow roll scenarios.  In chaotic scenarios, the equation of state 
just after preheating is between that of radiation ($p = \varepsilon/3$) and
matter where the matter dominated regime is reached only for late times.
The time scale over which the equation of state becomes matter dominated
depends on the distribution of created particles in momentum space in 
addition to the approximation scheme implemented.

\section{Evolution}

We focus our study of the evolution on radiation or matter dominated
cosmologies, as the case for de Sitter expansion has been studied
previously\cite{desitter}.  We write the scale factor as $a(t)=(t/t_0)^n$ with
$n=1/2$ and $n=2/3$ corresponding to radiation and matter dominated
backgrounds respectively.  Note that the value of $t_0$ determines the
initial Hubble constant since 
$$
H(t_0)=\frac{\dot{a}(t_0)
}{a(t_0)}=\frac{n}{t_0}.
$$
We now solve the system of Eqs.~(\ref{hartphieq})--(\ref{fluct}) in the
Hartree approximation and in the
large $N$ limit.  We begin by presenting an early time analysis of the
slow roll scenario.  We then undertake a thorough numerical investigation of
various cases of interest.  For the symmetry broken case, we also provide an
investigation of the late time behavior of the zero mode and the quantum
fluctuations.

It will also be convenient to define the number operator $N(t)$.  We
do so using conformal time quantization, as the particles so defined
will most closely match the definition of particles in flat
spacetime\cite{frwbig}.  We have, at zero temperature,
\begin{equation}
N_k(t) = \frac{a^2(t)}{4} \left| \frac{f_k(t)}{f_k(t_0)}\right|^2
\left[ 1+ \frac{1}{\omega^2_k(t_0)} \left| 
\frac{ \dot{f}_k(t) + H(t) f_k(t)}{f_k(t)}
\right|^2 \right] - \frac{1}{2} \; . \label{partnumber}
\end{equation}
Examination of the large $k$ behavior of $f_k$ and $\dot{f}_k$ shows
that $N_k$ is ${\cal O}(1/k^4)$ for large $k$, so that the total
number of particles
\begin{equation}
N(t) = \int \frac{d^3k}{(2\pi)^3} N_k
\label{totalpart}
\end{equation}
is finite.

In what follows, we will assume minimal coupling to
the curvature, $\xi_R = 0$.  In most of the cases of interest, 
${\cal R}/|m| \ll 1$, so that finite $\xi_R$ has little effect.

\subsection{Early Time Solutions for Slow Roll}

For early times in a slow roll scenario [$m^2=-1$, $\eta(\tau_0) \ll 1$], we
can neglect in Eq.~(\ref{hartphieq}) and in
Eq.~(\ref{hartukeq}) both the quadratic and cubic terms in $\phi(t)$ as well 
as
the quantum fluctuations $\langle\psi^2(t)\rangle_R $ [recall that
$\langle\psi^2(t_0)\rangle_R = 0 $]. Thus, the differential equations for the
zero mode (\ref{hartphieq}) and the mode functions
(\ref{hartukeq}) become linear equations. In terms of the scaled variables
introduced at the beginning of Sec.~4.2, 
with $ a(\tau)= (\tau/\tau_0)^n $ ($ n=2/3 $ for a matter dominated
cosmology while $ n=1/2 $ for a radiation dominated cosmology) we have:
\begin{eqnarray}
\ddot{\eta}(\tau)+\frac{3n}{\tau}\dot{\eta}(\tau)-\eta(\tau) & = & 0 \; ,
\label{earlyeta} \\ \nonumber \\
\left[\frac{d^2}{d\tau^2}+\frac{3n}{\tau}\frac{d}{d\tau}
+\frac{q^2}{(\tau/\tau_0)^{2n}}-1\right]f_q(\tau) & = & 0 \;. \label{earlyuk}
\end{eqnarray}

The solutions to the zero mode equation (\ref{earlyeta}) are
\begin{equation}
\eta(\tau)=c\; \tau^{-\nu}I_{\nu}(\tau)+d\; \tau^{-\nu}K_{\nu}(\tau) \; ,
\label{earlyetasoln}
\end{equation}
where $\nu \equiv (3n-1)/2$, and $I_{\nu}(\tau)$ and $K_{\nu}(\tau)$ are modified
Bessel functions.  The coefficients, $c$ and $d$, are determined by the 
initial
conditions on $\eta$.  For $\eta(\tau_0)=\eta_0$ and $\dot{\eta}(\tau_0)=0$, we 
have:
\begin{eqnarray}
c & = & \eta_0  \; \tau_0^{\nu+1}
\left[ \dot{K}_{\nu}(\tau_0)-\frac{\nu}{\tau_0}K_{\nu}(\tau_0)\right] \; ,
\label{coeffc} \\
d & = & -\eta_0  \; \tau_0^{\nu+1}
\left[\dot{I}_{\nu}(\tau_0)-\frac{\nu}{\tau_0}I_{\nu}(\tau_0)\right] \; .
\label{coeffd}
\end{eqnarray}
Taking the asymptotic forms of the modified Bessel functions, we find that 
for
intermediate times $\eta(\tau)$ grows as
\begin{equation}
\eta(\tau) \stackrel{\tau \gg 1}{=} {c \over {\sqrt{2 \pi}}}
\; \tau^{-3n/2}\; e^\tau\left[1-\frac{9n^2-6n}{8\tau}+
O\left({1\over{\tau^2}}\right)\right].
\label{asymeta}
\end{equation}
We see that $\eta(\tau)$ grows very quickly in time, and the approximations
(\ref{earlyeta}) and (\ref{earlyuk}) will quickly break down.  For the case
shown in Fig.~\ref{fig1} (with $n=2/3$, $h(\tau_0)=0.1$, $\eta(\tau_0)=10^{-7}$, 
and $\dot{\eta}(\tau_0)=0$),
we find that this approximation is valid up to $\tau-\tau_0 \simeq 10$.

The equations for the mode functions ({\ref{earlyuk}) can be solved in closed
form for the modes in the case of a radiation dominated cosmology with 
$n=1/2$.
The solutions are
\begin{equation}
f_q(\tau) = c_q \; e^{-\tau}\, 
U\left(\frac34-\frac{q^2 \tau_0}{2},\frac32,2\tau\right) 
+ d_q \; e^{-\tau}\, M\left(\frac34-\frac{q^2 \tau_0}{2},\frac32,2\tau\right).
\label{earlyuksoln}
\end{equation}
Here, $U(\cdot)$ and $M(\cdot)$ are confluent hypergeometric functions
\cite{aands} (in another common notation, $M(\cdot) \equiv\; _1F_1(\cdot)$),
and the $c_q$ and $d_q$ are coefficients determined by the initial conditions
(\ref{initcond}) on the modes.  The solutions can also be written in terms of
parabolic cylinder functions.

For large $ \tau $ we have the asymptotic form
\begin{eqnarray}
f_q(\tau) & \stackrel{\tau \gg 1}{=} & d_q \; e^\tau (2\tau)^{-(3/4+q^2 \tau_0/2)}
\frac{\sqrt{\pi}}{2\;\Gamma\left(\frac34-\frac{q^2 \tau_0}{2}\right)}
\left[1+O\left({1\over{\tau}}\right)\right] \nonumber \\
& + & c_q \; e^{-\tau}\,  (2\tau)^{(-3/4+q^2 \tau_0/2)}\left[1+O
\left({1\over{\tau}}\right)\right]\;
. \label{asymuk}
\end{eqnarray}
Again, these expressions only apply for intermediate times before the
nonlinearities have grown significantly.

\subsection{Numerical Analysis}
We now present the numerical analysis of the dynamical evolution of
scalar fields in time dependent, matter and radiation dominated cosmological
backgrounds.  We use initial values of the Hubble constant such that 
$h(\tau_0) \geq 0.1$.  For expansion
rates much less than this value the evolution will look similar to
Minkowski space, which has been studied in great detail elsewhere
\cite{us1,mink,cooper}.  As will be seen, the equation of state found
numerically is, in the majority of cases, that of cold matter.  We therefore
use matter dominated expansion for the evolution in much of the analysis that
follows.  While it presents some inconsistency at late times, 
the evolution in radiation
dominated universes remains largely unchanged, although there is greater
initial growth of quantum fluctuations due to the scale factor growing more
slowly in time.  Using the large $N$ and Hartree approximations to study 
theories with continuous and discrete symmetries respectively, we treat three
important cases.  They are
\begin{enumerate}
\item $m^2<0$, $\eta(\tau_0)\ll 1$, 
\item $m^2<0$, $\eta(\tau_0)\gg 1$, and
\item $m^2>0$, $\eta(\tau_0)\gg 1$.
\end{enumerate}

In presenting the figures, we have shifted the origin of time such that
$\tau \to \tau'=\tau-\tau_0$.  This places the initial time, $\tau_0$, at the origin.
In these shifted coordinates, the scale factor is given by 
$$
a(\tau)=\left(\frac{\tau+\tilde{\tau}}{\tilde{\tau}}\right)^n,
$$ 
where, once again, $n=2/3$ and $n=1/2$ in matter and radiation dominated
backgrounds respectively, and the value of $\tilde{\tau}$ is determined by the 
Hubble constant at the initial time:
$$
h(\tau_0=0)=\frac{n}{\tilde{\tau}}.
$$

{\bf Case 1: $m^2<0$, $\eta(\tau_0)\ll 1$}.  This is the case of an early 
universe
phase transition in which there is little or no biasing in the initial
configuration (by biasing we mean that the initial conditions break the 
$ \eta
\to -\eta $ symmetry).  The transition occurs from an initial temperature 
above
the critical temperature, $T>T_c$, which is quenched at $t_0$ to the
temperature $T_f \ll T_c$.  This change in temperature due to the rapid
expansion of the universe is modeled here by an instantaneous change in the
mass from an initial value $m_i^2=T^2/T_c^2-1$ to a final value
$m_f^2=-1$.  We will use the value $m_i^2=1$ in what follows.
This quench approximation is necessary since the low momentum
frequencies (\ref{freq}) appearing in our initial conditions 
(\ref{initcond}) 
are complex for negative mass squared and small $\eta(\tau_0)$.  An alternative
choice is to use initial frequencies (\ref{modinit}) which were used
in Ch.~4.  We find that different choices of initial
conditions change the quantitative value of the particle number by a few 
percent, but leave the qualitative results unchanged.

We plot the the zero mode $\eta(\tau)$, the equal time correlator $g\Sigma(\tau)$, 
the total number of produced particles $gN(\tau)$  [see
Eq.~(\ref{totalpart})], the number of particles
$gN_q(\tau)$ as a function of wavenumber for both intermediate and
late times [see Eq.~(\ref{partnumber})], and the ratio of the pressure
and energy densities $p(\tau)/\varepsilon(\tau)$ (giving the equation of state).

Figure \ref{fig1}a-e shows these quantities in the large $N$ approximation for a 
matter
dominated cosmology with an initial condition on the zero mode given by
$\eta(\tau_0\! =\! 0)=10^{-7}$, $\dot{\eta}(\tau_0\! =\! 0)=0$ and for an initial
expansion rate of $h(\tau_0)=0.1$.  This choice for
the initial value of $\eta$ stems from the fact that the quantum fluctuations
only have time to grow significantly for initial values satisfying
$\eta(\tau_0) \ll  \sqrt{g}$; for values $\eta(\tau_0) \gg \sqrt{g}$ the evolution
is essentially classical.  This result is clear from the intermediate time
dependence of the zero mode and the low momentum mode functions given by 
the expressions (\ref{asymeta}) and (\ref{asymuk}) respectively.

\begin{figure}
\epsfig{file=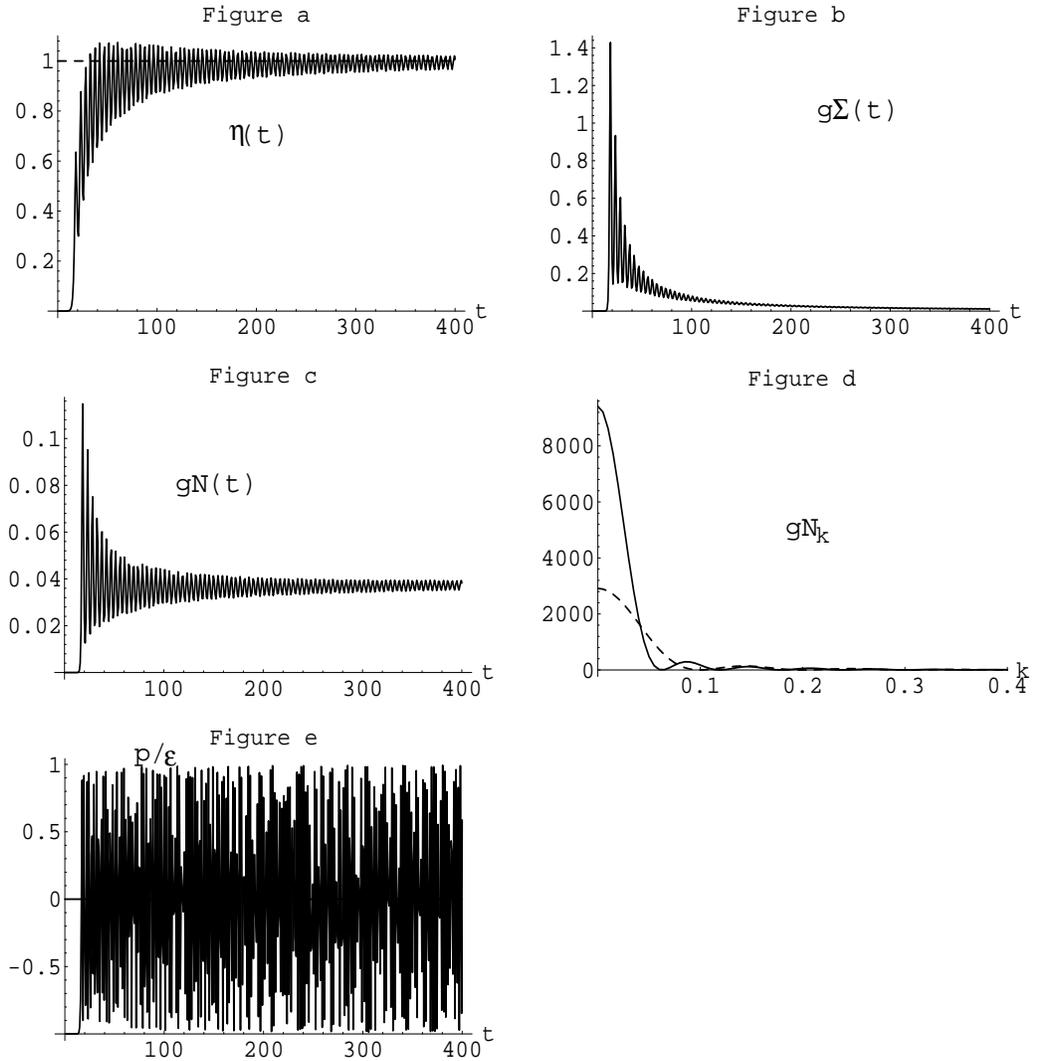,width=6in,height=6in}
\caption{Symmetry broken, slow roll, large $N$, matter dominated
evolution of (a) the zero mode $\eta(\tau)$ vs. $\tau$, (b) the quantum fluctuation
operator $g\Sigma(\tau)$ vs. $\tau$, (c) the number of particles $gN(\tau)$ vs. $\tau$,
(d) the particle distribution $gN_q(\tau)$ vs. $q$ at $\tau=149.1$ (dashed line)
and $\tau=398.2$ (solid line),  and (e) the ratio of the 
pressure and energy density
$p(\tau)/\varepsilon(\tau)$ vs. $\tau$ for the parameter values $m^2=-1$, $\eta(\tau_0) =
10^{-7}$, $\dot{\eta}(\tau_0)=0$, $g = 10^{-12}$, $h(\tau_0) = 0.1$. \label{fig1}}
\end{figure}

After the initial
growth of the fluctuation $g\Sigma$ (Fig.~\ref{fig1}b) we see that the zero mode
(Fig.~\ref{fig1}a) approaches the value given by the minimum of the tree level
potential, $\eta=1$, while $g\Sigma$ decays for late times as
$$
g\Sigma \simeq {{\cal C}\over{a^2(\tau)}} ={{\cal C}\over{\tau^{4/3}}} \; .
$$
For these late times, the Ward identity corresponding to the $O(N)$ 
symmetry of
the field theory is satisfied, enforcing the condition [see
(\ref{sumrule})] 
\begin{equation}
-1 + \eta^2(\tau) + g\Sigma(\tau) = 0.
\label{ward}
\end{equation} 
Hence, the zero mode approaches the classical minimum as
$$
\eta^2(\tau) \simeq 1 -{{\cal C}\over{a^2(\tau)}} \; .
$$ 

Figure \ref{fig1}c depicts the number of particles produced.  After an initial burst 
of 
particle production, the number of particles settles down to a relatively
constant value.  Notice that the number of particles produced is 
approximately
of order $1/g$.
In Fig.~\ref{fig1}d, we show the number of particles as a function of the
wavenumber, $q$.  For intermediate times we see the simple structure depicted
by the dashed line in the figure, while for late times this quantity becomes
concentrated more at low values of the momentum $q$.  

Finally, Fig.~\ref{fig1}e shows that the field begins with a de Sitter equation of 
state $p=-\varepsilon$ but evolves quickly to a state dominated by ordinary 
matter, 
with an equation of state (averaged over the oscillation time scale) $p=0$.
This last result is a bit surprising as one expects from the condition
(\ref{ward}) that the particles produced in the final state are massless
Goldstone bosons (pions) which should have the equation of state of 
radiation.
However, as shown in Fig.~\ref{fig1}d, the produced particles are of low momentum,
$q \ll 1$, and while the effective mass of the particles is zero to very 
high 
accuracy when averaged over the oscillation time scale, the effective mass 
makes small oscillations about zero so that the dispersion relation for these
particles differs from that of radiation.  In addition, since the produced
particles have little energy, the contribution to the energy density from
the zero mode, which contributes to a cold matter equation of state, remains
significant.

In Fig.~\ref{fig2}a-e we show the same situation depicted in Fig.~\ref{fig1} using the
Hartree approximation.  The initial condition on the zero mode is $\eta(\tau_0\!
=\! 0)=\sqrt{3}\cdot 10^{-7}$; the factor of $\sqrt{3}$ appears due to the
different scaling in the zero mode equation, (\ref{hartphieq}), which causes 
the minimum of the tree level effective potential
in the Hartree approximation to have a value of $\eta=\sqrt{3}$.  Again, the
Hubble constant has the value $h(\tau_0)=0.1$.  Here, we see
again that there is an initial burst of particle production as $g\Sigma$
(Fig.~\ref{fig2}b) grows large.  However, the zero mode 
(Fig.~\ref{fig2}a) quickly reaches the
minimum of the potential and the condition
\begin{equation}
-1 + \eta^2(\tau)/3 + g\Sigma(\tau) = 0
\label{ward2}
\end{equation}
is approximately satisfied by forcing the value of $g\Sigma$ quickly to zero.
There are somewhat fewer particles produced here compared to the large $N$ 
case, 
and the distribution of particles is more extended.  
Since the effective mass of the particles is nonzero, we expect a matter 
dominated
equation of state (Fig.~\ref{fig2}e) for later times.  
The fact that the Hartree approximation does not
satisfy Goldstone's theorem means that the resulting particles must be 
massive, 
explaining why somewhat fewer particles are produced.

\begin{figure}
\epsfig{file=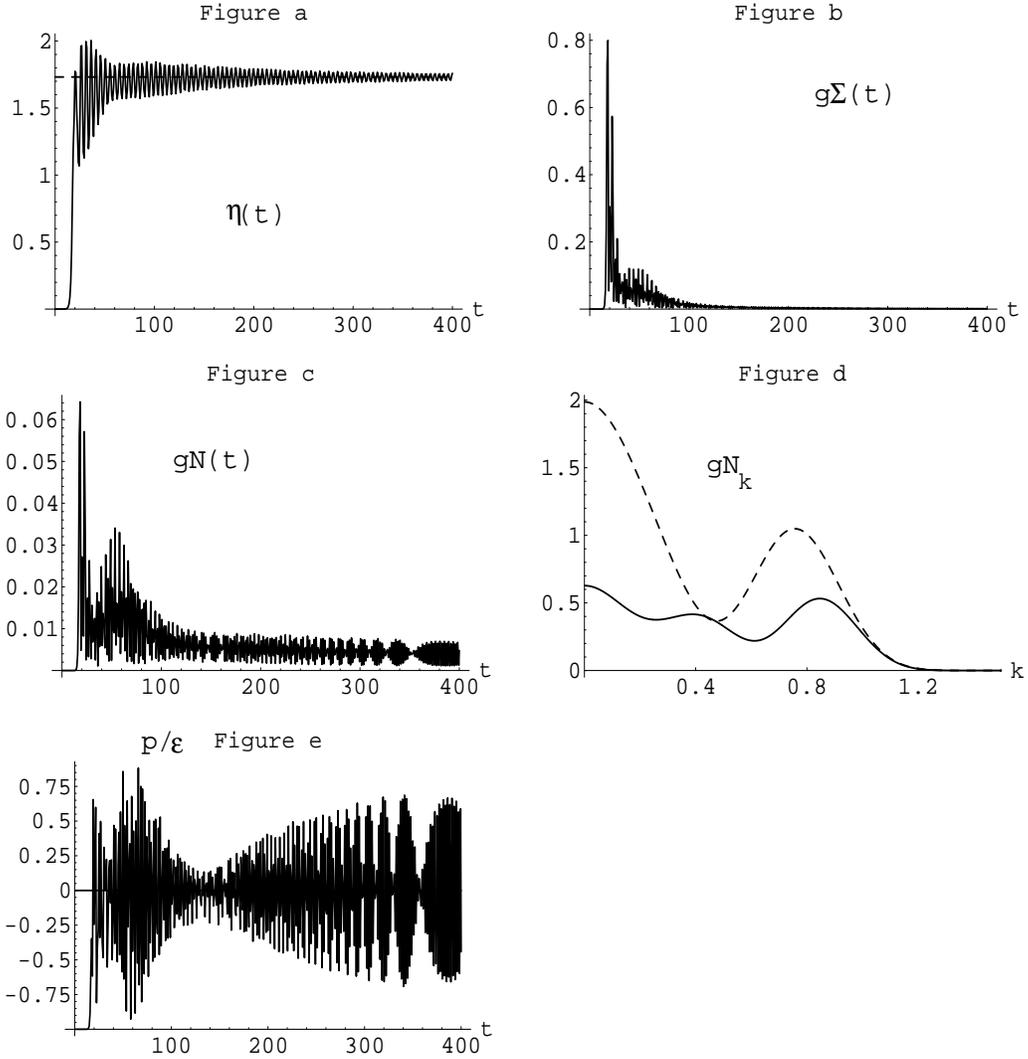,width=6in,height=6in}
\caption{Symmetry broken, slow roll, Hartree, matter dominated
evolution of (a) the zero mode $\eta(\tau)$ vs. $\tau$, (b) the quantum fluctuation
operator $g\Sigma(\tau)$ vs. $\tau$, (c) the number of particles $gN(\tau)$ vs. $\tau$,
(d) the particle distribution $gN_q(\tau)$ vs. $q$ at $\tau=150.7$ (dashed line)
and $\tau=396.1$ (solid line),  and (e) the ratio of the 
pressure and energy density
$p(\tau)/\varepsilon(\tau)$ vs. $\tau$ for the parameter values $m^2=-1$, $\eta(\tau_0) =
3^{1/2}\cdot 10^{-7}$, $\dot{\eta}(\tau_0)=0$, $g = 10^{-12}$, $h(\tau_0) =
0.1$. \label{fig2}}
\end{figure}

Finally, we show the special case in which
there is no initial biasing in the field, $\eta(\tau_0\! =\! 0)=0$, 
$\dot{\eta}(\tau_0\! =\! 0)=0$, and $h(\tau_0)=0.1$ in
Fig.~\ref{fig3}a-d.  
With such an initial condition, the Hartree approximation and the 
large $N$ limit are equivalent.
The zero mode remains zero for all time, so that the quantity
$ g\Sigma(\tau) $ (Fig.~\ref{fig3}a) satisfies the sum rule (\ref{ward}) by reaching 
the value one without decaying for late times. 
Notice that many more particles are produced in this case (Fig.~\ref{fig3}b); the 
growth
of the particle number for late times is due to the expansion of the 
universe.  
The particle distribution (Fig.~\ref{fig3}c) is similar to that of the slow roll 
case in 
figure 1.  The equation of state (Fig.~\ref{fig3}d) is likewise similar.

\begin{figure}
\epsfig{file=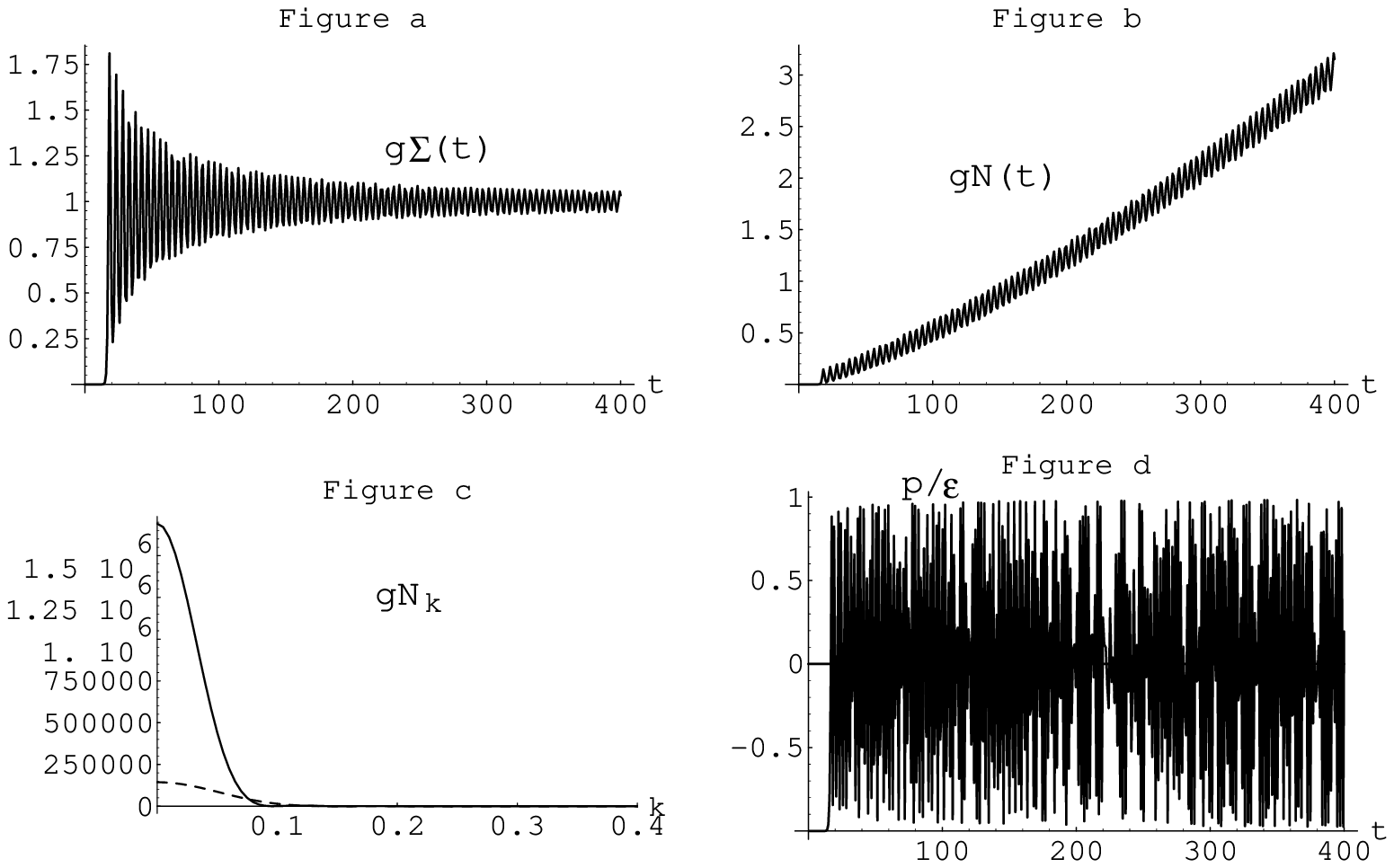,width=6in,height=6in}
\caption{Symmetry broken, no roll, matter dominated
evolution of (a) the quantum fluctuation operator $g\Sigma(\tau)$ vs. $\tau$, 
(b) the number of particles $gN(\tau)$ vs. $\tau$,
(c) the particle distribution $gN_q(\tau)$ vs. $q$ at $\tau=150.1$ (dashed line)
and $\tau=397.1$ (solid line),  and (d)
the ratio of the pressure and energy density $p(\tau)/\varepsilon(\tau)$ 
vs. $\tau$ for
the parameter values $m^2=-1$, $\eta(\tau_0) = 0$, $\dot{\eta}(\tau_0)=0$, $g =
10^{-12}$, $h(\tau_0) = 0.1$. \label{fig3}}
\end{figure}

In each of these cases of slow roll dynamics, increasing the Hubble constant
has the effect of slowing the growth of both $\eta$ and $g\Sigma$.  The 
equation
of state will be that of a de Sitter universe for a longer period before
moving to a matter dominated equation of state.  Otherwise, the dynamics is
much the same as in Figs.~\ref{fig1}--\ref{fig3}.

{\bf Case 2: $m^2<0$, $\eta(\tau_0)\gg 1$}.  We now examine the case of 
preheating
occurring in a chaotic inflationary scenario with a symmetry broken 
potential.
In chaotic inflation, the zero mode begins with a value $\eta(\tau)\gg 1$.  
During
the de Sitter phase, $h \gg 1$, and the field initially evolves classically,
dominated by the first order derivative term appearing in the zero mode
equation [see (\ref{hartphieq})].  Eventually, the zero 
mode
rolls down the potential, ending the de Sitter phase and beginning the
preheating phase.  We consider the field dynamics in the FRW universe
where preheating occurs after  the end of inflation. 
We thus take the initial temperature to be zero, $T=0$. 

Figure \ref{fig4} shows our results for the quantities, $\eta(\tau)$, $g\Sigma(\tau)$,
$gN(\tau)$, $gN_q(\tau)$, and $p(\tau)/\varepsilon(\tau)$ for the evolution in
the large $N$ approximation within a {\em radiation} dominated gravitational
background with $h(\tau_0)=0.1$.  The initial condition on the zero mode is 
chosen 
to have the representative value $\eta(\tau_0\!=\! 0)=4$ with 
$\dot{\eta}(\tau_0\! =\! 0)=0$.  
Initial values of the zero mode much smaller than this will not produce
significant growth of quantum fluctuations; initial values larger than this
produces qualitatively similar results, although the resulting number of 
particles will be greater and the time it takes for the zero mode to settle
into its asymptotic state will be longer. 

\begin{figure}
\epsfig{file=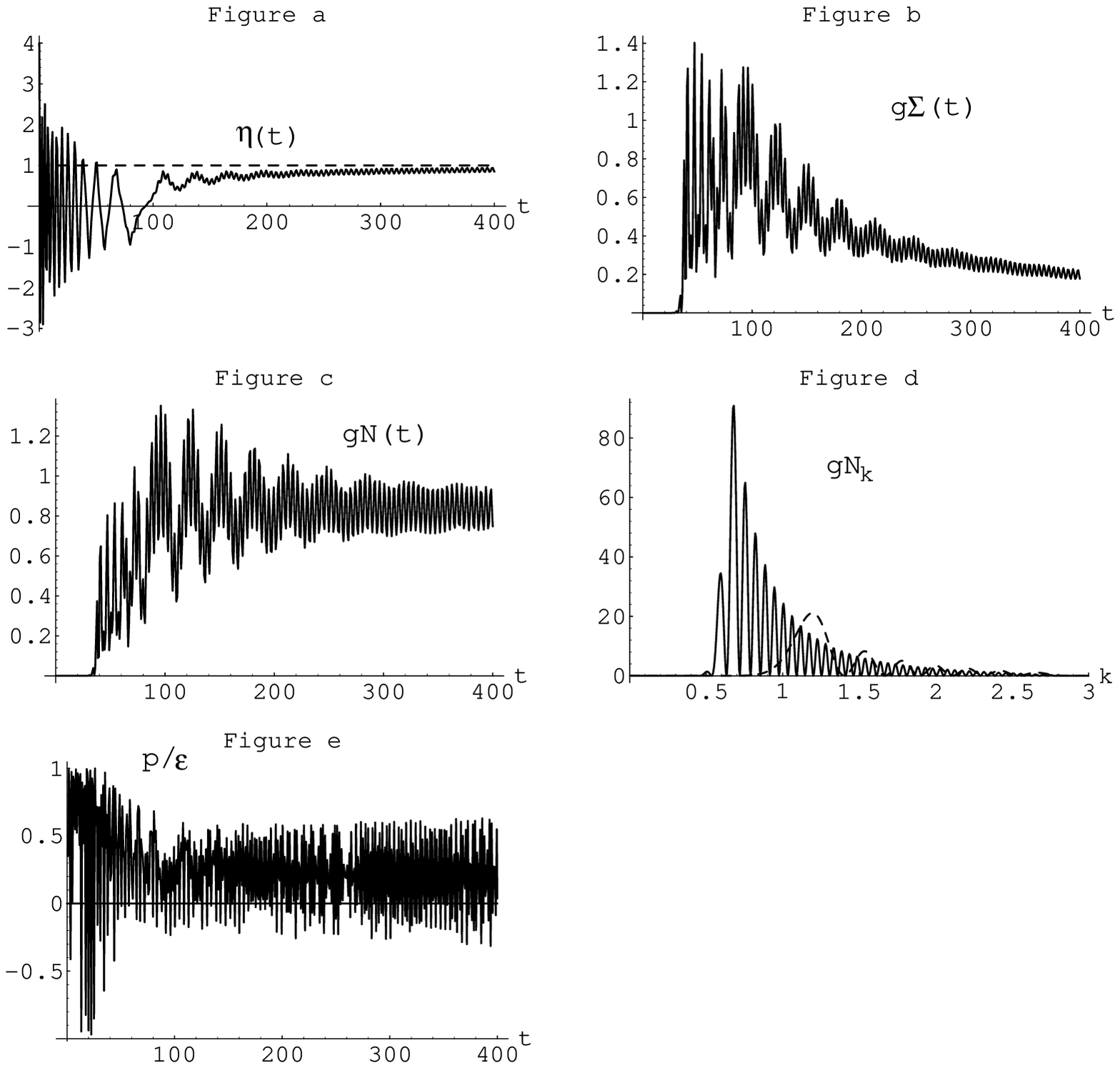,width=6in,height=6in}
\caption{Symmetry broken, chaotic, large $N$, radiation dominated
evolution of (a) the zero mode $\eta(\tau)$ vs. $\tau$, (b) the quantum fluctuation
operator $g\Sigma(\tau)$ vs. $\tau$, (c) the number of particles $gN(\tau)$ vs. $\tau$,
(d) the particle distribution $gN_q(\tau)$ vs. $q$ at $\tau=76.4$ (dashed line)
and $\tau=392.8$ (solid line),  and (e) the ratio of the pressure and energy
density $p(\tau)/\varepsilon(\tau)$ vs. $\tau$ for the parameter values $m^2=-1$,
$\eta(\tau_0) = 4$, $\dot{\eta}(\tau_0)=0$, $g = 10^{-12}$, $h(\tau_0) =
0.1$. \label{fig4}}
\end{figure}

We see from Fig.~\ref{fig4}a that the zero
mode oscillates rapidly, while the amplitude of the oscillation decreases due
to the expansion of the universe.  This oscillation induces particle 
production
through the process of parametric amplification (Fig.~\ref{fig4}c) and causes the
fluctuation $g\Sigma$ to grow (Fig.~\ref{fig4}b).  Eventually, the zero mode loses
enough energy that it is restricted to one of the two minima of the tree 
level
effective potential.  The subsequent evolution closely follows that of Case 1
above with $g\Sigma$ decaying in time as $1/a^2(\tau) \sim 1/\tau$ with $\eta$ 
given by
the sum rule (\ref{ward}).  The spectrum (Fig.~\ref{fig4}d) indicates a single 
unstable
band of particle production dominated by the modes $q=1/2$ to about $q=3$ 
for late times.  The structure within this band becomes more complex with 
time 
and shifts somewhat toward lower momentum modes.  Such a shift has already 
been 
observed in Minkowski spacetimes \cite{us1}. Figure \ref{fig4}e shows the equation of
state which we see to be somewhere between the relations for matter and
radiation for times out as far as $\tau=400$, but slowly moving to a matter
equation of state.  Since matter redshifts as $1/a^3(\tau)$ while radiation
redshifts as $1/a^4(\tau)$, the equation of state should eventually become 
matter
dominated.  Given the equation of state indicated by Fig.~\ref{fig4}e, we estimate 
that
this occurs for times of order $\tau=10^4$.  The reason the equation of state
in this case differs from that of cold matter as was seen in 
Figs.~\ref{fig1}--\ref{fig3} is 
that the particle distribution produced by parametric amplification is 
concentrated at higher momenta, $q \simeq 1$. 

Figure \ref{fig5} shows the corresponding case with a matter dominated background.  
The
results are qualitatively very similar to those described for Fig.~\ref{fig4} above.
Due to the faster expansion, the zero mode (Fig.~\ref{fig5}a) finds one of the two 
wells
more quickly and slightly less particles are produced.  For late times, the
fluctuation $g\Sigma$ (Fig.~\ref{fig5}b) decays as $1/a^2(\tau) \propto 1/\tau^{4/3}$.  
Again we 
see an equation of state (Fig.~\ref{fig5}e) which evolves from a state between that 
of
pure radiation or matter toward one of cold matter.

\begin{figure}
\epsfig{file=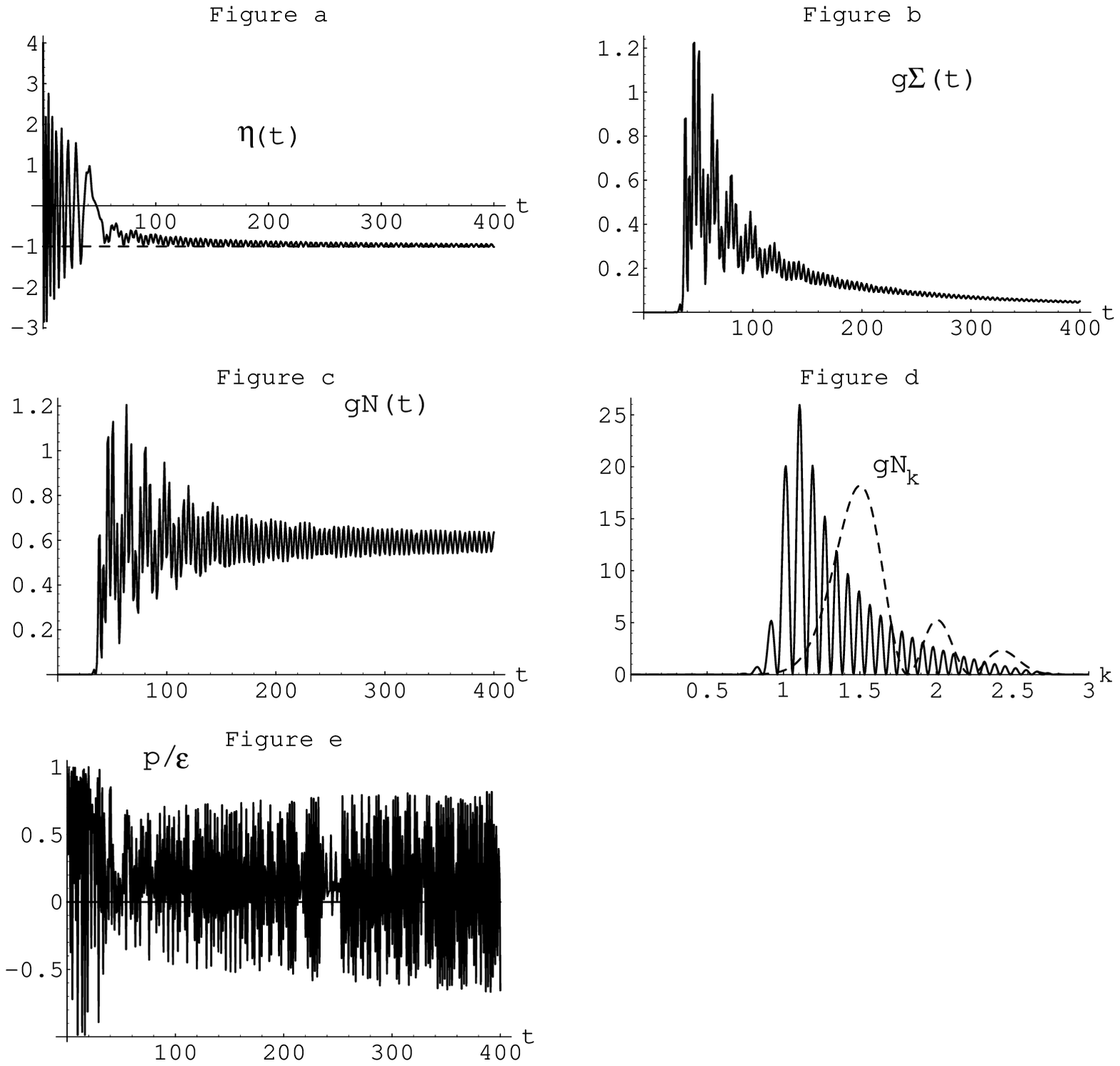,width=6in,height=6in}
\caption{Symmetry broken, chaotic, large $N$, matter dominated
evolution of (a) the zero mode $\eta(\tau)$ vs. $\tau$, (b) the quantum fluctuation
operator $g\Sigma(\tau)$ vs. $\tau$, (c) the number of particles $gN(\tau)$ vs. $\tau$,
(d) the particle distribution $gN_q(\tau)$ vs. $q$ at $\tau=50.8$ (dashed line)
and $\tau=399.4$ (solid line),  and (e) the ratio of the 
pressure and energy density
$p(\tau)/\varepsilon(\tau)$ vs. $\tau$ for the parameter values $m^2=-1$, $\eta(\tau_0) =
4$, $\dot{\eta}(\tau_0)=0$, $g = 10^{-12}$, $h(\tau_0) = 0.1$. \label{fig5}}
\end{figure}

The Hartree case is depicted in Fig.~\ref{fig6} for a matter dominated universe, 
with
the initial condition on the zero mode $\eta(\tau_0\! =\! 0)=4\sqrt{3}$.  Again,
the evolution begins in much the same manner as in the large $N$ 
approximation
with oscillation of the zero mode (Fig.~\ref{fig6}a), which eventually settles into 
one
of the two minima of the effective potential.  Whereas in the large $N$
approximation, the zero mode approaches the minimum asymptotically [as 
given by
(\ref{ward}) and our late time analysis below], in the Hartree 
approximation we
see that the zero mode finds the minimum quickly and proceeds to oscillate
about that value.  The two point correlator (Fig.~\ref{fig6}b) quickly evolves toward
zero without growing large.  Particle production in the Hartree approximation
(Figs.~\ref{fig6}c-d) is again seen to be inefficient compared to that of the 
large $N$ 
case above.  Figure \ref{fig6}e again shows that the equation of state is matter 
dominated 
for all but the earliest times.

\begin{figure}
\epsfig{file=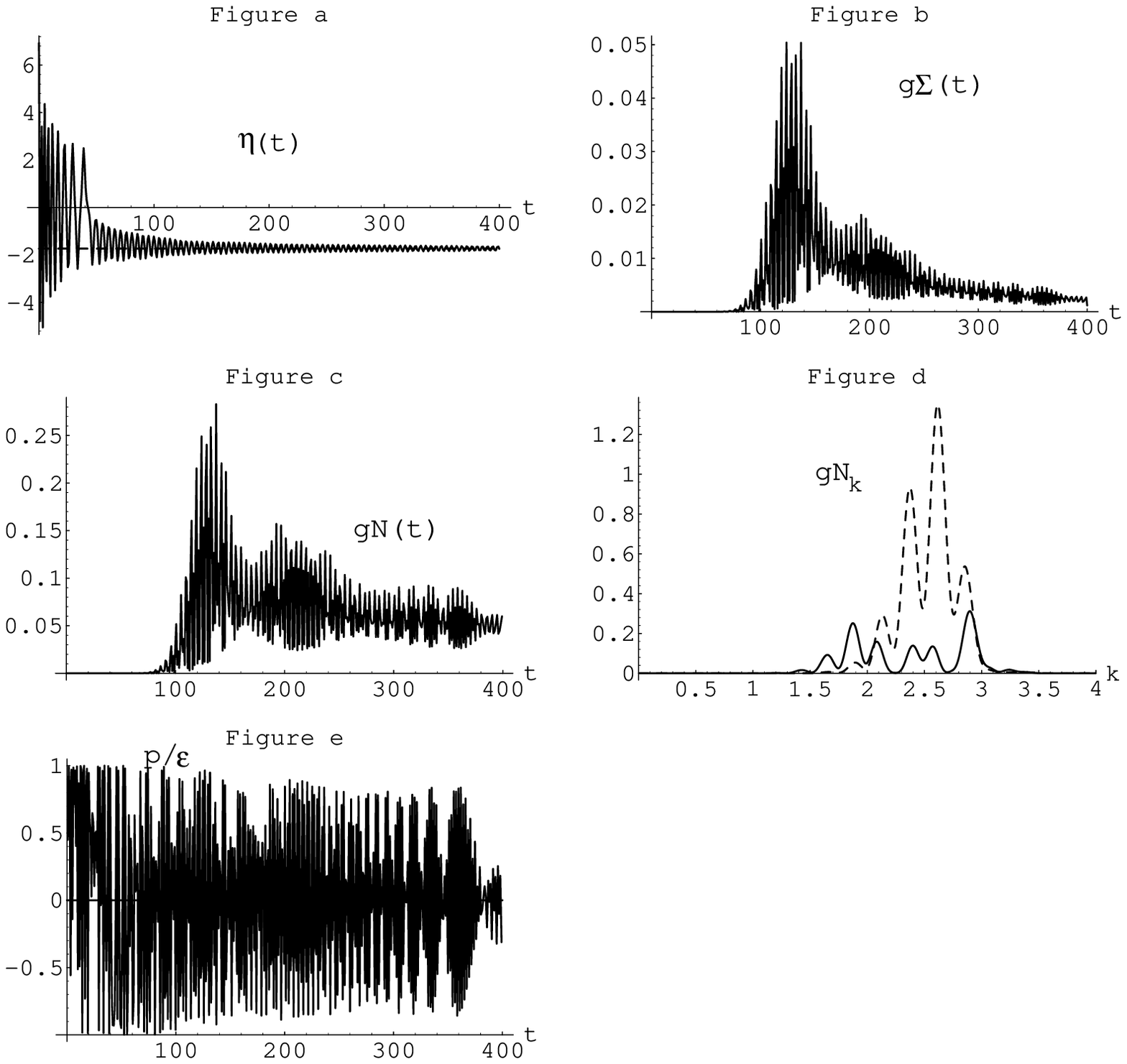,width=6in,height=6in}
\caption{Symmetry broken, chaotic, Hartree, matter dominated
evolution of (a) the zero mode $\eta(\tau)$ vs. $\tau$, (b) the quantum fluctuation
operator $g\Sigma(\tau)$ vs. $\tau$, (c) the number of particles $gN(\tau)$ vs. $\tau$,
(d) the particle distribution $gN_q(\tau)$ vs. $q$ at $\tau=151.3$ (dashed line)
and $\tau=397.0$ (solid line),  and (e) the ratio of the 
pressure and energy density
$p(\tau)/\varepsilon(\tau)$ vs. $\tau$ for the parameter values $m^2=-1$, $\eta(\tau_0) =
4\cdot 3^{1/2}$, $\dot{\eta}(\tau_0)=0$, $g = 10^{-12}$, $h(\tau_0) =
0.1$. \label{fig6}}
\end{figure}

A larger Hubble constant prevents significant particle production unless the 
initial amplitude of the zero mode is likewise increased such that the 
relation
$\eta(\tau_0) \gg h(\tau_0)$ is satisfied.  For very large amplitude $\eta(\tau_0) 
\gg 1$, 
to the extent that the mass term can be neglected and while the quantum 
fluctuation 
term has not grown to be large, the equations of motion (\ref{hartphieq}) 
and (\ref{hartukeq}) are scale invariant 
with the scaling $\eta \to \mu \eta$, $h \to \mu h$, $\tau \to \tau/\mu$, and 
$q \to \mu q$, where $\mu$ is an arbitrary scale.

For completeness, we show the case of the evolution with initial values of 
the 
Hubble constant given by $h(\tau_0) = 5$ and $h(\tau_0) = 2$ respectively in 
Figs.~\ref{fig7} and \ref{fig8} using radiation dominated expansion. 
Here, we have used the large $N$ approximation and have made an appropriate
increase in the initial value of the zero mode such that the fluctuations 
(Figs.~\ref{fig7}b and \ref{fig8}b) 
grow significantly [we have chosen $\eta(\tau_0)=40$ in Fig.~\ref{fig7} and
$\eta(\tau_0)=16$ in Fig.~\ref{fig8}].  While the dynamics looks much like that of 
Figs.~\ref{fig4}--\ref{fig6} above, we point out that the particle 
distribution (Figs.~\ref{fig7}d and \ref{fig8}d)
is extended to higher values with the result being that the equation of 
state (Figs.~\ref{fig7}e and \ref{fig8}e) is weighted more toward that of radiation.

\begin{figure}
\epsfig{file=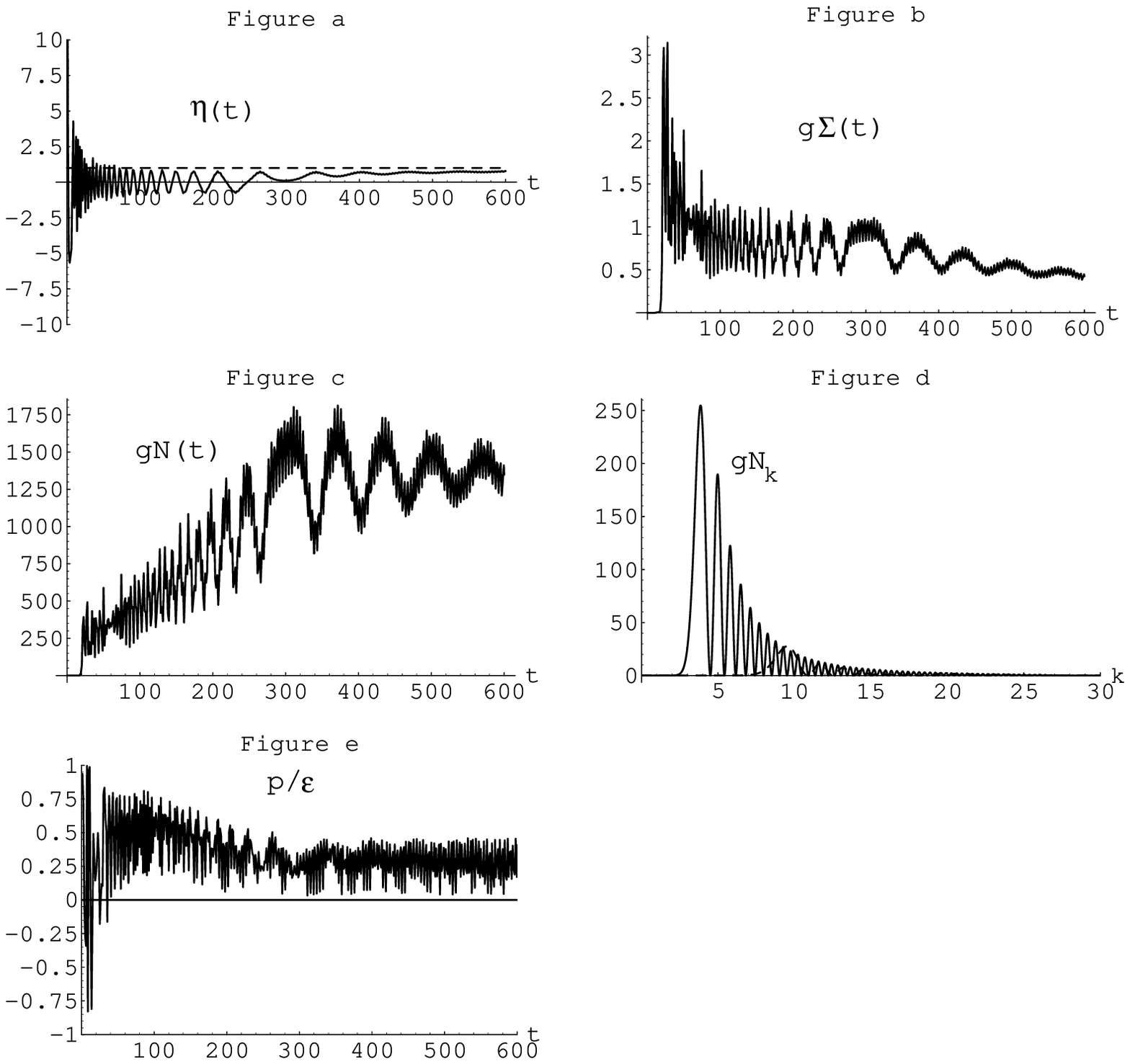,width=6in,height=6in}
\caption{Symmetry broken, chaotic, large $N$, radiation dominated
evolution of (a) the zero mode $\eta(\tau)$ vs. $\tau$, (b) the quantum fluctuation
operator $g\Sigma(\tau)$ vs. $\tau$, (c) the number of particles $gN(\tau)$ vs. $\tau$,
(d) the particle distribution $gN_q(\tau)$ vs. $q$ at $\tau=118.9$ (dashed line)
and $\tau=394.7$ (solid line),  and (e) the ratio of the 
pressure and energy density
$p(\tau)/\varepsilon(\tau)$ vs. $\tau$ for the parameter values $m^2=-1$, $\eta(\tau_0) =
40$, $\dot{\eta}(\tau_0)=0$, $g = 10^{-12}$, $h(\tau_0) =
5.0$. \label{fig7}}
\end{figure}

\begin{figure}
\epsfig{file=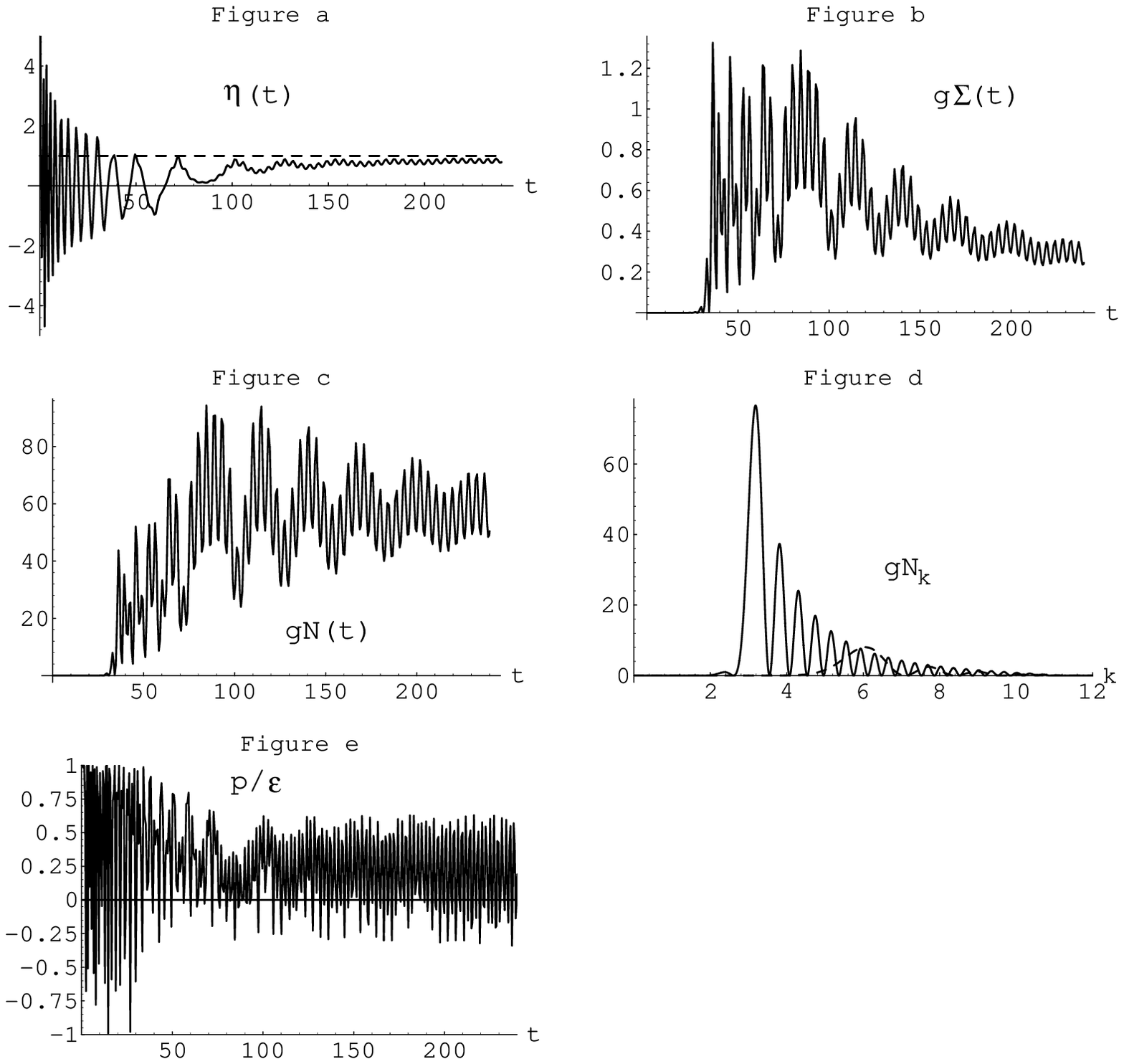,width=6in,height=6in}
\caption{Symmetry broken, chaotic, large $N$, radiation dominated
evolution of (a) the zero mode $\eta(\tau)$ vs. $\tau$, (b) the quantum fluctuation
operator $g\Sigma(\tau)$ vs. $\tau$, (c) the number of particles $gN(\tau)$ vs. $\tau$,
(d) the particle distribution $gN_q(\tau)$ vs. $q$ at $\tau=55.1$ (dashed line)
and $\tau=194.2$ (solid line),  and (e) the ratio of the 
pressure and energy density
$p(\tau)/\varepsilon(\tau)$ vs. $\tau$ for the parameter values $m^2=-1$, $\eta(\tau_0) =
16$, $\dot{\eta}(\tau_0)=0$, $g = 10^{-12}$, $h(\tau_0) =
2.0$. \label{fig8}}
\end{figure}

{\bf Case 3: $m^2>0$, $\eta(\tau_0)\gg 1$}.  The final case we examine is that 
of
a simple chaotic scenario with a positive mass term in the Lagrangian.  
Again,
preheating can begin only after the inflationary phase of exponential 
expansion; 
this allows us to take a zero temperature initial state for 
the FRW stage.

Figure \ref{fig9} shows this situation in the large $N$ approximation for a matter
dominated cosmology.  The zero mode, $\eta(\tau)$, oscillates in time while
decaying in amplitude from its initial value of $\eta(\tau_0\! =\! 0)=5$,
$\dot{\eta}(\tau_0\! =\! 0)=0$ (Fig.~\ref{fig9}a), while the quantum fluctuation,
$g\Sigma$, grows rapidly for early times due to parametric resonance
(Figs.~\ref{fig9}b).  We choose here an initial condition on the zero mode which
differs from that of Figs.~\ref{fig4}--\ref{fig5} 
above since there is no significant growth
of quantum fluctuations for smaller initial values.  From Fig.~\ref{fig9}d, we see
that there exists a single unstable band at values of roughly $q=1$ to
$q=3$, although careful examination reveals that the unstable band 
extends all the way to $q=0$.  The equation of state is depicted by the 
quantity $p(\tau)/\varepsilon(\tau)$ in Fig.~\ref{fig9}e. As expected in this massive 
theory, the equation of state is matter dominated.

\begin{figure}
\epsfig{file=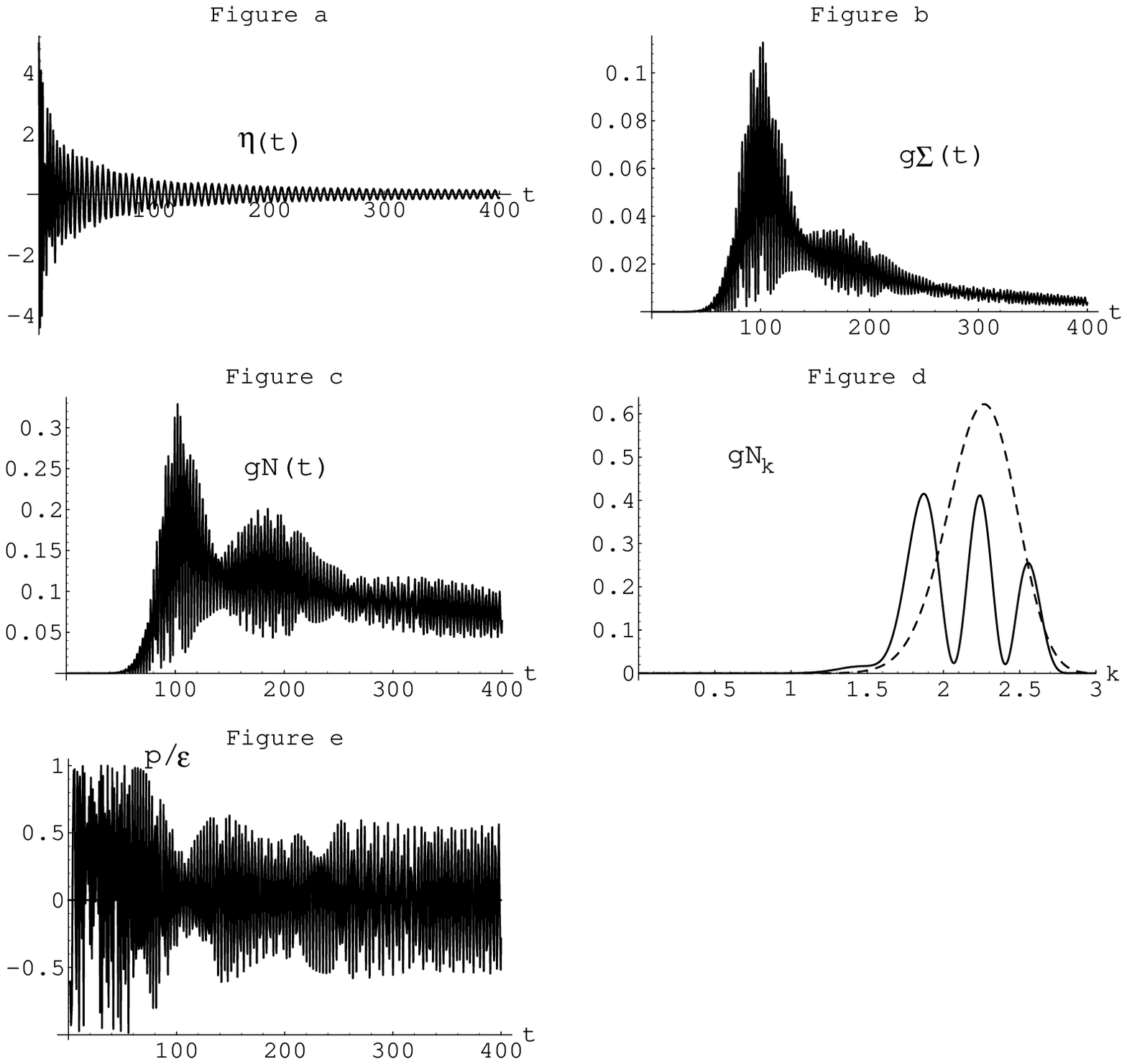,width=6in,height=6in}
\caption{Symmetry unbroken, chaotic, large $N$, matter dominated
evolution of (a) the zero mode $\eta(\tau)$ vs. $\tau$, (b) the quantum fluctuation
operator $g\Sigma(\tau)$ vs. $\tau$, (c) the number of particles $gN(\tau)$ vs. $\tau$,
(d) the particle distribution $gN_q(\tau)$ vs. $q$ at $\tau=77.4$ (dashed line)
and $\tau=399.7$ (solid line),  and (e) the ratio of the 
pressure and energy density
$p(\tau)/\varepsilon(\tau)$ vs. $\tau$ for the parameter values $m^2=+1$, $\eta(\tau_0) =
5$, $\dot{\eta}(\tau_0)=0$, $g = 10^{-12}$, $h(\tau_0) = 0.1$. \label{fig9}}
\end{figure}

The final case is the Hartree approximation, shown in Fig.~\ref{fig10}.  Here,
parametric amplification is entirely inefficient when expansion of the 
universe
is included and we require an initial condition on the zero mode of
$\eta(\tau_0\! =\! 0)=12\sqrt{3}$ to provide even meager growth of quantum 
fluctuations.  We have used a matter dominated gravitational background with 
$h(\tau_0)=0.1$.  We see that while the zero mode oscillates (Fig.~\ref{fig10}a), 
there is little growth in quantum fluctuations (Fig.~\ref{fig10}b) and few particles
produced (Fig.~\ref{fig10}c).  Examining the particle distribution (Fig.~\ref{fig10}d), 
it is found that the bulk of these particles is produced within a single  
resonance band extending from $q \simeq 15$ to $q \simeq 16$.  
This resonance develops at early time during the large amplitude oscillation 
of the zero mode.  These results are explained by a simple resonance band
analysis described below.

\begin{figure}
\epsfig{file=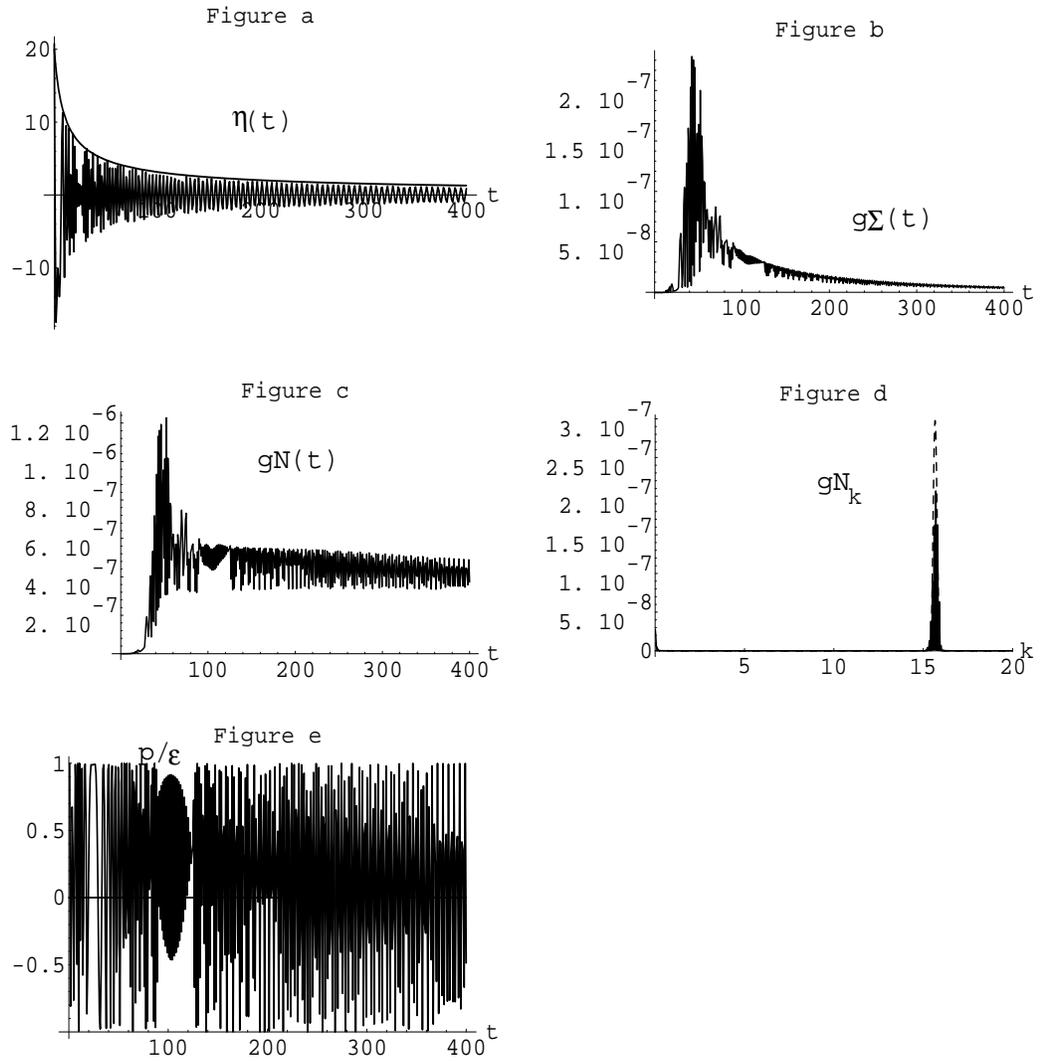,width=6in,height=6in}
\caption{Symmetry unbroken, chaotic, Hartree, matter dominated
evolution of (a) the zero mode $\eta(\tau)$ vs. $\tau$, (b) the quantum fluctuation
operator $g\Sigma(\tau)$ vs. $\tau$, (c) the number of particles $gN(\tau)$ vs. $\tau$,
(d) the particle distribution $gN_q(\tau)$ vs. $q$ at $\tau=50.5$ (dashed line)
and $\tau=391.2$ (solid line),  and (e) the ratio of the 
pressure and energy density
$p(\tau)/\varepsilon(\tau)$ vs. $\tau$ for the parameter values $m^2=+1$, $\eta(\tau_0) =
12\cdot 3^{1/2}$, $\dot{\eta}(\tau_0)=0$, $g = 10^{-12}$, $h(\tau_0) =
0.1$. \label{fig10}}
\end{figure}

At first glance, it is not entirely clear why there are so many more particles
produced in the large $N$ case of Fig.~\ref{fig9} than in the Hartree case of 
Fig.~\ref{fig10}.
Since in the present case the Hubble time is long compared to the oscillation
time scale of the zero mode, $h \ll 1$, we would expect a forbidden band for
early times at the location given approximately by the Minkowski results
provided in Ref. \cite{erice}.  In fact, we find this to be the case.  
However, we know from previous studies that in Minkowski space a similar 
number
of particles is produced in both the Hartree and large $N$ case 
\cite{mink,erice}.

The solution to this problem is inherent in the band structure of the two 
cases
when combined with an understanding of the dynamics in an expanding 
spacetime.
First, we note that, for early times when $g\Sigma \ll 1$, the zero mode 
is well fit by the function $\eta(\tau)=\eta_0 f(\tau)/a(\tau)$ where $f(\tau)$ is an 
oscillatory function taking on values from $-1$ to $1$.  This is clearly seen
from the envelope function $\eta_0/a(\tau)$ shown in Fig.~\ref{fig10}a (recall that 
$g\Sigma \ll 1$ during the entire evolution in this case).  Second, the 
momentum that appears in the equations for the modes (\ref{hartukeq}) is the
{\em physical} momentum $q/a(\tau)$.  We therefore write the approximate 
expressions for the locations of the forbidden bands in FRW by using the
Minkowski results of \cite{erice} with the substitutions 
$\eta_0^2 \to \gamma\eta_0^2/a^2(\tau)$ (where the factor of $\gamma$ accounts
for the difference in the definition of the non-linear coupling between
this study and \cite{erice}) and $q^2 \to q^2/a^2(\tau)$.  

Making these substitutions, we find for the location in comoving momentum 
$q$ 
of the forbidden band in the large $N$ (Fig.~\ref{fig9}) and Hartree 
(Fig.~\ref{fig10}) cases:
\begin{eqnarray}
0 \leq & q^2 & \leq \frac{\eta_0^2}{2}, \; \; (\mbox{large N}) \\
\frac{\eta_0^2}{2}+3a^2(\tau) \leq & q^2 & \leq
a^2(\tau)\left(\sqrt{\frac{\eta_0^4}{3a^4(\tau)}+\frac{2\eta_0^2}{a^2(\tau)}+4}+
1\right). 
\; \; (\mbox{Hartree}) \label{hartband}
\end{eqnarray}
The important feature to notice is that while the location of the unstable
band (to a first approximation) in the case of the continuous $O(N)$ theory
is the same as in Minkowski and does not change in time, the location of 
the band is time dependent in the discrete theory described by the 
non-perturbative Hartree approximation.

While $\eta_0/a(\tau) \gg 1$, the Hartree relation reduces to
\begin{equation}
\frac{\eta_0^2}{2} \leq q^2 \leq \frac{\eta_0^2}{\sqrt{3}}.
\end{equation}
This is the same as the Minkowski result for large amplitude, and one finds
that this expression accurately predicts the location of the resonance band
of Fig.~\ref{fig10}d.  However, with time the band shifts its location toward higher
values of comoving momentum as given by (\ref{hartband}), cutting off 
particle
production in that initial band.  There is continuing particle production 
for higher modes, but since the Floquet index is decreased due to the reduced
amplitude of the zero mode, since there is no enhancement of production
of particles in these modes (as these modes begin with at most of order $1$
particles), and because the band continues to shift to higher momenta while 
becoming smaller in width, this particle production never becomes 
significant.

We emphasize that this result is not an artifact of the approximations used
but rather reflects an important difference between the behaviors of the
theory with discrete symmetry described by the Hartree approximation and the 
theory with continuous symmetry described by the large $N$ limit.  
In the case of a continuous 
symmetry, the dynamics must satisfy the appropriate Ward identities
corresponding to that symmetry.  The Hartree approximation does not 
respect these identities and is therefore only appropriate for describing
the discrete symmetry.  Likewise, the discrete case is not well described
by the large $N$ approximation.  Mathematically, these differences show
up in the form of the differing band structure described above, and we
conclude that preheating is inefficient in simple models
in which the relevant field obeys an unbroken discrete symmetry.

As in the symmetry broken case of Figs.~\ref{fig4}--\ref{fig6}, 
the equations of motion for
large amplitude and relatively early times are approximately scale invariant.
In Figs.~\ref{fig11}--\ref{fig12} we show the case of the large $N$ 
evolution in a radiation
dominated universe with initial Hubble constants of $h(\tau_0)=5$ and 
$h(\tau_0)=2$ 
respectively with appropriately scaled initial values of the zero mode of
$\eta(\tau_0)=40$ and $\eta(\tau_0)=16$.  Again, the qualitative dynamics remains 
largely unchanged from the case of a smaller Hubble constant.

\begin{figure}
\epsfig{file=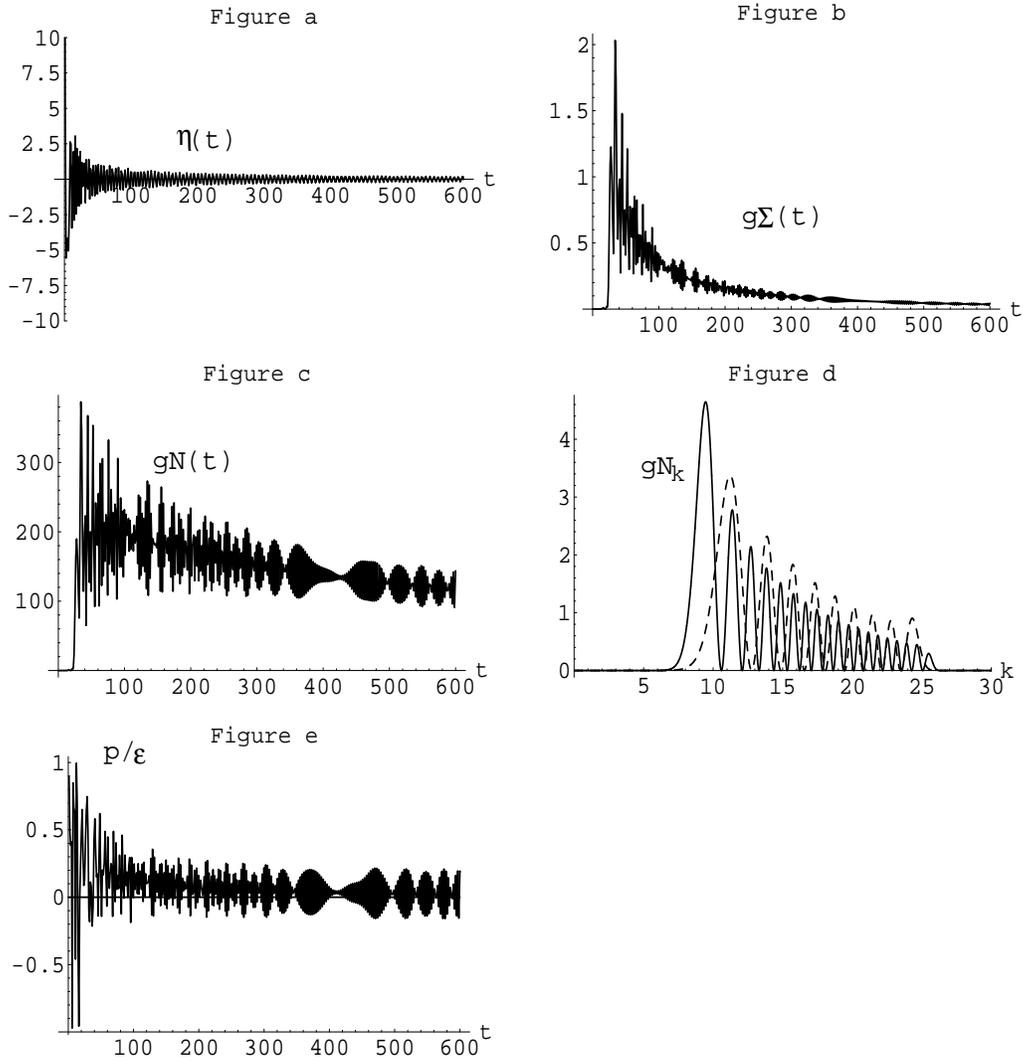,width=6in,height=6in}
\caption{Symmetry unbroken, chaotic, large $N$, radiation dominated
evolution of (a) the zero mode $\eta(\tau)$ vs. $\tau$, (b) the quantum fluctuation
operator $g\Sigma(\tau)$ vs. $\tau$, (c) the number of particles $gN(\tau)$ vs. $\tau$,
(d) the particle distribution $gN_q(\tau)$ vs. $q$ at $\tau=117.3$ (dashed line)
and $\tau=393.6$ (solid line),  and (e) the ratio of the 
pressure and energy density
$p(\tau)/\varepsilon(\tau)$ vs. $\tau$ for the parameter values $m^2=+1$, $\eta(\tau_0) =
40$, $\dot{\eta}(\tau_0)=0$, $g = 10^{-12}$, $h(\tau_0) = 5.0$. \label{fig11}}
\end{figure}

\begin{figure}
\epsfig{file=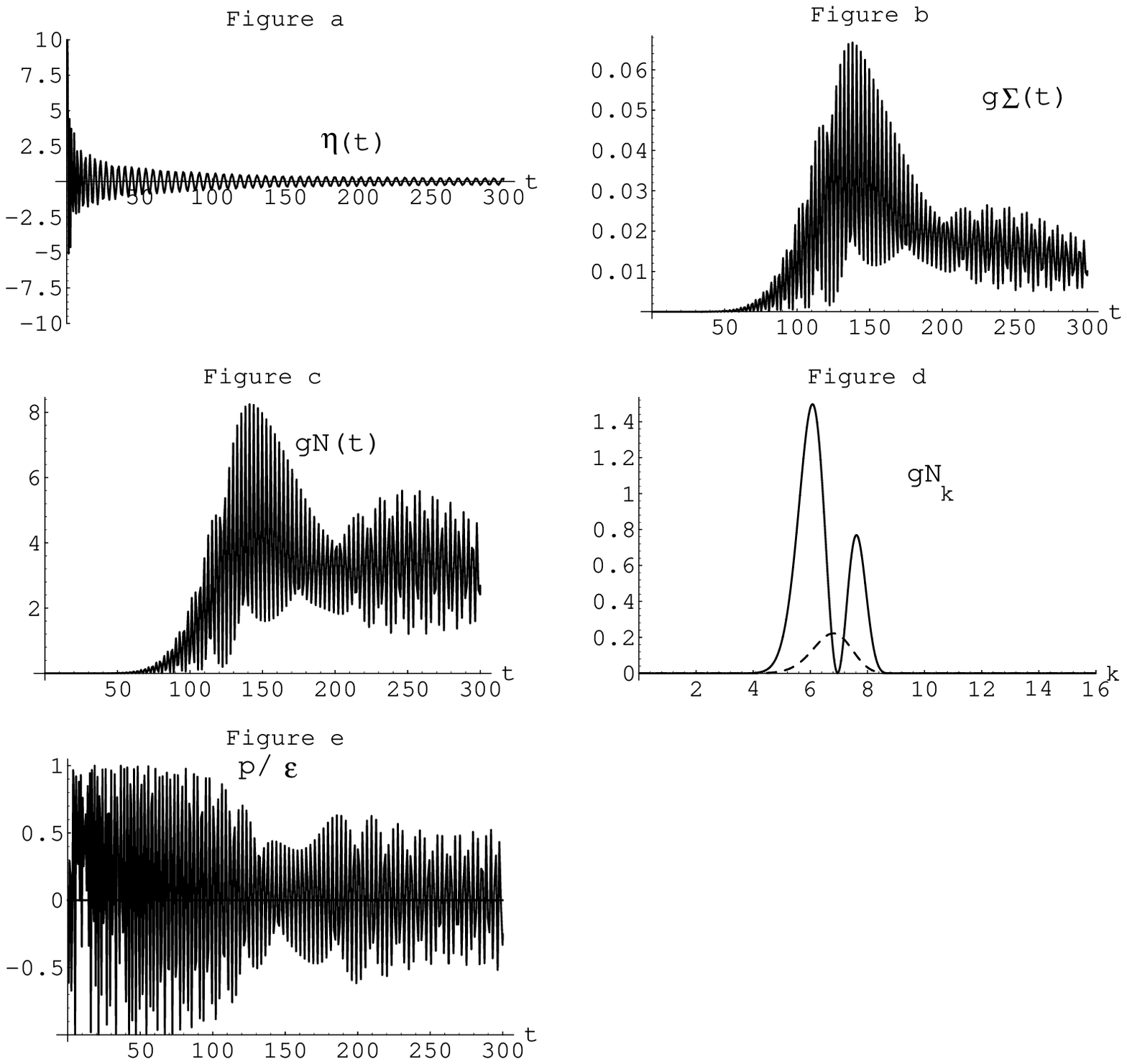,width=6in,height=6in}
\caption{Symmetry unbroken, chaotic, large $N$, radiation dominated
evolution of (a) the zero mode $\eta(\tau)$ vs. $\tau$, (b) the quantum fluctuation
operator $g\Sigma(\tau)$ vs. $\tau$, (c) the number of particles $gN(\tau)$ vs. $\tau$,
(d) the particle distribution $gN_q(\tau)$ vs. $q$ at $\tau=102.1$ (dashed line)
and $\tau=251.6$ (solid line),  and (e) the ratio of the 
pressure and energy density
$p(\tau)/\varepsilon(\tau)$ vs. $\tau$ for the parameter values $m^2=+1$, $\eta(\tau_0) =
16$, $\dot{\eta}(\tau_0)=0$, $g = 10^{-12}$, $h(\tau_0) = 2.0$. \label{fig12}}
\end{figure}

\subsection{Late Time Behavior}
We see clearly from the numerical evolution that in the case of a symmetry
broken potential, the late time large $N$ solutions obey the sum rule
(\ref{sumrule}).
Because of this sum rule, we can write down the
analytical expressions for the late time behavior of the fluctuations and the
zero mode.  Using (\ref{sumrule}), the mode equation (\ref{hartukeq}) becomes
\begin{equation}
\left[\frac{d^2}{dt^2}+3\frac{\dot{a}(t)}{a(t)}\frac{d}{dt}+
\frac{k^2}{a^2(t)}
\right]f_k(t) = 0.
\label{lateuk}
\end{equation}
This equation can be solved exactly if we assume a power law dependence for 
the
scale factor $a(t) = (t/t_0)^n$.  The solution is
\begin{equation}
f_k(t) = c_k \; t^{(1-3n)/2} \; J_{\frac{1-3n}{2-2n}}\left(\frac{k t_0^n
	t^{1-n}}{n-1}\right) + 
	d_k \; t^{(1-3n)/2} \; Y_{\frac{1-3n}{2-2n}}\left(\frac{k t_0^n
	t^{1-n}}{n-1}\right), 
\label{latesoln}
\end{equation}
where $J_{\nu}$ and $Y_{\nu}$ are Bessel and Neumann functions respectively, 
and the constants $c_k$ and $d_k$ 
carry dependence on the initial conditions and on the dynamics up to the 
point
at which the sum rule is satisfied.  

These functions have several important properties.  In particular, in 
radiation
or matter dominated universes, $n<1$, and for values of wavenumber satisfying
$k \gg t^{-(1-n)}/t_0^n$, the mode functions decay in time as 
$1/a(t) \sim t^{-n}$.  Since
the sum rule applies for late times, $\tau-\tau_0 \gg 1$, 
we see
that all values of $k$ except a very small band about $k=0$ redshift as
$1/a(t)$.  The $k=0$ mode, however, remains constant in time, explaining the
support evidenced in the numerical results for values of small $k$ (see
Figs.~\ref{fig1} and \ref{fig3}).  
These results mean that the quantum fluctuation has a late time
dependence of $\langle \psi^2(t)\rangle_R \sim 1/a^2(t)$.  The late time
dependence of the zero mode is given by this expression combined with the sum
rule (\ref{sumrule}).  These results are accurately reproduced by our 
numerical
analysis.  Note that qualitatively this late time dependence is independent 
of
the choice of initial conditions for the zero mode, except that there is no
growth of modes near $k=0$ in the case in which particles are produced via
parametric amplification (Figs.~\ref{fig4} and \ref{fig5}).

For the radiation $ n= \frac12 $ and matter dominated $ n = \frac23 $
universes, Eq.~(\ref{latesoln}) reduces to elementary functions:
\begin{eqnarray}
a(t) \; f_k(t) &=&  c_k \; e^{2ik t_0^{1/2} t^{1/2} } + 
d_k\; e^{-2ik t_0^{1/2} t^{1/2} }  
\;  \; \mbox{(RD) } \; ,\cr \cr
a(t) \; f_k(t) &=&  c_k \; e^{3ik t_0^{2/3} t^{1/3} } \; 
\left[ 1 + {i \over {3k t_0^{2/3}
t^{1/3}}}\right] \nonumber \\
&+& d_k\; e^{-3ik t_0^{2/3} t^{1/3}}\left[ 1 - {i \over {3k
t_0^{2/3} t^{1/3}}}\right] \; \; \mbox{ (MD) }.
\end{eqnarray}

It is also of interest to examine the $n>1$ case.  Here, the modes of 
interest
satisfy the condition $k \ll t^{n-1}/t_0^n$ for late times.  These modes 
are constant
in time and one sees that the modes are {\it frozen}.  In the case of a de
Sitter universe, we can formally take the limit $n \to \infty$ and we see 
that
{\it all} modes become frozen at late times.  This case was studied in 
detail 
in Ch.~4.

\section{Conclusions}

We have shown that there can be significant particle production through 
quantum
fluctuations after inflation.  However, this production is somewhat sensitive
to the expansion of the universe. From our analysis of the equation of state,
we see that the late time dynamics is given by a matter dominated cosmology.
We have also shown that the quantum fluctuations of the inflaton decay for 
late
times as $1/a^2(t)$, while in the case of a symmetry broken inflationary 
model,
the inflaton field moves to the minimum of its tree level potential.  The
exception to this behavior is the case when the inflaton begins exactly at 
the
unstable extremum of its potential for which the fluctuations grow out to the
minimum of the potential and do not decay.  Initial production of particles 
due
to parametric amplification is significantly greater in chaotic scenarios 
with
symmetry broken potentials than in the corresponding theories with positive
mass terms in the Lagrangian, given similar initial conditions on the zero 
mode
of the inflaton.

Since there are a number of articles in the literature treating the problem 
of
preheating, it is useful to review the unique features of the present work.
First, we have treated the problem {\em dynamically}, without using the
effective potential (an equilibrium construct) to determine the evolution.
Second, we have provided consistent non-perturbative calculations of the
evolution to bring out some of the most relevant aspects of the late time
behavior.  In particular, we found that the quantum backreaction naturally
inhibits catastrophic growth of fluctuations and provides a smooth transition
to the late time regime in which the quantum fluctuations decay as the zero
mode approaches its asymptotic state.  Third, the dynamics studied obeys the
constraint of covariant conservation of the energy momentum tensor.

As we have seen, significant particle production may occur due to parametric
amplification after inflation.  However, this production depends strongly
on the expansion rate and is effectively cut off for values of the expansion
rate $h(t_0) \geq 1$.

\chapter{Conclusions}

Presently, inflation is the most promising paradigm for solving
the horizon, flatness, and relic problems, as well as
the fluctuation problem.  However, it remains a paradigm in search of
foundation with literally thousands of models, none of which are
particularly well motivated by an overall theory of particle physics.
As new observational data on cosmological parameters become available,
we should begin to see the elimination of large classes of these
models, and there is certainly the potential that a small class of
models will gain favor.  The ideal situation, of course, would be for
such clues as to the nature of inflation to provide guidance as to the
nature of particle physics in general at energies approaching the
Grand Unified scale.

This is an ambitious goal and will not only require unprecedented
observational tools such as the MAP and PLANCK satellites, but also
complete utilization of all available theoretical tools.  Observations
may provide values of cosmological parameters to an accuracy of one
percent within the next decade, and it will be necessary to have
confidence in theoretical predictions of these parameters for various
cosmological models to a similar accuracy.  It therefore becomes
necessary to question the methods that have been utilized during the
past 20 years as these methods were required to provide only order of
magnitude predictions.

Hence, we have undertaken this study of the non-equilibrium aspects of
inflationary theory, and we have discovered a number of important
properties of inflationary dynamics.  Of particular note is the
construction of a classical background field from the quantum inflaton
in new inflation models, which allows us to make confident predictions
of the primordial density perturbations.  Also of primary importance
is the discovery that explosive particle production may not be
significant if the expansion rate immediately after inflation is of
the order or larger than the inflaton mass scale.  In addition, we
have shown how the symmetry properties of the underlying inflationary
model may have a significant influence on the importance of explosive
particle production in an expanding spacetime.

While we have used the powerful tools of non-equilibrium field theory
to study one of the most exciting areas in cosmology, we note that
there are a number of other problems to which these techniques are
particularly well suited.  Baryogenesis is a striking example, since
it cannot take place at all in equilibrium.  A thorough out of
equilibrium study is bound to provide new insight into this process,
which results in the production of the bulk of the matter with which
we are familiar.

Another process taken from cosmology is topological defect formation.
Topological defects represent the only known competition to inflation
as the source of primordial density perturbations.  And while they are
presently out of favor as candidates for seeding structure, they will
remain important because they are almost certain to have formed during
the evolution of the universe.  They may be a source of dark matter,
or the necessity of mechanisms which reduce their number and
importance may provide important clues about the history of our
universe.  

The potential uses of non-equilibrium field theory outside of
cosmology are just as numerous, with many applications in condensed
matter physics as well as particle physics, most of which remain
unexplored.

\vskip .2in

This is an exciting time for cosmology and physics.  We are now just
beginning to discover the answers to some of the most profound
questions that can be asked:
\begin{itemize}
\item Is the universe open or flat?
\item What kind of physics is there beyond the Standard Model of Particles?
\item What is the nature of the dark matter?
\item Is the universe accelerating?
\item What would it mean to have a significant cosmological constant?
\item How old is the universe?
\item What does this all mean for the physics of the early universe?
\end{itemize}
There is only one thing for certain: there will be surprises!

\end{document}